# Phase transitions and amorphization of $M_2AgF_4$ (M = Na, K, Rb) compounds at high pressure

Jakub Gawraczyński[1]*, Łukasz Wolański[1], Adam Grzelak[1]*, Zoran Mazej[2], Viktor V. Struzhkin[3] and Wojciech Grochala[1]

[1]*Centre of New Technologies, University of Warsaw, Banacha 2C, 02-097, Warsaw, Poland*
[2]*Department of Inorganic Chemistry and Technology, Jožef Stefan Institute, Jamova cesta 39, SI-1000, Ljubljana, Slovenia*
[3]*Center for High Pressure Science and Technology Advanced Research, Shanghai 201203, China*

**Abstract**

We report the results of Raman spectroscopy high-pressure studies of alkali metal fluoroargentates ($M_2AgF_4$, where M=Na, K, Rb) associated with theoretical and x-ray diffraction studies for the K member of the series. Theoretical density functional calculations predict two structural phase transitions for $K_2AgF_4$: one from low pressure monoclinic $P2_1/c$ (beta) phase to intermediate-pressure tetragonal $I\bar{4}2d$ structure at 6 GPa, and another to high-pressure triclinic $P\bar{1}$ phase at 58 GPa. However, Raman spectroscopy and X-ray diffraction data indicate that both polymorphic forms of $K_2AgF_4$ as well as two other fluoroargentate phases studied undergo amorphization at pressure as low as several GPa.

1. Introduction

Alkaline metal fluoroargentate $M_2AgF_4$ family (M = Na, K, Rb) [1]are a group of compounds analogous in many properties to important $La_2CuO_4$ oxocuprate, [2,3] which is a precursor of the first known oxocuprate superconductor [4]. Specifically, $[AgF_2]$ layers are isoelectronic with $[CuO_2]$ planes, both hosting formally one hole in the $d^{10}$ set of the transition metal [5–7]. Ambient-pressure crystal structures and magnetic properties of these compounds have been recently explored in thorough experimental and computational studies [8,9].

At ambient pressure, double fluoroargentates crystallize in different space groups depending on the alkali metal cation embedded in the structure. $Na_2AgF_4$ adopts a monoclinic $P2_1/c$ post-perovskite structure [10], while $Rb_2AgF_4$ adopts an orthorhombic layered double-perovskite structure [11]. Under ambient pressure and room temperature, $K_2AgF_4$ can – depending on the route of synthesis – adopt either the lower-enthalpy $P2_1/c$ structure (β-$K_2AgF_4$ polymorph [12]), or the higher-enthalpy metastable orthorhombic $Cmce$ structure (disordered α-$K_2AgF_4$) [1]. The effect of high-pressure on double fluoroargentates, or an effect of chemical pressure from the smaller alkaline metal cations (Na, K), have been predicted to lead to structural phase transition to more closely-packed β structure [12].

The purpose of this study is to determine the behavior of $Na_2AgF_4$ and isostructural β-$K_2AgF_4$, as well as α-$K_2AgF_4$ and isostructural $Rb_2AgF_4$, at high external pressure, using a combination of theoretical and experimental methods.

2. Experimental methods
2.1. Synthesis

Powder samples of K and Rb fluoroargentates were prepared at high temperature using anhydrous alkaline metal fluorides and silver difluoride as described in the literature [1]. Platinum boats enclosed in nickel reactors were used to handle reactive specimen. The solid substrates were loaded in argon-filled glove box, with residual water content lower than 2 ppm. $AgF_2$, which was used as a substrate in all syntheses, was prepared using a previously described method [13]. β-K$_2$AgF$_4$ polymorph was obtained via thermal annealing of the α-form at 180°C for 6 hrs [12].

Sodium analogue was obtained using a somewhat similar synthetic pathway as the one reported for high-purity KAgF$_3$ [14] i.e. via a trivalent silver intermediate.

$$2\ MF + AgF_2 \xrightarrow{T} M_2AgF_4 \quad \text{(M= K, T=480 °C; M=Rb, T=400 °C)}$$

$$2\ NaF + AgF_2 \xrightarrow{aHF,\ KrF_2} NaF + NaAgF_4 \xrightarrow{400\ °C - 550\ °C,\ vacuum} Na_2AgF_4$$

$$\alpha\text{-}K_2AgF_4 \xrightarrow{180\ °C,\ 6\ hours} \beta\text{-}K_2AgF_4$$

### 2.2. Instrumental methods

All Raman spectra were measured using T64000 spectrometer with LN2-cooled CCD detector. A SpectraPhysics Ar,Kr gas laser was used to provide a 514.5 nm excitation line. A confocal microscope with 200 μm aperture was used in all experiments. 300 l/mm diffraction grating was used in every experiment. Laser power lower than 5 mW was used. Rayleigh-scattered light was cut off using low-pass edge filter.

Pressure was determined in all cases using Ruby2020 gauge [15], and in the XRD experiments additionally crosschecked with pressure calculated using position of (111) reflection of gold [16]. Very decent agreement was observed and ruby scale was consistently applied. IIAS diamond anvil pair with 250 μm or 300 μm culets were used. Stainless steel 250 μm thick gaskets were indented by compression up to about 20 GPa, after which 100 μm diameter holes were drilled using tungsten carbide drills. Thin slices of FEP (fluorinated ethylene-propylene) foil were used as pressure medium in all measurements, except for Na$_2$AgF$_4$, where tightly pressed dry NaF powder (Sigma Aldrich) was used for this purpose. All samples were loaded in argon-filled glove box, with residual water content lower than 1 ppm. Raman measurements were conducted for samples enclosed in Almax Diacell SymmDAC60, whereas XRD diffraction patterns were obtained for samples enclosed in Almax One20DAC.

Laser heating of the samples was not used, since they tend to decompose with elimination of F$_2$; moreover, when laser heated, samples are extremely reactive with respect to diamond and gasket.

XRD diffraction patterns were collected at 293 K using SuperNova Single Source Rigaku Oxford diffractometer with laboratory source, an Ag lamp (λ = 0.56087 Å). Due to small intensity of signals from the sample, the scans were conducted for 2θ<45°.

### 3. Computational methods

Computational exploration of structures of K$_2$AgF$_4$ at high pressure was conducted using the following method. First, candidates for high-pressure polymorphs in the 0-100 GPa range were selected by learning algorithms implemented in XtalOpt r11.0 [17,18] and additionally by modifying the proposed high-pressure structures of Ag$_3$F$_4$ [19] via substitution of Ag(I) with K(I); such substitution is justified because Pauling ionic radii of silver and potassium cations are quite similar (Ag(I): 1.26 Å, K(I): 1.33 Å). DFT (PBE) geometry optimization with cut-off energy of 950 eV and self-consistent-field convergence criterion of 10$^{-6}$ eV per atom was then carried out, yielding a set of candidate structures. (VASP software was utilized for this purpose [20–24].) Further optimization of geometry using DFT+U method (U = 5.5 eV, J = 1 eV) [25] with PBE functional adapted for solids (PBEsol [26]) was carried out for a range of different ferromagnetic and antiferromagnetic models. Finally, one minimum-enthalpy structure was selected for each external pressure point. This structure was then used for additional single point calculations for several different magnetic models, to estimate the strength of magnetic superexchange using DFT+U method (*cf.* ESI). A typical density of the k-point grid was 0.04 Å$^{-1}$.

Since learning algorithms produce *P*1 structures, symmetry-recognition routines were applied. Space groups for unit cells presented in Supplemental Material (SM) were determined with accuracy of 0.05 Å.

## 4. Results
### 4.1. Computational results

We begin by discussing theoretical results obtained for $K_2AgF_4$. Potassium compound was selected for the theoretical study because it is the only fluoroargentate(II) which exhibits polymorphism in the absence of external pressure. Therefore, it may be used to validate accuracy of computational approach and conclusions from its study may be qualitatively applied to systems with smaller (Na) or larger (Rb) alkali metal cations.

Learning algorithms, used for structure prediction of $K_2AgF_4$ at ambient and elevated pressure up to 100 GPa, produced a large number of structures. However, many of those corresponded (within error margins) to the same few structure types. Moreover, only a few structures were relevant to the phase diagram in terms of their enthalpy. Therefore, only the most important five polymorph candidates (labeled from **A** to **E**) were described here; label of each structure also contains letter f (for ferromagnetic) or af (for antiferromagnetic ordering) as typical for each structure in its magnetic ground state. Their crystal structures and structural parameters are shown and listed in Electronic Supplement (ES).

*Inter alia*, using manual feed of XtalOpt we have considered the *A*mmm form which was proposed as a high-pressure structure of $K_2CuF_4$ [27,28], but its enthalpy was always large with respect to those reported here.

The five structures mentioned (*cf.* ES for .cif files) correspond to:

A – the layered double perovskite corresponding to the lowest-energy ordered variant of the disordered experimental α-$K_2AgF_4$ polymorph

B – the monoclinic post-perovskite structure corresponding to experimental β-$K_2AgF_4$ polymorph

C – a chain structure hosting $Ag_2F_7$ dimers interconnected to another chain via F anions

D – another chain structure with a more complex arrangement of $AgF_4$ squares and $AgF_2$ dumbbells

E – structure originating from $Ag(I)_2Ag(II)F_4$ [19] by Ag(I)→K(I) substitution.

Since only three of these structures (A, B, E) are relevant to experiment in terms of possible phase transitions, they are shown jointly in fig. 1 (structures corresponding to 0 GPa are shown). The unit cell vector and volumes of the high-pressure structures are given as a percent value of the 0 GPa theoretical structures (*cf.* SM). The relative enthalpy of the five structures in the 0-100 GPa range is presented in fig. 2.

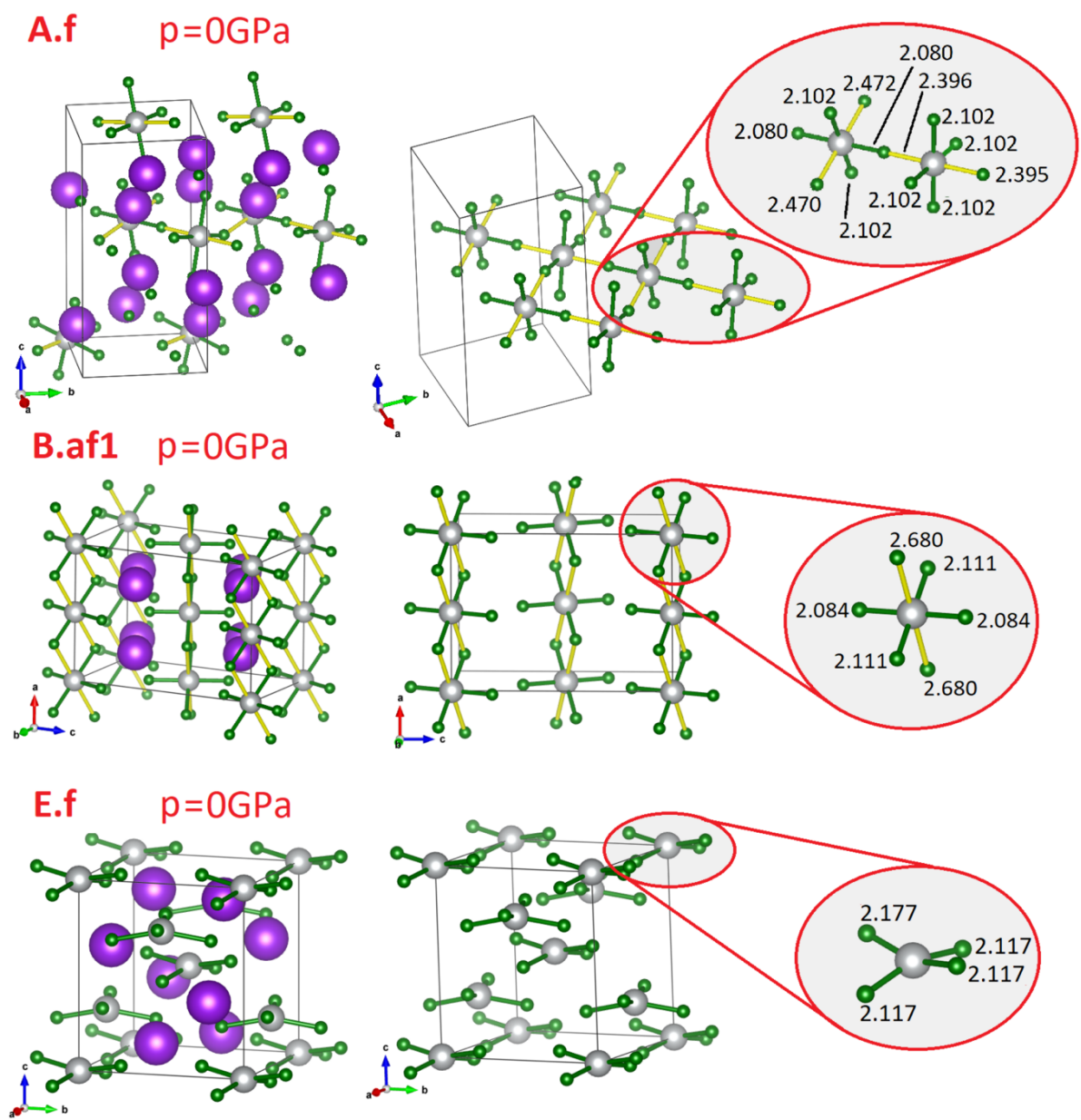

Fig. 1. Crystal structures of three important polymorphs of $K_2AgF_4$ derived from DFT computations. Color code: gray = Ag, green = F, purple = K. Green lines indicate presence of Ag-F bonds shorter than or equal to 2.2 Å, whereas yellow lines show Ag-F distances between 2.2 and 2.8 Å. Only $AgF_4$ sublattice is shown in the right panel, with potassium atoms removed for clarity.

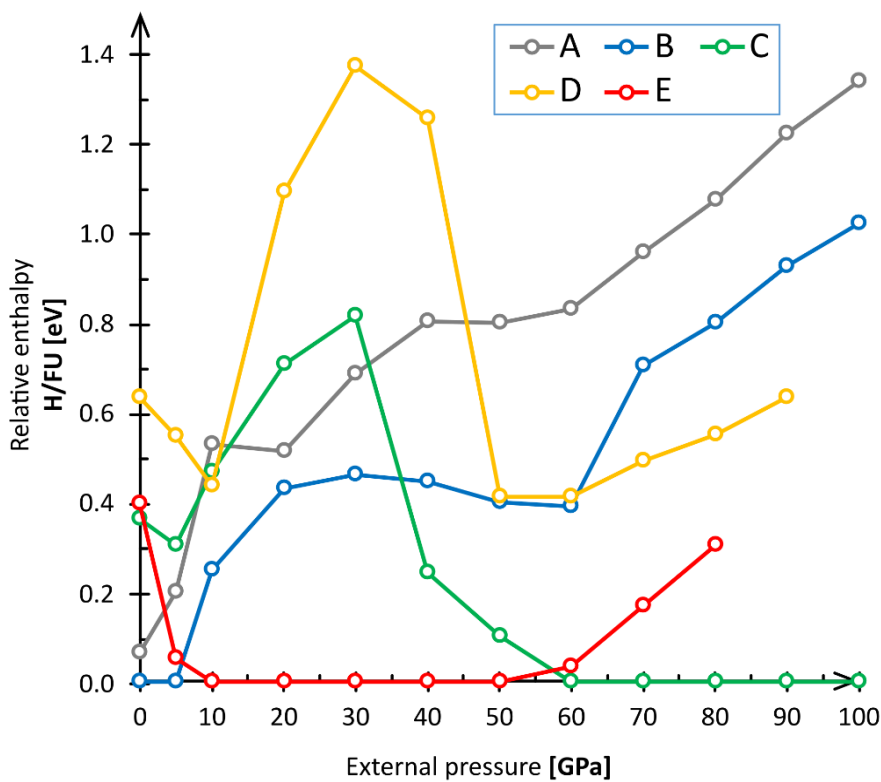

Fig. 2. Relative enthalpy of the five important polymorphs of $K_2AgF_4$ in the 0-100 GPa pressure range. Enthalpy of the most stable polymorph at any given pressure is taken as reference (0 eV) at that pressure.

DFT calculations correctly grasp the key structural features of the α and β polymorphs of $K_2AgF_4$, the associated magnetic properties, as well as their respective stability at 0 GPa [9]. α-$K_2AgF_4$ hosts an antiferrodistortive ordering of tilted $AgF_6$ octahedra, which leads to ferromagnetic ground state [9,29]. On the other hand, the β-$K_2AgF_4$ polymorph features infinite chains composed of isolated $AgF_4$ units which are stacked in such manner that the magnetic superexchange is very weak, and the antiferromagnetic and ferromagnetic solutions are close in energy [29]. Moreover, β-form is more stable at 0 GPa, as evidenced by the facile structural collapse of metastable α-$K_2AgF_4$ in a properly designed experiment [12]. Since β-form is more suited than α-form to accommodate K(I) cation, it is expected that the former should prevail over the latter as pressure is increased. This is indeed what calculations show, the enthalpy of the α-form rising fast with the pressure increase.

β-form is predicted here to be the ground state of $K_2AgF_4$ up to ca. 6 GPa, when it should be substituted by $Ag_2AgF_4$-type polymorph (E); E form, in turn, should be stable to at least 58 GPa, when C form should prevail in enthalpy. Theoretical calculations predict also that pressure increase should lead to progressive cross-linking of the structural elements features in all polymorphs studied, as it is customary at elevated pressure [30–32].

### 4.2. Experimental results

We begin our analysis with the case of α-$K_2AgF_4$, for which both Raman spectra and x-ray diffraction patterns (XRDPs) were obtained as a function of pressure. An analogous set of results was also obtained for β-$K_2AgF_4$.

Raman spectra of both polymorphs of $K_2AgF_4$ are presented in fig. 2, while XRDPs of the α-form are shown in fig. 3.

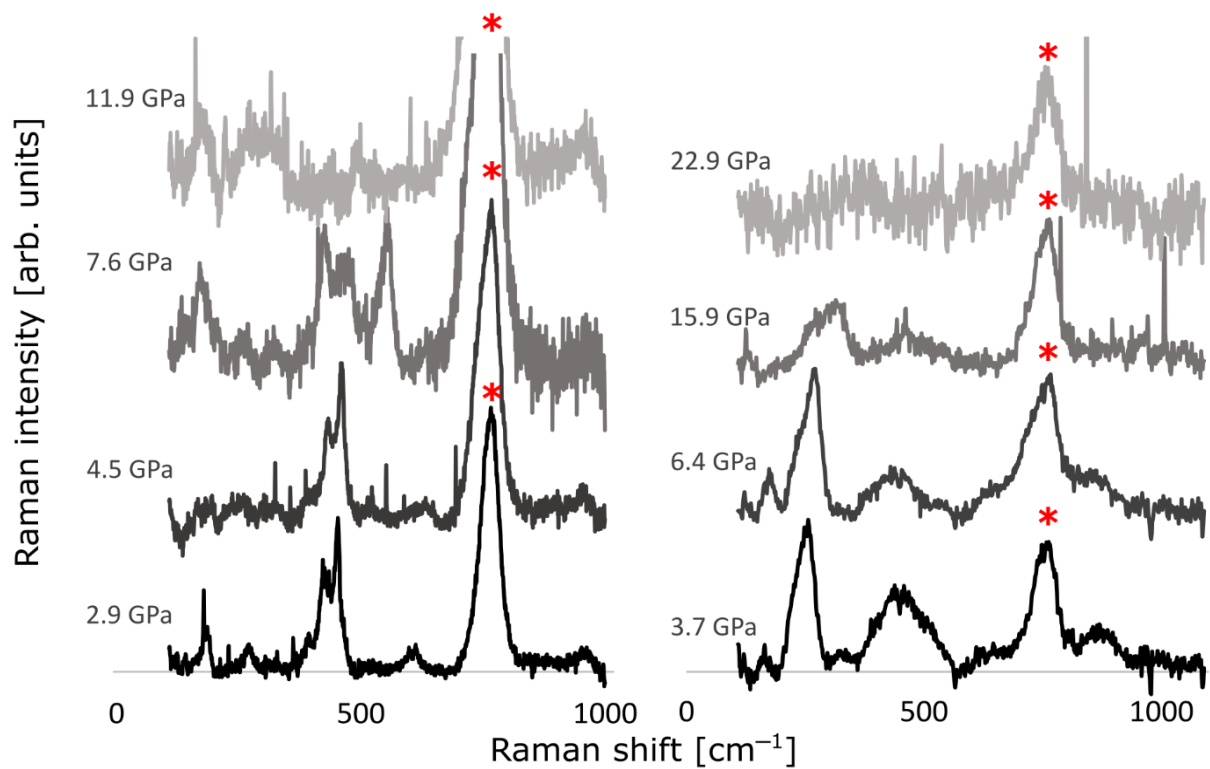

Fig. 2. Raman spectra of α-$K_2AgF_4$ (left) and β-$K_2AgF_4$ (right) measured with 514 nm laser at increasing pressure, with pressure values labelled next to each spectrum. Red asterisk indicates bands originating from FEP used as pressure medium.

Because the sample of α-$K_2AgF_4$ was loaded into DAC along with FEP slices used as an inert hydrostatic medium, the measured Raman spectrum contains bands from both materials. More precisely, the broad, strong band near 760 cm$^{-1}$ is caused by symmetric stretching of C-C-C chains in the polymer, while a shoulder band at 395 cm$^{-1}$ originates from bending of the polymer skeleton – both bands are indicated with red asterisks in respective figures. Further analysis of these FEP bands is omitted here.

Raman spectrum of α-$K_2AgF_4$ at a rather low pressure of 2.9 GPa is reminiscent of that at 1 atm (*cf*. Supporting Information to ref. [12]). The ambient pressure spectrum is predominated by three bands coming from vibrational fundamentals, at 320 cm$^{-1}$ (weak), 415 cm$^{-1}$ (strong) and 476 cm$^{-1}$ (very strong) [[12]. The weakest of these bands is not seen in the spectrum measured at 2.9 GPa but the strongest two are clearly visible at 427 cm$^{-1}$ and 453 cm$^{-1}$. The upshift of the former band by 12 cm$^{-1}$ and the downshift of the latter by 23 cm$^{-1}$ clearly originate from the impact of external pressure. These bands further migrate to 433 cm$^{-1}$ and 461 cm$^{-1}$ (both showing a small upshift) respectively, at 4.5 GPa. A more dramatic effect is seen at 7.6 GPa, when three bands appear in the spectra, at 413 cm$^{-1}$, 470 cm$^{-1}$, and 547 cm$^{-1}$. While the former two might be reminiscent of the bands seen for α-$K_2AgF_4$, the highest-Raman shift band certainly signifies the appearance of a new phase. Its wavenumber is unusually large and it suggests the presence of very short Ag–F bonds, whose stretching mode could give rise to this band.

One is tempted to associate the appearance of the high-wavenumber band with either β-$K_2AgF_4$ (which is certainly more stable then α-$K_2AgF_4$) or even with the **E** structure, predicted to be minimum of enthalpy at this pressure. However, Raman spectra of the β-$K_2AgF_4$ form (fig.2) show that the main broad Raman doublet detected for this form at 437 and 460 cm$^{-1}$ (420 and 486 cm$^{-1}$ at ambient pressure) does not stiffen as pressure is raised, hence it cannot be responsible for the appearance of the band at 547 cm$^{-1}$. Simultaneously, theoretical analysis of Ag–F bond lengths seen for the **A**, **B** and **E** structures suggest that it is polymorph **A**, which hosts the shortest Ag–F bonds, that is observed here. Therefore, the appearance of the 547 cm$^{-1}$ band cannot be explained by the presence of polymorph **E**. The origin of the latter band is not clear at present.

Unfortunately, further analysis of the evolution of the Raman bands for α-K$_2$AgF$_4$ and β-K$_2$AgF$_4$ at higher pressures was precluded by very low signal-to-noise ratio in the spectra measured at 11.9 GPa and 22.9 GPa, respectively. The disappearance of bands indicative of K$_2$AgF$_4$ phases likely reflects amorphization or even decomposition of both samples.

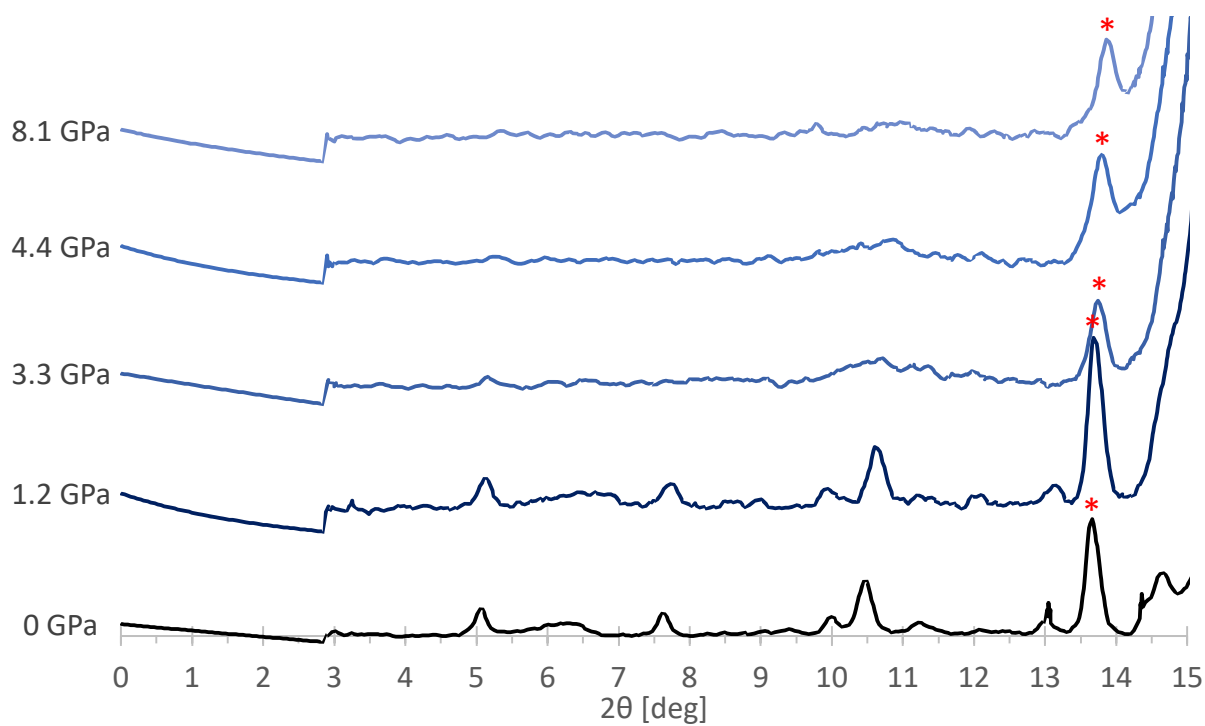

Fig. 3. X-ray diffraction patterns of α-K$_2$AgF$_4$ at increasing pressure, pressure values being labelled next to each pattern. Radiation with wavelength λ=0.56087 Å was applied. The reflexes marked with an asterisk are derived from gold powder used as an additional pressure gauge.

In order to elucidate the nature of the phase transition taking place between 4.5 and 7.6 GPa, we have attempted to obtain additional insight from the evolution of XRDPs of α-K$_2$AgF$_4$ at 2θ > 14° with pressure (fig. 3). Unfortunately, XRDPs are dominated by contributions from DAC elements (diamond, stainless steel gasket) and gold – pressure gauge. Furthermore, due to the size and position of the beamstop in our experimental setup, patterns have been cut off at 2θ < 3°. Therefore, usable information can be extracted from the patterns only for the narrow 3–14° 2θ range. This of course increases the difficulty of the analysis, although some basic information about the crystal structure, i.e. cell parameters, could possibly be deduced. This is because for α-K$_2$AgF$_4$ form at 0 GPa, the five strong reflections are expected in the available 2θ range – these are (002), (111), (020), (200) and (113) (fig. 3). Positions of those reflections can be then used to calculate the cell vectors and volume in the orthogonal system.

In the case of α-K$_2$AgF$_4$ sample, our measurement inside DAC at 0 GPa properly reproduced the expected diffraction pattern of this compound at 1 atm [12]. Increasing pressure to 1.2 GPa yielded a very similar pattern with positions slightly shifted to higher 2θ values, as expected for pressure-induced compression. Precise calculation of the unit cell vectors is complicated by broadening of the reflections. Diffraction patterns measured at still higher pressures of 3.3, 4.4 and 8.1 GPa do not improve the analysis, as most of the α-$K_2AgF_4$ reflections fade away in intensity, and they are hardly discernible at the largest pressure applied here. This observation is consistent with the Raman data and suggests that α-$K_2AgF_4$ likely undergoes amorphization at pressure above *ca.* 8 GPa. In view of this discouraging result, and as indicated by equally poor-quality Raman spectra of the β-form, its XRDPs were not studied.

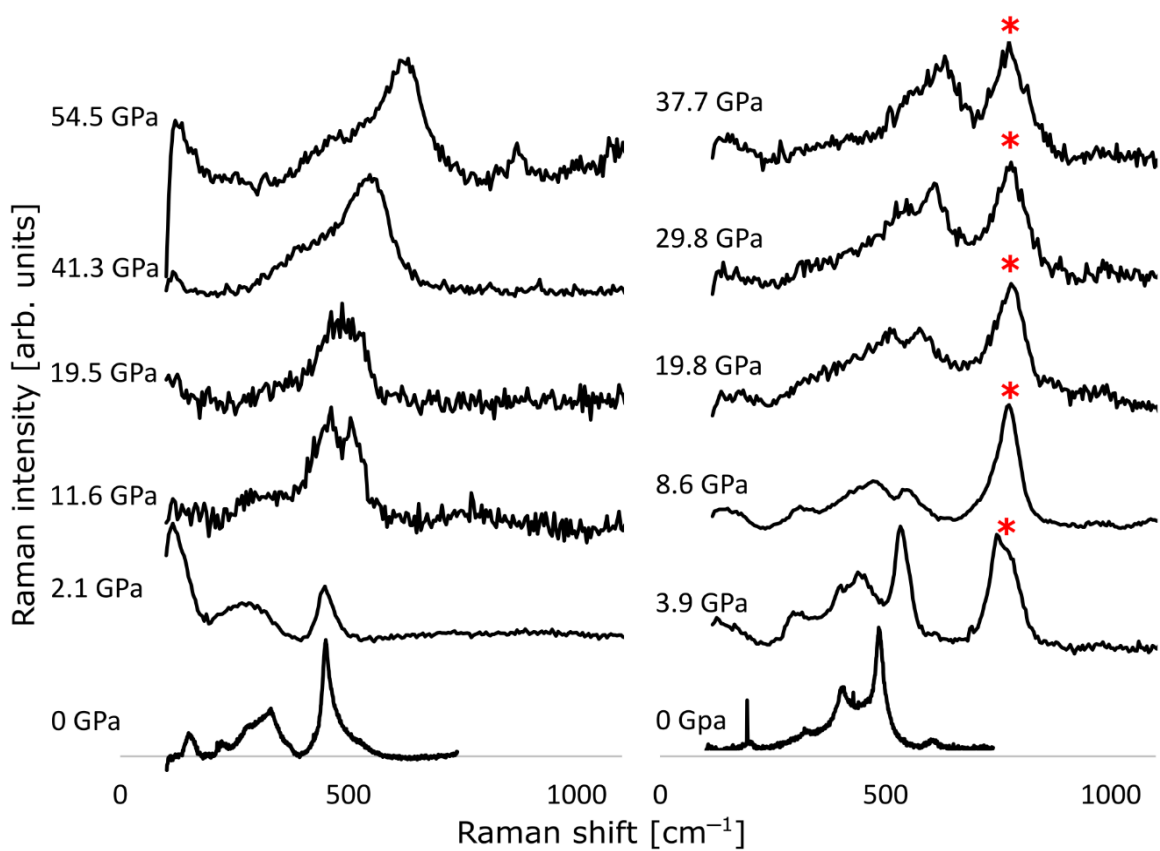

Fig. 4. Raman spectra of $Na_2AgF_4$ (left) and $Rb_2AgF_4$ (right) measured with 514 nm laser at increasing pressure, the pressure values being labelled next to each spectrum. Red asterisk indicates bands originating from FEP used as pressure medium.

Two analogs of $K_2AgF_4$, *i.e.* disodium and dirubidium salts, were also studied using Raman scattering spectroscopy.

The most prominent band in the ambient-pressure Raman spectrum of $Na_2AgF_4$ (fig. 4) is present at 450 cm$^{-1}$. As we have seen, analogous bands are present in Raman spectra of other members of $M_2AgF_4$ family, and the 450 cm$^{-1}$ band corresponds to the totally symmetric vibrations of $[AgF_4]^{2-}$ square subunits. Several other, much weaker bands can be found at 153, 222, 273 (sh), 330, 370 (sh) and 524 (sh) cm$^{-1}$. Compression of the $Na_2AgF_4$ sample leads to a clear decrease in spectral quality associated also with band broadening, and disappearance of all weaker features; only the broad 276 and 458 cm$^{-1}$ features are seen. Between 2.1 GPa and 11.6 GPa the signal-to-noise ratio for all bands decreases even further; two main bands are seen at 463 and 505 cm$^{-1}$. The appearance of the new band might indicate a structural phase transition. Subsequent measurements at even higher pressures up to 54.5 GPa lead to additional broadening of the bands, indicating a possible amorphization of the fluoroargentate. At the highest recorded pressure for which spectra were recorded, only two extremely broad bands are seen, at 640 cm$^{-1}$ and a shoulder at 467 cm$^{-1}$. They likely correspond to the stretching and deformation vibrations, respectively, of the [AgF$_4$]$^{2-}$ units in a disordered structure.

Rb$_2$AgF$_4$ spectrum at ambient pressure (fig. 4) contains two strong/very strong bands at 405 and 485 cm$^{-1}$, as well as several weaker bands at: 193, 320, and 605 cm$^{-1}$. The former two bands correspond to analogous bands seen at 415 cm$^{-1}$ (strong) and 476 cm$^{-1}$ (very strong) in Raman spectrum of α-K$_2$AgF$_4$. Compression of the sample from 0 to 3.9 GPa leads to substantial stiffening of the main bands from 485 to 538 cm$^{-1}$ and from 405 to about 443 cm$^{-1}$. Stiffening of the fundamentals is quite large (> 10%) for a relatively small pressure increment, which together with the fact that the Ag–F bonds are quite incompressible [33] hints at the occurrence of a structural phase transition below 3.9 GPa. Such scenario would be further supported by the appearance of several weaker spectral features in the Ag–F stretching region. Unfortunately, progressive compression up to 37.7 GPa has a similar effect as already seen for the other members of the M$_2$AgF$_4$ family – that is, an immense band broadening which might again suggest amorphization of a sample.

## 5. Conclusions

Summarizing the research done, there is – at the first glance – an apparent discrepancy between theoretical and experimental results. Theory suggest the occurrence of at least two structural pressure-induced phase transitions in the pressure range up to 60 GPa for K$_2$AgF$_4$. Based on rules of thumb of the high-pressure research [31,32], one may expect that these transitions will be present also for Na and Rb analogues, occurring at higher and lower pressures, respectively, than for potassium salt. On the other hand, experiments point to a progressive amorphization of all samples as pressure is increased.

One explanation for this discrepancy could be that the energy barrier for structural transitions predicted here is too large to be overcome at studied p/T conditions. Indeed, it can be noticed that e.g. the low-pressure **B → E** structural transition is associated with breaking up of [AgF$_4$] infinite chains and reorganization of the local [AgF$_4$] square subunits, which adopt a strained (not flat) orientation in the B polymorph. Since the heavy atom sublattice is also strongly affected by the transition, its energy barrier could be large indeed. Note that our experiments were conducted without laser heating of the samples, since such heating would inevitably lead to reaction between extremely reactive Ag(II) salts and DAC elements. As a consequence, phases observed in experiment may not corresponds to the most thermodynamically stable ones (*i.e.* ones which are predicted using theoretical approach). Computations of amorphous systems are certainly possible using programs adapted for periodic systems (e.g. [34]). Regretfully, they require extremely large supercells ("quasi-amorphous periodic systems") and they are currently beyond the possibilities of our supercomputer resources.

It is clear that there are many factors at play which determine high-pressure behavior of fluoroargentates(II), and this will supposedly lead to further experimental and computational studies aimed at elucidating the structural features of the amorphous phases.

**Acknowledgements**

JG would like to thank Polish National Science Centre (NCN) for the Preludium 14 project (2017/27/N/ST5/01066). ZM acknowledges the financial support from the Slovenian Research Agency (research core funding No. P1-0045; Inorganic Chemistry and Technology). Computations has been conducted using supercomputers of the Interdisciplinary Centre for Mathematical and Computational Modelling, ICM, University of Warsaw, within grant no. GA83-34 (SAPPHIRE).

**Phase transitions and amorphization of $M_2AgF_4$ (M = Na, K, Rb) compounds at high pressure**

Jakub Gawraczyński[1]\*, Łukasz Wolański[1], Adam Grzelak[1]\*, Zoran Mazej[2], Viktor V. Struzhkin[3] and Wojciech Grochala[1]

[1]Centre of New Technologies, University of Warsaw, Banacha 2C, 02-097, Warsaw, Poland
[2]Department of Inorganic Chemistry and Technology, Jožef Stefan Institute, Jamova 39, SI-1000, Ljubljana, Slovenia
[3]Center for High Pressure Science and Technology Advanced Research, Shanghai 201203, China

## SUPPLEMENTARY MATERIAL

Structures and crystallographic information files for five computationally studied polymorphs of $K_2AgF_4$

Color legend: gray = Ag, green = F, purple = K). Green lines indicate presence of Ag-F bonds shorter than or equal to 2.2 Å, whereas yellow lines show Ag-F distances between 2.2 and 2.8 Å. Space groups for unit cells were determined with accuracy of 0.05 Å. The unit cell dimensions and volumes of the high-pressure structures are also given as a percent value of the 0 GPa structures.

## A.f    p = 0GPa

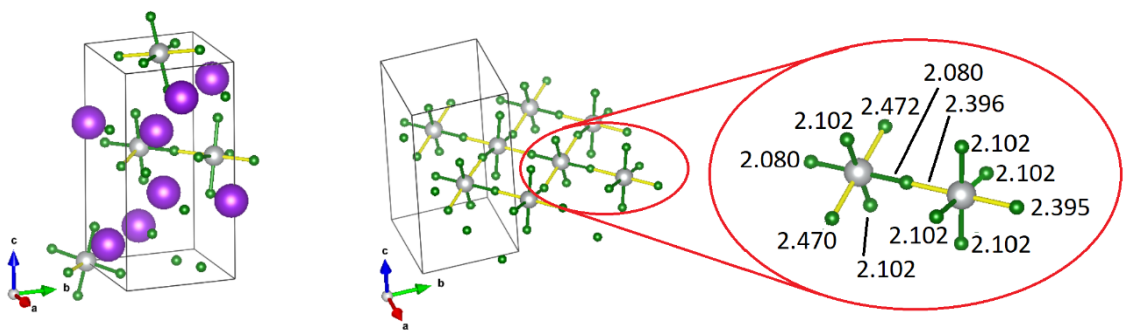

## A.af1    p = 50GPa

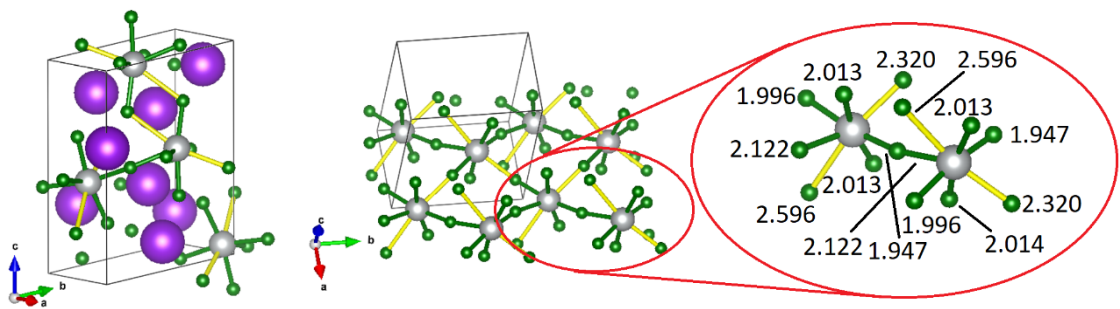

| External pressure [GPa] | 0 | 50 |
|---|---|---|
| Magnetic model | A.f | A.af1 |
| Z | 4 | 4 |
| Symmetry | $P2_1/c$ | $P2_1/c$ |
| $a$ [Å] | 6.328 | 4.800 (75.9%) |
| $b$ [Å] | 6.332 | 6.273 (99.1%) |
| $c$ [Å] | 12.482 | 9.793 (78.5%) |
| $V$ [Å³] | 500.08 | 286.23 (57.2%) |
| $\alpha$ [°] | 90.0 | 90.0 |
| $\beta$ [°] | 90.0 | 76.1 |
| $\gamma$ [°] | 90.0 | 90.0 |

## B.af1  p = 0GPa

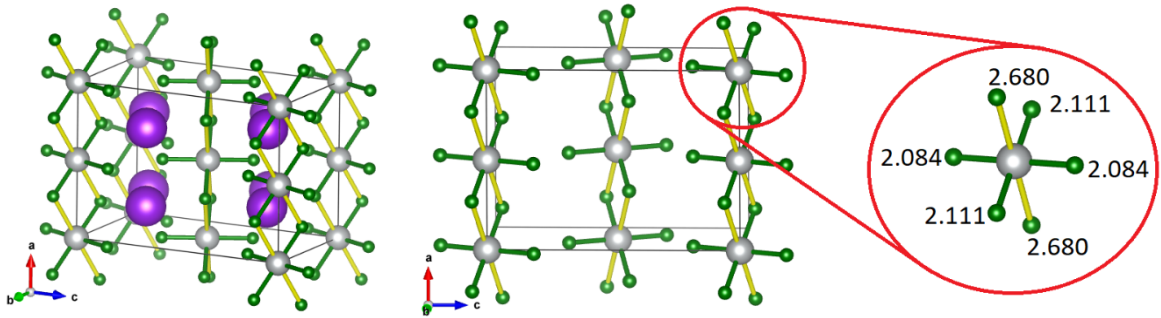

## B.af1  p = 50GPa

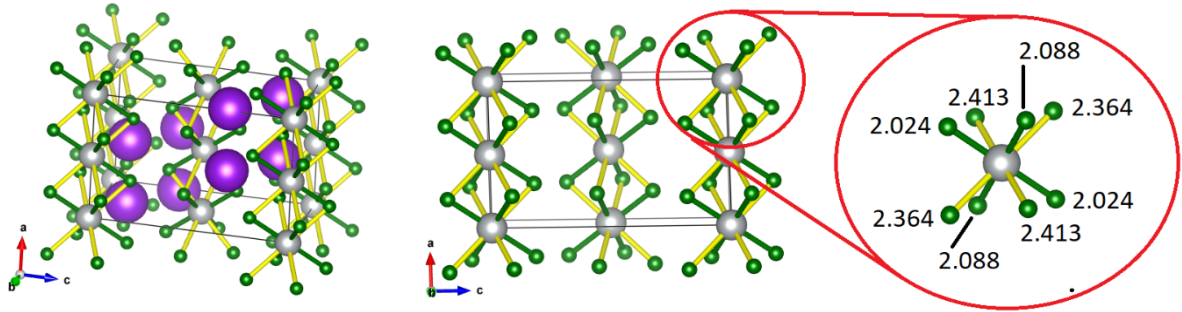

| External pressure [GPa] | 0 | 50 |
|---|---|---|
| Magnetic model | B.af1 | B.af1 |
| Z | 4† | 4† |
| Symmetry | $P2_1/c$ | $P-1$ |
| a [Å] | 7.428 | 5.431 (73.1%) |
| b [Å] | 10.218 | 8.865 (86.8%) |
| c [Å] | 6.377 | 5.941 (93.2%) |
| V [Å³] | 483.21 | 283.46 (58.7%) |
| α [°] | 90.0 | 84.3 |
| β [°] | 88.1 | 95.1 |
| γ [°] | 90.0 | 91.2 |

†Data presented for the supercell – the antiferromagnetic ordering could not be properly described for the for Z=2 cell.

## C.af1   p = 0GPa

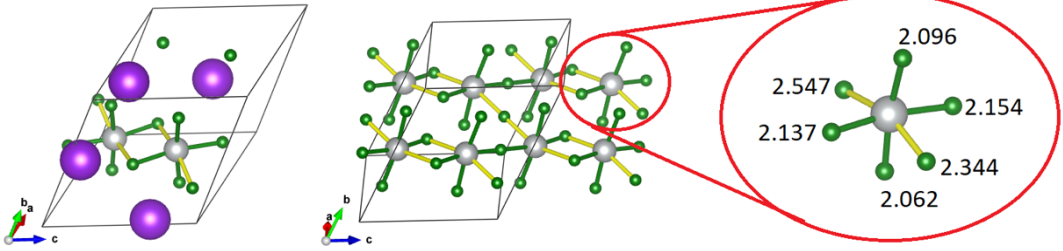

## C.f   p = 50GPa

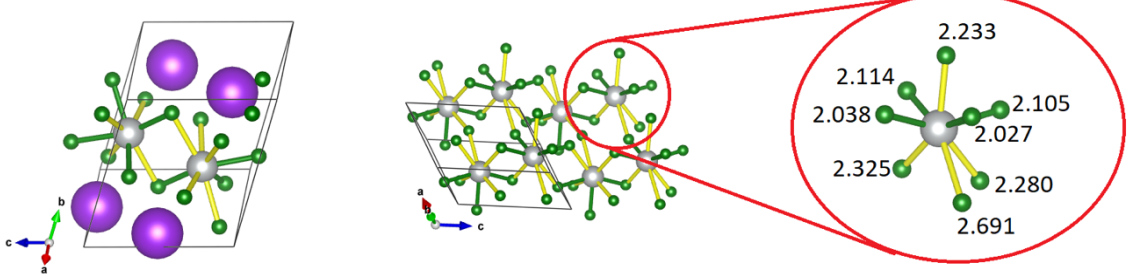

| External pressure [GPa] | 0 | 50 |
|---|---|---|
| Magnetic model | C.af1 | C.f |
| Z | 2 | 2 |
| Symmetry | *P-1* | *P1* |
| *a* [Å] | 6.033 | 4.697 (77.9%) |
| *b* [Å] | 7.098 | 6.020 (84.8%) |
| *c* [Å] | 7.083 | 6.021 (85.0%) |
| *V* [Å³] | 239.38 | 138.76 (58.0%) |
| α [°] | 66.1 | 113.6 |
| β [°] | 67.3 | 103.4 |
| γ [°] | 63.1 | 105.6 |

### D.f   p = 0GPa

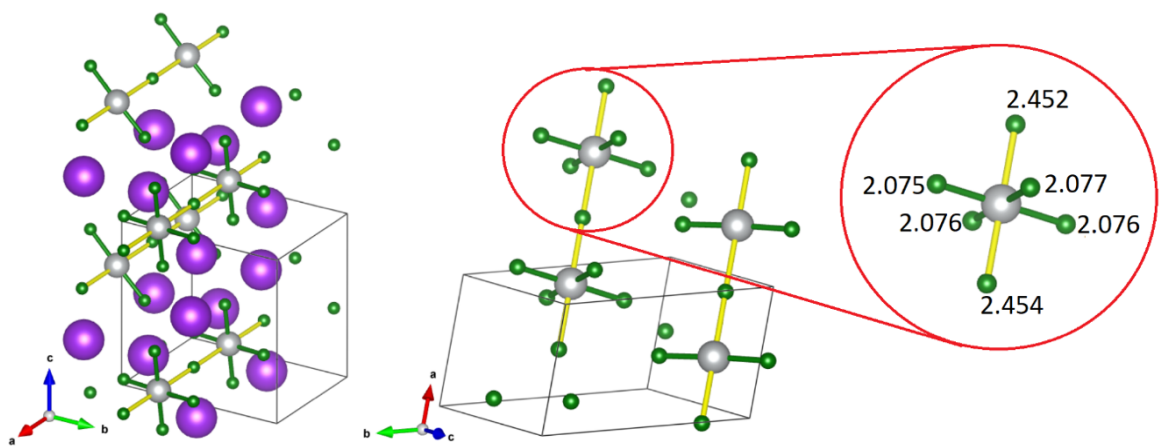

### D.af1  p = 50GPa

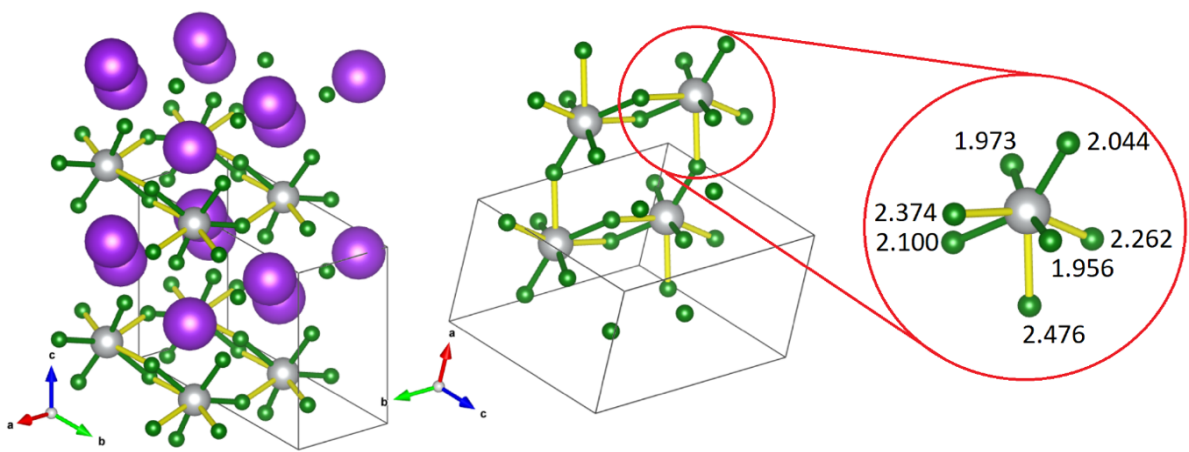

| External pressure [GPa] | 0 | 50 |
|---|---|---|
| Magnetic model | D.f | D.af1 |
| Z | 2 | 2 |
| Symmetry | P1 | P-1 |
| $a$ [Å] | 4.906 | 4.366 (90.0%) |
| $b$ [Å] | 7.521 | 6.555 (87.1%) |
| $c$ [Å] | 7.530 | 6.277 (83.4%) |
| $V$ [Å³] | 234.85 | 142.38 (60.6%) |
| $\alpha$ [º] | 100.8 | 116.8 |
| $\beta$ [º] | 100.9 | 95.5 |
| $\gamma$ [º] | 109.0 | 111.0 |

## E.f p = 0GPa

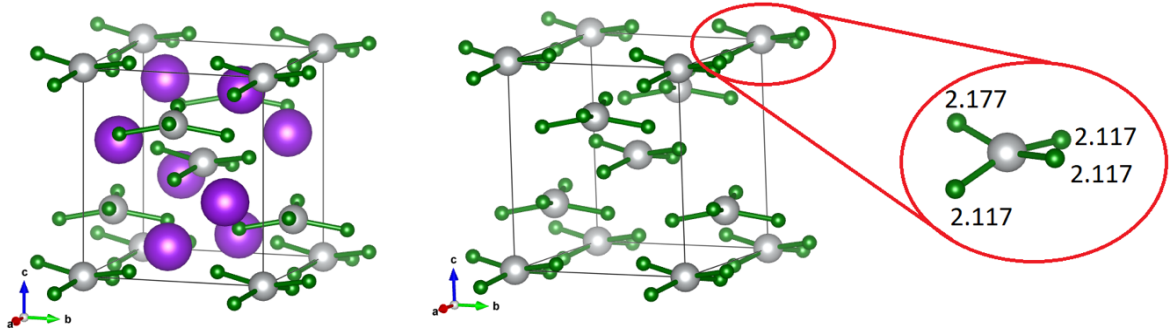

## E.f p = 50GPa

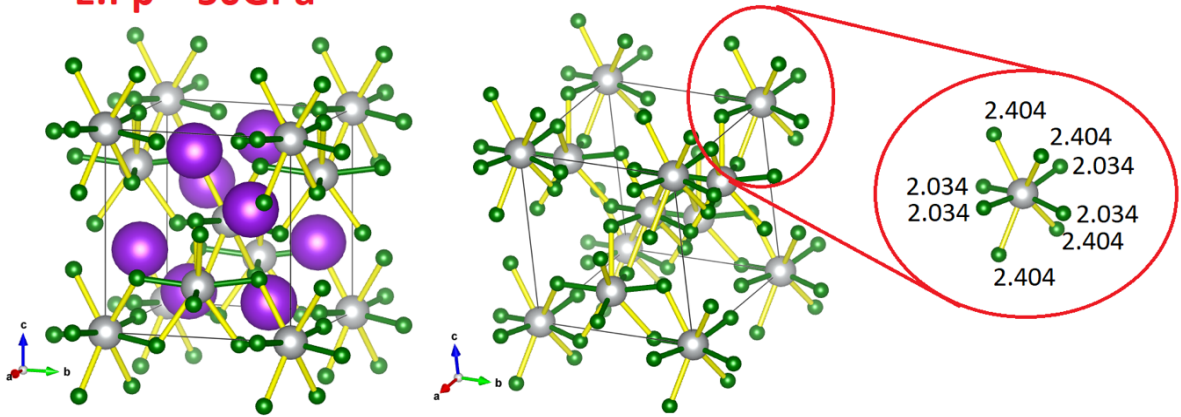

| External pressure [GPa] | 0 | 50 |
| --- | --- | --- |
| Magnetic model | E.f | E.f |
| Z | 4 | 4 |
| Symmetry | I-42d | I-42d |
| a [Å] | 7.296 | 6.450 (88.4%) |
| b [Å] | 7.296 | 6.450 (88.4%) |
| c [Å] | 8.205 | 6.875 (83.8%) |
| V [Å³] | 436.70 | 286.04 (65.5%) |
| α [º] | 90.0 | 90.0 |
| β [º] | 90.0 | 90.0 |
| γ [º] | 90.0 | 90.0 |

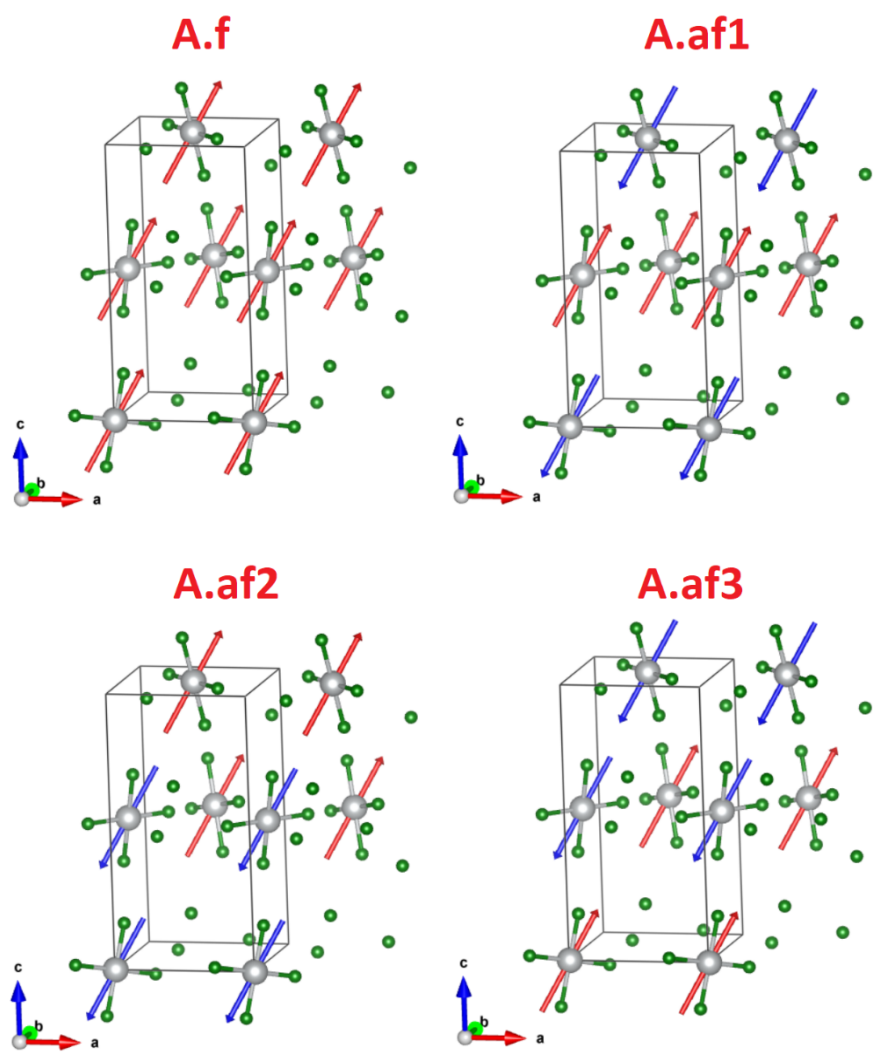

**Figure S1.** Investigated magnetic variants of **K$_2$AgF$_4$ A** structures: ferromagnetic (**A.f**) and antiferromagnetic ones (**A.af1**, **A.af2** and **A.af3**). Only Ag-F sublattices are shown. Ag-F bonds are indicated for interatomic distances not longer than 2.2 Å. All visualized example structures are for p = 0 GPa.

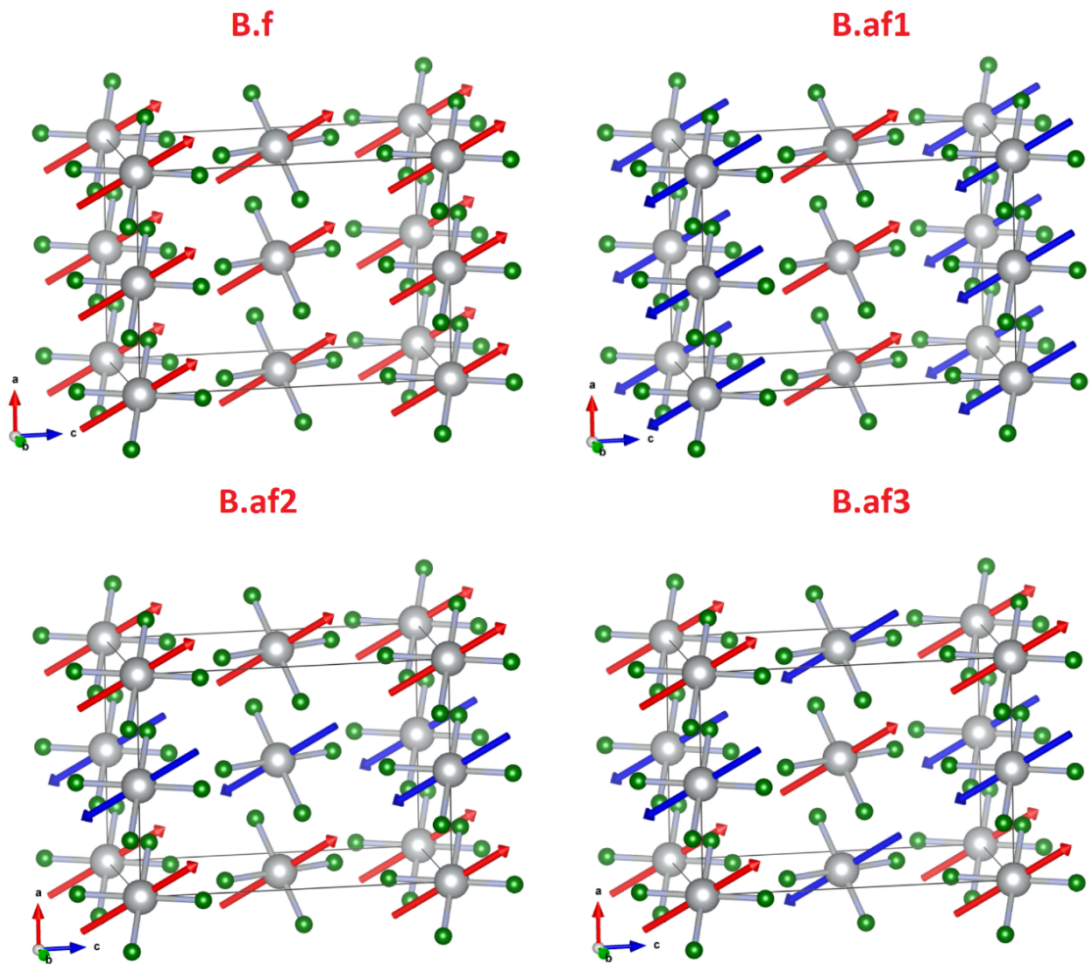

**Figure S2.** Investigated magnetic variants of **K$_2$AgF$_4$ B** structures: ferromagnetic (**B.f**) and antiferromagnetic ones (**B.af1**, **B.af2** and **B.af3**). Only Ag-F sublattices are shown. Ag-F bonds are indicated for interatomic distances not longer than 2.2 Å. All visualized example structures are for p = 0 GPa.

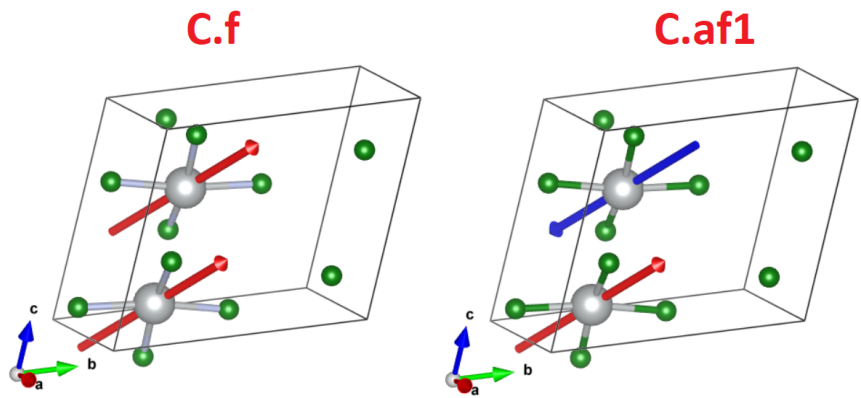

**Figure S3.** Investigated magnetic variants of **K$_2$AgF$_4$ C** structures: ferromagnetic (**C.f**) and antiferromagnetic one (**C.af1**). Only Ag-F sublattices are shown. Ag-F bonds are indicated for interatomic distances not longer than 2.2 Å. All visualized example structures are for p = 0 GPa.

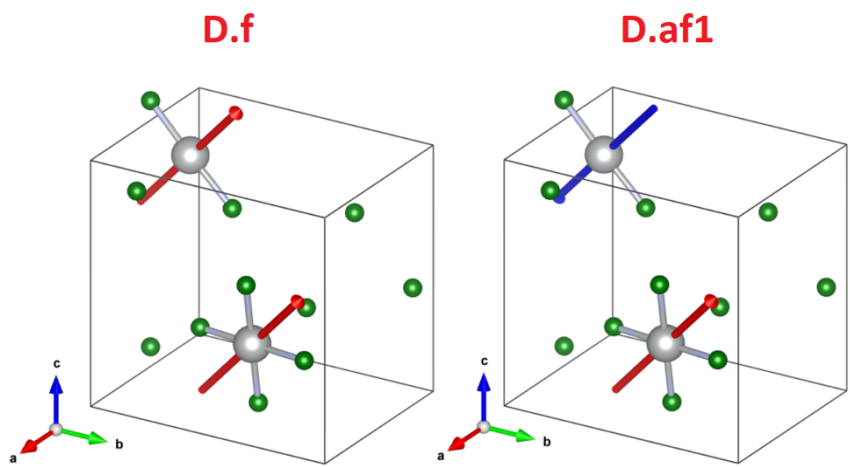

**Figure S4.** Investigated magnetic variants of **K$_2$AgF$_4$ D** structures: ferromagnetic (**D.f**) and antiferromagnetic one (**D.af1**). Only Ag-F sublattices are shown. Ag-F bonds are indicated for interatomic distances not longer than 2.2 Å. All visualized example structures are for p = 0 GPa.

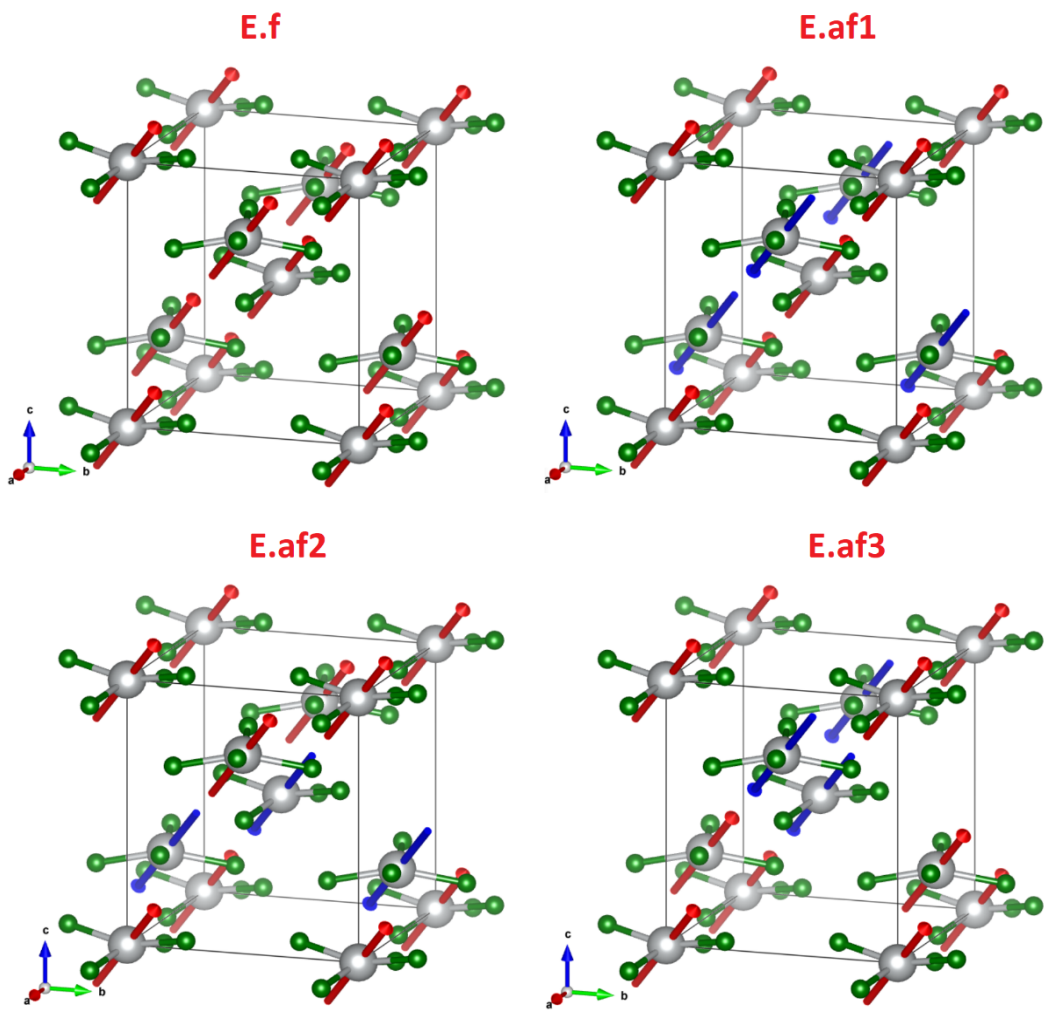

**Figure S5.** Investigated magnetic variants of **K$_2$AgF$_4$ E** structures: ferromagnetic (**E.f**) and antiferromagnetic ones (**E.af1**, **E.af2** and **E.af3**). Only Ag-F sublattices are shown. Ag-F bonds are indicated for interatomic distances not longer than 2.2 Å. All visualized example structures are for p = 0 GPa.

**Appendix S1.** CIF files of structures discussed in this study.

```
## DFT(PBEsol)+U K2AgF4 A p=000GPa magn.: f
data_1
_audit_creation_method   'vasp2cif'
_cell_length_a    6.32772727043
_cell_length_b    6.33155419994
_cell_length_c    12.4819750792
_cell_angle_alpha    90.0449569363
_cell_angle_beta     90.0023632874
_cell_angle_gamma    89.9998888167

_symmetry_space_group_name_H-M   'P 1'
loop_
_atom_site_label
_atom_site_type_symbol
_atom_site_fract_x
_atom_site_fract_y
_atom_site_fract_z
_atom_site_occupancy
Ag1   Ag   0.005554689763624   0.000090965482804   0.000043499674507   1.0
Ag2   Ag   0.005665989890118   0.499725054385921   0.500064678773409   1.0
Ag3   Ag   0.505551513097716   0.999914760130198   0.499958456800951   1.0
Ag4   Ag   0.505667464933424   0.500289529304280   0.999927112788812   1.0
F5    F    0.563342142504102   0.941477017779963   0.336723580900030   1.0
F6    F    0.447953561766328   0.058403713461015   0.663216138457617   1.0
F7    F    0.449164165633307   0.443622821689357   0.163352895045630   1.0
F8    F    0.562379255064713   0.556923819186946   0.836501896800705   1.0
F9    F    0.063370479421799   0.058475789062073   0.163277114626166   1.0
F10   F    0.947947325929100   0.941642510989339   0.836792902957445   1.0
F11   F    0.062352461696254   0.443136785965484   0.663493512082365   1.0
F12   F    0.949158892832934   0.556315371117526   0.336639035019625   1.0
F13   F    0.733469530252574   0.770042823446273   0.537424519699339   1.0
F14   F    0.276757233997900   0.228950350071465   0.462439573372951   1.0
F15   F    0.233621186297503   0.229865131187315   0.962580369042828   1.0
F16   F    0.776677748456322   0.771121473237880   0.037561861932334   1.0
F17   F    0.273536802645504   0.732905633078529   0.000068350790785   1.0
F18   F    0.737126139144803   0.266923395833647   0.999605399636767   1.0
F19   F    0.773602091871712   0.266978589364688   0.499931602625299   1.0
F20   F    0.236948893267550   0.733205334321255   0.500377859427039   1.0
K21   K    0.528284164922166   0.976651402499370   0.859665888402801   1.0
K22   K    0.529112412982684   0.522627882910753   0.360040371707704   1.0
K23   K    0.482381105182251   0.476235155420197   0.639930180536693   1.0
K24   K    0.482999980106362   0.023665699717251   0.140310072819542   1.0
K25   K    0.983006436289149   0.976339707770550   0.359689805105954   1.0
K26   K    0.028246401949390   0.023348966089451   0.640331400629370   1.0
K27   K    0.982392388543890   0.523050543694653   0.860066413887015   1.0
K28   K    0.029130501556823   0.477377312801842   0.139942686456298   1.0

## DFT(PBEsol)+U K2AgF4 A p=5GPa magn.: f
data_1
_audit_creation_method   'vasp2cif'
_cell_length_a    6.10970065899
_cell_length_b    6.11836679007
_cell_length_c    12.0615427383
_cell_angle_alpha    90.0336466132
_cell_angle_beta     90.0014116733
_cell_angle_gamma    89.9996215016

_symmetry_space_group_name_H-M   'P 1'
loop_
_atom_site_label
_atom_site_type_symbol
_atom_site_fract_x
_atom_site_fract_y
_atom_site_fract_z
_atom_site_occupancy
```

```
Ag1   Ag   0.005611322984208   0.000181782544944   0.000018288965068   1.0
Ag2   Ag   0.005649213953785   0.499789565081304   0.499993318918617   1.0
Ag3   Ag   0.505597887648245   0.999822867882423   0.499980233004369   1.0
Ag4   Ag   0.505668510621848   0.500210707241571   0.000002141823100   1.0
F5    F    0.565273536196829   0.938959640230068   0.332959866515758   1.0
F6    F    0.445476288119992   0.060641702143287   0.666970877235561   1.0
F7    F    0.447316753171630   0.441942845076877   0.167566135706258   1.0
F8    F    0.563116682225344   0.558467331178308   0.832398024460961   1.0
F9    F    0.065302986895970   0.061048951063275   0.167040481881788   1.0
F10   F    0.945494448191877   0.939337336155598   0.833026092795553   1.0
F11   F    0.063610902984998   0.441515879927451   0.667596267479283   1.0
F12   F    0.947275751710482   0.558070626195498   0.332429344336044   1.0
F13   F    0.739205082011094   0.766520024973056   0.542019776117573   1.0
F14   F    0.272287668238736   0.233320450657348   0.457886724446992   1.0
F15   F    0.239229983474497   0.233466391054864   0.957981081863129   1.0
F16   F    0.772291416334778   0.766701063434149   0.042109967594629   1.0
F17   F    0.269318259271016   0.737582159181011   0.000252589307277   1.0
F18   F    0.741977786627598   0.262916170627917   0.999708788216186   1.0
F19   F    0.769332236457019   0.262398193819624   0.499336868871404   1.0
F20   F    0.241953535468053   0.737104792371125   0.500290845950603   1.0
K21   K    0.526424895438936   0.979027939285707   0.860293881731039   1.0
K22   K    0.526321545250046   0.521168291220350   0.360932930544735   1.0
K23   K    0.484671210989019   0.478651489423351   0.639089330230935   1.0
K24   K    0.484535806706577   0.021349964100187   0.139691363100231   1.0
K25   K    0.984519337342234   0.978663195963909   0.360302068160067   1.0
K26   K    0.026405341575100   0.020960811934113   0.639703289467347   1.0
K27   K    0.984700952466821   0.521343974373379   0.860905564420224   1.0
K28   K    0.026331217643269   0.478836772859333   0.139071036855250   1.0

## DFT(PBEsol)+U K2AgF4 A p=10GPa magn.: f
data_1
_audit_creation_method   'vasp2cif'
_cell_length_a     5.42567301651
_cell_length_b     6.47170965112
_cell_length_c     11.7137349569
_cell_angle_alpha    89.8158785068
_cell_angle_beta     89.9881451278
_cell_angle_gamma    89.9912882987

_symmetry_space_group_name_H-M   'P 1'
loop_
_atom_site_label
_atom_site_type_symbol
_atom_site_fract_x
_atom_site_fract_y
_atom_site_fract_z
_atom_site_occupancy
Ag1   Ag   0.005497318043035   0.000115676949617   0.000021403773227   1.0
Ag2   Ag   0.005468161452201   0.500007344440883   0.500020228330343   1.0
Ag3   Ag   0.505523677666516   0.999882798328347   0.499968885536128   1.0
Ag4   Ag   0.505428663212867   0.500000834536017   0.999970916855218   1.0
F5    F    0.511993955007033   0.913408521608563   0.330593104268945   1.0
F6    F    0.499774424037013   0.086378113490549   0.669314586954027   1.0
F7    F    0.509816651075277   0.412235419031032   0.169561021121248   1.0
F8    F    0.501883019212733   0.587781029197700   0.830388661806718   1.0
F9    F    0.011634406430230   0.086530570453213   0.169392408447811   1.0
F10   F    0.000080861828849   0.913676673089300   0.830673415433676   1.0
F11   F    0.001318728565736   0.412268219231890   0.669618448098861   1.0
F12   F    0.009917689755127   0.587701888438069   0.330454073643071   1.0
F13   F    0.739519613401054   0.762556571770888   0.540498987467932   1.0
F14   F    0.271401758529971   0.237298914762820   0.459513367561296   1.0
F15   F    0.239565459937860   0.237404778978422   0.959570272097364   1.0
F16   F    0.771307744871406   0.762728151571636   0.040393110083804   1.0
F17   F    0.271325115253482   0.737005389091834   0.039777068652016   1.0
F18   F    0.739514438148545   0.263007336979395   0.960145969473972   1.0
F19   F    0.771348986401989   0.263016032005838   0.460292835697990   1.0
F20   F    0.239534412403623   0.736972600273542   0.539766923431379   1.0
```

```
K21   K    0.507138918012257    0.963894781043390    0.860164972676927    1.0
K22   K    0.504148441632879    0.536839404595324    0.359745666436665    1.0
K23   K    0.506804222523601    0.462909529614885    0.640268398054304    1.0
K24   K    0.503948630362719    0.036050461783808    0.139781982193256    1.0
K25   K    0.004129265266668    0.963977020980736    0.360211992805815    1.0
K26   K    0.007004608642127    0.036082945179800    0.639825930583626    1.0
K27   K    0.006950729605373    0.537075076734358    0.859731709047254    1.0
K28   K    0.003956058719825    0.463194835838178    0.140260839467106    1.0

## DFT(PBEsol)+U K2AgF4 A p=20GPa magn.: f
data_1
_audit_creation_method    'vasp2cif'
_cell_length_a    5.23316016824
_cell_length_b    6.27525457313
_cell_length_c    11.3622979195
_cell_angle_alpha    104.769450737
_cell_angle_beta    98.9691816204
_cell_angle_gamma    100.545305453

_symmetry_space_group_name_H-M    'P 1'
loop_
_atom_site_label
_atom_site_type_symbol
_atom_site_fract_x
_atom_site_fract_y
_atom_site_fract_z
_atom_site_occupancy
Ag1   Ag   0.004972422127504    0.999768288070051    0.000044136173926    1.0
Ag2   Ag   0.005229388378993    0.500274763386730    0.500091724646840    1.0
Ag3   Ag   0.505960874800593    0.000080177935655    0.499805638868994    1.0
Ag4   Ag   0.505265072166753    0.499770907424535    0.000023961335785    1.0
F5    F    0.630808978207563    0.175608556629724    0.689204210846020    1.0
F6    F    0.381137459005449    0.824385355173610    0.310342843204720    1.0
F7    F    0.471991297419670    0.532602189965135    0.829007818357768    1.0
F8    F    0.538650478485442    0.466484699210642    0.170942724653722    1.0
F9    F    0.891525265160286    0.851225516922033    0.807999838621249    1.0
F10   F    0.118731180493281    0.148752025497562    0.192117191451945    1.0
F11   F    0.944002584313083    0.491601554336784    0.322105019372113    1.0
F12   F    0.067493838668970    0.509288976934666    0.678146444944192    1.0
F13   F    0.687393536692299    0.204813435681082    0.413181640194089    1.0
F14   F    0.324402258935907    0.795538190513039    0.586502311412923    1.0
F15   F    0.272650219692641    0.794517565960626    0.011746143433014    1.0
F16   F    0.737401700234414    0.204969420577786    0.988038677043642    1.0
F17   F    0.212377302528922    0.226456250757093    0.905101826684893    1.0
F18   F    0.798168923456009    0.773040891549448    0.095159741427605    1.0
F19   F    0.801160177830937    0.775315561297206    0.515169929802958    1.0
F20   F    0.210786001925966    0.225559229449102    0.485067388867181    1.0
K21   K    0.591184084061196    0.094880197570118    0.179767434338157    1.0
K22   K    0.585659662355537    0.532993348607432    0.638159589615432    1.0
K23   K    0.425945557027909    0.467518238107962    0.362005665707303    1.0
K24   K    0.418932158778790    0.904355217940414    0.820114170664730    1.0
K25   K    0.092976923076088    0.113410211848115    0.674031197876117    1.0
K26   K    0.919516160506448    0.887458451099623    0.326143824059993    1.0
K27   K    0.081291707676241    0.521486573497858    0.137906265121893    1.0
K28   K    0.929284925993120    0.477844214055948    0.862072721272796    1.0

## DFT(PBEsol)+U K2AgF4 A p=30GPa magn.: af2
data_1
_audit_creation_method    'vasp2cif'
_cell_length_a    4.92720281434
_cell_length_b    6.16935041327
_cell_length_c    11.3857171092
_cell_angle_alpha    75.2808318667
_cell_angle_beta    82.3064150179
_cell_angle_gamma    101.348094665
```

```
_symmetry_space_group_name_H-M    'P 1'
loop_
_atom_site_label
_atom_site_type_symbol
_atom_site_fract_x
_atom_site_fract_y
_atom_site_fract_z
_atom_site_occupancy
Ag1   Ag    0.005547941398243    0.000046524888534   -0.000003960745461   1.0
Ag2   Ag    0.005522077993327    0.499986545774716    0.500003175565505   1.0
Ag3   Ag    0.505553160446385   -0.000021118727813    0.499988249063226   1.0
Ag4   Ag    0.505536752810444    0.499996077105243    0.000001712662777   1.0
F5    F     0.558500962078585   -0.024997471036652    0.329722820768661   1.0
F6    F     0.452585337291591    0.024980044074818    0.670257309145734   1.0
F7    F     0.396372093876065    0.332106660275056    0.202576481620399   1.0
F8    F     0.614721290708484    0.667895889550661    0.797427868008812   1.0
F9    F     0.953008029169119    0.026222567816612    0.170026683656536   1.0
F10   F     0.058064194957628    0.973816311694132    0.829974089874181   1.0
F11   F     0.896535517796629    0.331263676774122    0.702766992754261   1.0
F12   F     0.114571827462199    0.668722286993611    0.297246646735795   1.0
F13   F     0.729353051644929    0.699003742426570    0.511201728621681   1.0
F14   F     0.281769130460552    0.300953783738417    0.488794014865562   1.0
F15   F     0.302605027588854    0.285134670328488    0.906569769835564   1.0
F16   F     0.708541188841867    0.714916299215490    0.093427901483208   1.0
F17   F     0.229140211470323    0.698786266905019    0.011202842096723   1.0
F18   F     0.781971705079804    0.301237897689855    0.988799797388673   1.0
F19   F     0.802211683317522    0.285001067201760    0.406806416207643   1.0
F20   F     0.208916868030643    0.714375891409685    0.593175668631398   1.0
K21   K     0.577894073297293    0.021575976459179    0.860952274610630   1.0
K22   K     0.585461930584447    0.602219824455886    0.318229309101098   1.0
K23   K     0.425636728769460    0.397789523683001    0.681775448351347   1.0
K24   K     0.433224425912092    0.978427260732811    0.139026057664045   1.0
K25   K     0.077823331677493    0.022055148019912    0.360841306707673   1.0
K26   K     0.933205018083709    0.977889184761788    0.639169368908276   1.0
K27   K     0.085785387748966    0.601626230180170    0.818106261366238   1.0
K28   K     0.925342011503347    0.398390157608945    0.181890945049808   1.0

## DFT(PBEsol)+U K2AgF4 A p=40GPa magn.: f
data_1
_audit_creation_method   'vasp2cif'
_cell_length_a    4.7954347275
_cell_length_b    6.06967110704
_cell_length_c    11.2653396376
_cell_angle_alpha    75.4170533577
_cell_angle_beta     82.9294584736
_cell_angle_gamma    102.143613272

_symmetry_space_group_name_H-M    'P 1'
loop_
_atom_site_label
_atom_site_type_symbol
_atom_site_fract_x
_atom_site_fract_y
_atom_site_fract_z
_atom_site_occupancy
Ag1   Ag    0.005318522709985    0.000080002285181    0.999972824217536   1.0
Ag2   Ag    0.005415516581726    0.499775871955448    0.500022369350084   1.0
Ag3   Ag    0.505641983308264    0.999963021653621    0.499939439135811   1.0
Ag4   Ag    0.505451065572762    0.500168013324354    0.000055695324424   1.0
F5    F     0.560261686273651    0.977330417287786    0.328500023279196   1.0
F6    F     0.450981404200596    0.022474392269584    0.671412435970350   1.0
F7    F     0.393719220625924    0.329353591317878    0.205694243307488   1.0
F8    F     0.617452664233611    0.670552820163624    0.794464539835321   1.0
F9    F     0.950727753877335    0.023659356174906    0.171256234690829   1.0
F10   F     0.060396567913704    0.976660664236505    0.828640668061032   1.0
F11   F     0.894443762133229    0.328573783095073    0.705687614324357   1.0
F12   F     0.116722154860721    0.671412644192643    0.294305239510233   1.0
```

```
F13   F   0.720415882773493   0.696644274361737   0.511460084517949   1.0
F14   F   0.290660837087908   0.303078017977384   0.488660117613690   1.0
F15   F   0.308665708885451   0.288546901121006   0.903858028926143   1.0
F16   F   0.702213192170100   0.711591354128961   0.096322408764493   1.0
F17   F   0.220117026664200   0.696733798502659   0.011387027211662   1.0
F18   F   0.790912947615180   0.303769963138816   0.988547340370679   1.0
F19   F   0.808224862618312   0.289055324112338   0.403747327799976   1.0
F20   F   0.203171440369214   0.710760093949702   0.596066549642068   1.0
K21   K   0.576267500704597   0.018308151102128   0.862462478132782   1.0
K22   K   0.586537968980911   0.604261029797355   0.317383779964586   1.0
K23   K   0.424480137023688   0.395457035537157   0.682770774903258   1.0
K24   K   0.434729774151006   0.981957232408201   0.137447999909220   1.0
K25   K   0.076071562693699   0.018716697684415   0.362421642691723   1.0
K26   K   0.935050075517665   0.980867265753579   0.637567977927717   1.0
K27   K   0.087500563699319   0.603883973841038   0.817185617206119   1.0
K28   K   0.923854176753747   0.396365228626940   0.182716697411285   1.0

## DFT(PBEsol)+U K2AgF4 A p=50GPa magn.: af1
data_1
_audit_creation_method   'vasp2cif'
_cell_length_a    4.80006804017
_cell_length_b    6.27344079166
_cell_length_c    9.79322979863
_cell_angle_alpha    90.0010068335
_cell_angle_beta     76.0735286408
_cell_angle_gamma    90.0000548112

_symmetry_space_group_name_H-M   'P 1'
loop_
_atom_site_label
_atom_site_type_symbol
_atom_site_fract_x
_atom_site_fract_y
_atom_site_fract_z
_atom_site_occupancy
Ag1   Ag   0.960131026469753   0.928253764098717   0.054718094712678   1.0
Ag2   Ag   0.960183829808316   0.571755523710537   0.554713760320585   1.0
Ag3   Ag   0.528717394669711   0.071756837764474   0.445280403531322   1.0
Ag4   Ag   0.528670783418563   0.428235509221891   0.945281822776688   1.0
F5    F    0.419440370026312   0.037879627780205   0.655933249208286   1.0
F6    F    0.641969555175629   0.171503828534767   0.244079952002308   1.0
F7    F    0.641930933783319   0.328504834847435   0.744077603646034   1.0
F8    F    0.419439685175833   0.462105976156368   0.155935527923261   1.0
F9    F    0.069484231671776   0.962200796058090   0.844076121806812   1.0
F10   F    0.846908484628360   0.828409857164768   0.255895598693015   1.0
F11   F    0.069484549774949   0.537812996236147   0.344071057839394   1.0
F12   F    0.846958243171648   0.671582252141556   0.755895416490978   1.0
F13   F    0.232495465840473   0.851691430897410   0.437784237688829   1.0
F14   F    0.775988308168077   0.289911868231791   0.494705414070977   1.0
F15   F    0.775959161509863   0.210094794929584   0.994701648076920   1.0
F16   F    0.232477791529231   0.648325952241476   0.937792303733826   1.0
F17   F    0.712958885329069   0.710035659677667   0.005287881745310   1.0
F18   F    0.256413849468833   0.148262397055465   0.062225649685042   1.0
F19   F    0.256436469282822   0.351725490939183   0.562223103586039   1.0
F20   F    0.712981424900581   0.789954757562218   0.505292853676439   1.0
K21   K    0.369870418252495   0.825137992766427   0.193771276014468   1.0
K22   K    0.369919949240653   0.674844041667112   0.693776920015327   1.0
K23   K    0.548667861879521   0.534171826469985   0.369603177458866   1.0
K24   K    0.548661789101457   0.965824233092165   0.869599847008980   1.0
K25   K    0.940260459873375   0.034207644090809   0.630418209674511   1.0
K26   K    0.118991034642114   0.174755404443121   0.306222670650284   1.0
K27   K    0.940251284110866   0.465794526824784   0.130414566375599   1.0
K28   K    0.118946589096380   0.325260165395845   0.806221671587195   1.0

## DFT(PBEsol)+U K2AgF4 A p=60GPa magn.: af1
data_1
```

```
_audit_creation_method   'vasp2cif'
_cell_length_a    4.72600259085
_cell_length_b    6.20656207085
_cell_length_c    9.66684962898
_cell_angle_alpha    90.0011671149
_cell_angle_beta     76.3079160666
_cell_angle_gamma    89.9999078886

_symmetry_space_group_name_H-M    'P 1'
loop_
_atom_site_label
_atom_site_type_symbol
_atom_site_fract_x
_atom_site_fract_y
_atom_site_fract_z
_atom_site_occupancy
Ag1   Ag   0.959903855930167   0.929440530570113   0.056751210602064   1.0
Ag2   Ag   0.959966078886409   0.570570858889404   0.556747488645417   1.0
Ag3   Ag   0.528958553786239   0.070548627662135   0.443255592354090   1.0
Ag4   Ag   0.528902973842433   0.429442809995945   0.943258156445632   1.0
F5    F    0.419985599562896   0.037584473772336   0.655821889303963   1.0
F6    F    0.643726720963870   0.171217076226393   0.240777135264679   1.0
F7    F    0.643689011704479   0.328791116374399   0.740774804768988   1.0
F8    F    0.419980775311145   0.462398951652892   0.155824465958896   1.0
F9    F    0.068918300327696   0.962472110948104   0.844192855941404   1.0
F10   F    0.845170606510282   0.828689642965795   0.259211911964418   1.0
F11   F    0.068918893521035   0.537543496941261   0.344188433860896   1.0
F12   F    0.845227123494232   0.671301275598508   0.759213197128628   1.0
F13   F    0.229628223486152   0.849000332092421   0.438023384334719   1.0
F14   F    0.777587463055469   0.289115547004294   0.494375351287701   1.0
F15   F    0.777551544896880   0.210894187921992   0.994371312825457   1.0
F16   F    0.229601021971688   0.651018722894039   0.938031287580889   1.0
F17   F    0.711341376779175   0.710868347909686   0.005616548870641   1.0
F18   F    0.259268477428057   0.150991677110446   0.061973843293861   1.0
F19   F    0.259299595696581   0.348994544954078   0.561972148798098   1.0
F20   F    0.711373573177304   0.789120794311893   0.505620545322614   1.0
K21   K    0.369722467441960   0.825409145016137   0.193996468978348   1.0
K22   K    0.369782141837110   0.674571045711349   0.694001873108159   1.0
K23   K    0.548272506242090   0.532480185930048   0.369544280746884   1.0
K24   K    0.548265471037673   0.967515102315853   0.869540677408405   1.0
K25   K    0.940632843359577   0.032516511708945   0.630462724983569   1.0
K26   K    0.119176438390868   0.174529676265578   0.305996740510673   1.0
K27   K    0.940619496782313   0.467486906529072   0.130459515879836   1.0
K28   K    0.119128694576249   0.325486290726862   0.805996193831044   1.0

## DFT(PBEsol)+U K2AgF4 A p=70GPa magn.: af1
data_1
_audit_creation_method   'vasp2cif'
_cell_length_a    4.66061253326
_cell_length_b    6.14624151767
_cell_length_c    9.56524206621
_cell_angle_alpha    90.0009793635
_cell_angle_beta     76.5017827123
_cell_angle_gamma    90.0001582228

_symmetry_space_group_name_H-M    'P 1'
loop_
_atom_site_label
_atom_site_type_symbol
_atom_site_fract_x
_atom_site_fract_y
_atom_site_fract_z
_atom_site_occupancy
Ag1   Ag   0.959992482461831   0.929892640581542   0.058285950752598   1.0
Ag2   Ag   0.960034294250078   0.570117736024184   0.558281501714250   1.0
Ag3   Ag   0.529094610321694   0.070062646168500   0.441709574365841   1.0
Ag4   Ag   0.529060248500835   0.429930111432909   0.941711159845368   1.0
```

```
F5   F   0.420416033916049   0.037474176069938   0.655725018980938   1.0
F6   F   0.645234794796021   0.171148560361564   0.238237636592695   1.0
F7   F   0.645207180967236   0.328858178476271   0.738235030381198   1.0
F8   F   0.420419441784097   0.462513578545616   0.155725951232772   1.0
F9   F   0.068441520211024   0.962428614687959   0.844251175791463   1.0
F10  F   0.843709996778036   0.828860722136569   0.261770437732721   1.0
F11  F   0.068438137498844   0.537582380361462   0.344246631368162   1.0
F12  F   0.843752936613864   0.671132653132728   0.761770536328822   1.0
F13  F   0.226995908840022   0.847367819483202   0.438040627304278   1.0
F14  F   0.779497337409046   0.288619699288122   0.494237339267911   1.0
F15  F   0.779474431956702   0.211388480135549   0.994234578364429   1.0
F16  F   0.226982183665536   0.652647198321452   0.938046110912518   1.0
F17  F   0.709298952854467   0.711494019061634   0.005785745438715   1.0
F18  F   0.261787545171097   0.152769708410007   0.061939641368916   1.0
F19  F   0.261802639616262   0.347219039595406   0.561937634505485   1.0
F20  F   0.709318682903783   0.788497986987372   0.505787955418574   1.0
K21  K   0.369622909779558   0.825578865241886   0.194327798029778   1.0
K22  K   0.369665370709176   0.674406230686271   0.694331153054125   1.0
K23  K   0.547879725314138   0.531057421346378   0.369488412027679   1.0
K24  K   0.547878672416161   0.968937470137564   0.869486038616693   1.0
K25  K   0.940956838388699   0.031070248663967   0.630494377794478   1.0
K26  K   0.119358186324056   0.174569098952796   0.305710972767131   1.0
K27  K   0.940953526970902   0.468932858493151   0.130490859786377   1.0
K28  K   0.119325239580773   0.325441847216002   0.805710190256075   1.0

## DFT(PBEsol)+U K2AgF4 A p=80GPa magn.: af1
data_1
_audit_creation_method   'vasp2cif'
_cell_length_a    4.60188531339
_cell_length_b    6.09076756509
_cell_length_c    9.48267433064
_cell_angle_alpha    90.0014086992
_cell_angle_beta    76.6927930787
_cell_angle_gamma    90.0001357852

_symmetry_space_group_name_H-M    'P 1'
loop_
_atom_site_label
_atom_site_type_symbol
_atom_site_fract_x
_atom_site_fract_y
_atom_site_fract_z
_atom_site_occupancy
Ag1  Ag  0.960246335226501   0.929948852953752   0.059546667637058   1.0
Ag2  Ag  0.960274308776697   0.570064763106209   0.559540876042231   1.0
Ag3  Ag  0.528880068868221   0.069970783647659   0.440426079860790   1.0
Ag4  Ag  0.528854705976425   0.430016669197666   0.940429962618318   1.0
F5   F   0.420630798990446   0.037325852842115   0.655574995680892   1.0
F6   F   0.646306385456360   0.171214610800708   0.236198325836099   1.0
F7   F   0.646282231634709   0.328791751733671   0.736195808982752   1.0
F8   F   0.420640935812173   0.462664527419338   0.155575489188546   1.0
F9   F   0.068141254755683   0.962460937414235   0.844396273614367   1.0
F10  F   0.842652730339564   0.828916687891810   0.263828072117728   1.0
F11  F   0.068132978189532   0.537548082559470   0.344394267907906   1.0
F12  F   0.842680380212981   0.671077889060239   0.763829451364302   1.0
F13  F   0.224510355753724   0.846259113378942   0.437960456797987   1.0
F14  F   0.781822571169178   0.287993879398607   0.494241103677465   1.0
F15  F   0.781816509036690   0.212020169348635   0.994236428600754   1.0
F16  F   0.224509389675586   0.653758064361443   0.937967475019597   1.0
F17  F   0.706986060463831   0.712196815327124   0.005791307504322   1.0
F18  F   0.264300487355376   0.153934830345745   0.062020743442753   1.0
F19  F   0.264301638446334   0.346048301680010   0.562013726285453   1.0
F20  F   0.706991525571622   0.787789432901638   0.505795878571236   1.0
K21  K   0.369821714583108   0.825686827698072   0.194753483215747   1.0
K22  K   0.369849796790528   0.674302090976799   0.694758463981647   1.0
K23  K   0.547844061091087   0.529554560306505   0.369369220887255   1.0
K24  K   0.547847809366217   0.970441662290928   0.869367698449576   1.0
```

```
K25   K   0.940917762018902   0.029533211309654   0.630611589330098   1.0
K26   K   0.119230003426905   0.174583893265368   0.305285995407843   1.0
K27   K   0.940922750910197   0.470470044396050   0.130609153123958   1.0
K28   K   0.119204280101423   0.325425684387600   0.805281044853323   1.0

## DFT(PBEsol)+U K2AgF4 A p=90GPa magn.: af1
data_1
_audit_creation_method   'vasp2cif'
_cell_length_a    4.54851737199
_cell_length_b    6.0383678289
_cell_length_c    9.41431988205
_cell_angle_alpha    90.0014766934
_cell_angle_beta    76.8603463811
_cell_angle_gamma    90.000152513

_symmetry_space_group_name_H-M    'P 1'
loop_
_atom_site_label
_atom_site_type_symbol
_atom_site_fract_x
_atom_site_fract_y
_atom_site_fract_z
_atom_site_occupancy
Ag1   Ag   0.960615289394199   0.929730792356059   0.060578835439373   1.0
Ag2   Ag   0.960642678460464   0.570291127690732   0.560582209073928   1.0
Ag3   Ag   0.528477479420107   0.070224097797448   0.439395529198803   1.0
Ag4   Ag   0.528448661934838   0.429755498269671   0.939395031088000   1.0
F5    F    0.420928514973762   0.037304974800918   0.655437335130682   1.0
F6    F    0.647278869611282   0.171221126534174   0.234515169243008   1.0
F7    F    0.646998193307478   0.328788003446945   0.734516800137153   1.0
F8    F    0.420940378139493   0.462685537412240   0.155443126874493   1.0
F9    F    0.067859094930461   0.962499754652168   0.844545178075778   1.0
F10   F    0.841944638079670   0.828923951680849   0.265505761642461   1.0
F11   F    0.067848722107889   0.537510619685292   0.344538033233369   1.0
F12   F    0.841971078179890   0.671067427499663   0.765502235492872   1.0
F13   F    0.222072072667241   0.845583750988204   0.437748007750210   1.0
F14   F    0.784282445585388   0.287358907822896   0.494361999538374   1.0
F15   F    0.784293880503289   0.212657429361864   0.994356080213507   1.0
F16   F    0.222077808025800   0.654438378758298   0.937758288758182   1.0
F17   F    0.704531389905336   0.712799004397706   0.005662443549054   1.0
F18   F    0.266754970864868   0.154575461106868   0.062235241390546   1.0
F19   F    0.266748912724911   0.345399954020704   0.562222230677145   1.0
F20   F    0.704524548263081   0.787183067406680   0.505668628691151   1.0
K21   K    0.370178828847850   0.825646914565122   0.195230291749743   1.0
K22   K    0.370202113053450   0.674341152070500   0.695238145394265   1.0
K23   K    0.547735831099893   0.528094469009983   0.369232654901882   1.0
K24   K    0.547745066875467   0.971903644482805   0.869233333935417   1.0
K25   K    0.941038240480891   0.028056707034728   0.630755790425380   1.0
K26   K    0.118844168339304   0.174594393672457   0.304797794644083   1.0
K27   K    0.941049474885727   0.471945665723098   0.130755850626491   1.0
K28   K    0.118817479337962   0.325418177751822   0.804788013124647   1.0

## DFT(PBEsol)+U K2AgF4 A p=100GPa magn.: af1
data_1
_audit_creation_method   'vasp2cif'
_cell_length_a    4.49997704578
_cell_length_b    5.98980853074
_cell_length_c    9.35415417477
_cell_angle_alpha    90.0007778884
_cell_angle_beta    77.0351168814
_cell_angle_gamma    90.0002012403

_symmetry_space_group_name_H-M    'P 1'
loop_
_atom_site_label
_atom_site_type_symbol
```

```
_atom_site_fract_x
_atom_site_fract_y
_atom_site_fract_z
_atom_site_occupancy
Ag1   Ag   0.039074979424465   0.929533519804786   0.938460818054041   1.0
Ag2   Ag   0.039075582815390   0.570466419267068   0.438454587429028   1.0
Ag3   Ag   0.472031933256345   0.070455751504128   0.561541836656820   1.0
Ag4   Ag   0.472021820935016   0.429544234523552   0.061539999923417   1.0
F5    F    0.578798874629299   0.037392245402705   0.344723739005219   1.0
F6    F    0.352652299753659   0.171013381185437   0.767044020003498   1.0
F7    F    0.352648631221393   0.328982456520408   0.267037342985170   1.0
F8    F    0.578793973248380   0.462620214652446   0.844722364475662   1.0
F9    F    0.932301077561961   0.962626358591714   0.155271332821571   1.0
F10   F    0.158415539234519   0.828952997142249   0.732963631610540   1.0
F11   F    0.932296810702311   0.537735553335930   0.655267469428955   1.0
F12   F    0.158414684590418   0.671047892693178   0.232960081603782   1.0
F13   F    0.780518972099517   0.845079663631350   0.562339419771329   1.0
F14   F    0.212941018164645   0.286469628740737   0.505623674736650   1.0
F15   F    0.212918528513565   0.213529367096463   0.005634118914705   1.0
F16   F    0.780530654305547   0.654907142202432   0.062355438198919   1.0
F17   F    0.298198000320713   0.713530959395956   0.994362594694819   1.0
F18   F    0.730610313187119   0.154930190938301   0.937657374043609   1.0
F19   F    0.730614397064270   0.345062319238643   0.437667717881795   1.0
F20   F    0.298165037131859   0.786477115100068   0.494378957513102   1.0
K21   K    0.629452682485121   0.825460859830017   0.804115741659418   1.0
K22   K    0.629461045461753   0.674532526419269   0.304127118375530   1.0
K23   K    0.452172944395697   0.526697360039379   0.630920906640240   1.0
K24   K    0.452176943908271   0.973313988737972   0.130926914758684   1.0
K25   K    0.058929099143630   0.026699711187139   0.369071812268358   1.0
K26   K    0.881622311379795   0.174532590914894   0.695877765972500   1.0
K27   K    0.058929815479569   0.473310113841558   0.869068474681984   1.0
K28   K    0.881632169585764   0.325455458062223   0.195884725890632   1.0

## DFT(PBEsol)+U K2AgF4 B p=000GPa magn.: af1
data_1
_audit_creation_method   'vasp2cif'
_cell_length_a     7.428416729
_cell_length_b     6.37681818004
_cell_length_c     10.2182149887
_cell_angle_alpha    90.0003890992
_cell_angle_beta     90.0042190551
_cell_angle_gamma    88.1402359007

_symmetry_space_group_name_H-M    'P 1'
loop_
_atom_site_label
_atom_site_type_symbol
_atom_site_fract_x
_atom_site_fract_y
_atom_site_fract_z
_atom_site_occupancy
Ag1   Ag   0.500000000000000   0.000000000000000   0.000000000000000   1.0
Ag2   Ag   0.000000000000000   0.000000000000000   0.000000000000000   1.0
Ag3   Ag   0.499999928000001   0.500000024000002   0.500000000999997   1.0
Ag4   Ag   0.000000000000000   0.500000024000002   0.500000000999997   1.0
F5    F    0.293237908000002   0.219054182999997   0.947440942000000   1.0
F6    F    0.793237908000002   0.219054182999997   0.947440942000000   1.0
F7    F    0.206762091999998   0.780945901999999   0.052559060999997   1.0
F8    F    0.706762044000001   0.780945901999999   0.052559060999997   1.0
F9    F    0.206760625999998   0.280886713000001   0.447457086999996   1.0
F10   F    0.706760658000000   0.280886713000001   0.447457086999996   1.0
F11   F    0.293239321999998   0.719113503000003   0.552542870000003   1.0
F12   F    0.793239354000001   0.719113503000003   0.552542870000003   1.0
F13   F    0.492825764000003   0.125555878000000   0.188334722000001   1.0
F14   F    0.992825797000002   0.125555878000000   0.188334722000001   1.0
F15   F    0.007174112000001   0.874444263000001   0.811665327000000   1.0
F16   F    0.507174122000002   0.874444263000001   0.811665327000000   1.0
```

```
F17    F    0.007169464000000    0.374451088000001    0.688367296000003    1.0
F18    F    0.507169503999997    0.374451088000001    0.688367296000003    1.0
F19    F    0.492830544999997    0.625548959000000    0.311632660000001    1.0
F20    F    0.992830480999999    0.625548959000000    0.311632660000001    1.0
K21    K    0.240628331000003    0.418174211999997    0.181853177000001    1.0
K22    K    0.740628362999999    0.418174211999997    0.181853177000001    1.0
K23    K    0.259371782999999    0.581825836000000    0.818146872000000    1.0
K24    K    0.759371719000001    0.581825836000000    0.818146872000000    1.0
K25    K    0.259346604999998    0.081827023000002    0.681857657999998    1.0
K26    K    0.759346620999999    0.081827023000002    0.681857657999998    1.0
K27    K    0.240653401000003    0.918172865999999    0.318142391000002    1.0
K28    K    0.740653401000003    0.918172865999999    0.318142391000002    1.0

## DFT(PBEsol)+U K2AgF4 B p=5GPa magn.: f
data_1
_audit_creation_method   'vasp2cif'
_cell_length_a    3.58481761052
_cell_length_b    6.21441490457
_cell_length_c    9.68320725887
_cell_angle_alpha    90.0010759972
_cell_angle_beta    90.0000460419
_cell_angle_gamma    89.5800591271

_symmetry_space_group_name_H-M    'P 1'
loop_
_atom_site_label
_atom_site_type_symbol
_atom_site_fract_x
_atom_site_fract_y
_atom_site_fract_z
_atom_site_occupancy
Ag1   Ag   0.999999989999999    0.000000090000000    0.000000030000002    1.0
Ag2   Ag   0.999999899999999    0.499999989999999    0.499999989999999    1.0
F3    F    0.573037359502400    0.217846847137454    0.944368341058741    1.0
F4    F    0.426962660497602    0.782153242862546    0.055631618941262    1.0
F5    F    0.426978895150101    0.282164188201687    0.444364061928229    1.0
F6    F    0.573021014849899    0.717835971798312    0.555635938071771    1.0
F7    F    0.988481778572506    0.133062043913791    0.195257080996895    1.0
F8    F    0.011517961427494    0.866938126086209    0.804742959003101    1.0
F9    F    0.011556946307310    0.366946005470201    0.695261274210651    1.0
F10   F    0.988443053692690    0.633054004529799    0.304738675789344    1.0
K11   K    0.487896011314402    0.421356714335300    0.178385324777835    1.0
K12   K    0.512104238685598    0.578643275664706    0.821614655222163    1.0
K13   K    0.512186770825052    0.078651057248076    0.678389525321023    1.0
K14   K    0.487813269174944    0.921348812751927    0.321610484678978    1.0

## DFT(PBEsol)+U K2AgF4 B p=10GPa magn.: f
data_1
_audit_creation_method   'vasp2cif'
_cell_length_a    3.50089018455
_cell_length_b    6.14886686
_cell_length_c    9.30501143397
_cell_angle_alpha    89.9955634724
_cell_angle_beta    90.0004862263
_cell_angle_gamma    89.9910660488

_symmetry_space_group_name_H-M    'P 1'
loop_
_atom_site_label
_atom_site_type_symbol
_atom_site_fract_x
_atom_site_fract_y
_atom_site_fract_z
_atom_site_occupancy
Ag1   Ag    0.000003126603338    -0.000001335173779    -0.000005534760399    1.0
Ag2   Ag   -0.000001504052230    0.500003332610946    0.499999423281487    1.0
```

```
F3    F   0.566506805432239   0.216536288071983   0.941141010995494   1.0
F4    F   0.433493673249551   0.783463073771940   0.058859796425863   1.0
F5    F   0.433496339853342   0.283472155713051   0.441146678826202   1.0
F6    F   0.566503711675495   0.716527219723968   0.558853603206152   1.0
F7    F   0.984010528910886   0.137487970208417   0.200760404735800   1.0
F8    F   0.015988755044815   0.862513229955319   0.799242565791604   1.0
F9    F   0.015981384357716   0.362510599723660   0.700764582663956   1.0
F10   F   0.984017890095622   0.637488878008991   0.299236130891039   1.0
K11   K   0.484061504234326   0.421706914106770   0.174973800628102   1.0
K12   K   0.515937821767612   0.578292929073181   0.825026685866558   1.0
K13   K   0.515928327202337   0.078282250160894   0.674977274748565   1.0
K14   K   0.484071485624944   0.921716864044669   0.325023536699573   1.0

## DFT(PBEsol)+U K2AgF4 B p=20GPa magn.: af2
data_1
_audit_creation_method   'vasp2cif'
_cell_length_a    5.7805905342
_cell_length_b    6.1673393249
_cell_length_c    9.53479194639
_cell_angle_alpha    90.2567596433
_cell_angle_beta     90.0368347168
_cell_angle_gamma    82.7993392947

_symmetry_space_group_name_H-M    'P 1'
loop_
_atom_site_label
_atom_site_type_symbol
_atom_site_fract_x
_atom_site_fract_y
_atom_site_fract_z
_atom_site_occupancy
Ag1   Ag  0.500000000000000   0.000000000000000   0.000000000000000   1.0
Ag2   Ag  0.000000000000000   0.000000000000000   0.000000000000000   1.0
Ag3   Ag  0.499999967000001   0.499999981999999   0.500000020999998   1.0
Ag4   Ag  0.000000002999997   0.499999981999999   0.500000020999998   1.0
F5    F   0.290783054000002   0.280900273000000   0.935992904000003   1.0
F6    F   0.790783075000000   0.280900273000000   0.935992904000003   1.0
F7    F   0.209216988999998   0.719099806999999   0.064007044999997   1.0
F8    F   0.709216968000000   0.719099806999999   0.064007044999997   1.0
F9    F   0.208505080000002   0.217112532999998   0.437493355000001   1.0
F10   F   0.708505037999998   0.217112532999998   0.437493355000001   1.0
F11   F   0.291494917000001   0.782887586999998   0.562506638000002   1.0
F12   F   0.791494917000001   0.782887586999998   0.562506638000002   1.0
F13   F   0.295824949000000   0.088007886000000   0.173378663999998   1.0
F14   F   0.795824867000000   0.088007886000000   0.173378663999998   1.0
F15   F   0.204174940000001   0.911992263000002   0.826621377999999   1.0
F16   F   0.704175001999999   0.911992263000002   0.826621377999999   1.0
F17   F   0.205098202000002   0.415247987000001   0.673309521000000   1.0
F18   F   0.705098180999997   0.415247987000001   0.673309521000000   1.0
F19   F   0.294901830999997   0.584752016000003   0.326690472000003   1.0
F20   F   0.794901871999997   0.584752016000003   0.326690472000003   1.0
K21   K   0.031063453000002   0.416628652999997   0.151862884000003   1.0
K22   K   0.531063463000002   0.416628652999997   0.151862884000003   1.0
K23   K   0.468936671999998   0.583371350000000   0.848137145999999   1.0
K24   K   0.968936630999998   0.583371350000000   0.848137145999999   1.0
K25   K   0.469037999999998   0.084901975999998   0.653181400000001   1.0
K26   K   0.969037999999998   0.084901975999998   0.653181400000001   1.0
K27   K   0.030962033999998   0.915097944000003   0.346818667000001   1.0
K28   K   0.530962023000001   0.915097944000003   0.346818667000001   1.0

## DFT(PBEsol)+U K2AgF4 B p=30GPa magn.: af2
data_1
_audit_creation_method   'vasp2cif'
_cell_length_a    5.63742733
_cell_length_b    6.09005355833
_cell_length_c    9.19685173033
```

```
_cell_angle_alpha     89.9302749635
_cell_angle_beta      89.9905624392
_cell_angle_gamma     84.1405487056

_symmetry_space_group_name_H-M    'P 1'
loop_
_atom_site_label
_atom_site_type_symbol
_atom_site_fract_x
_atom_site_fract_y
_atom_site_fract_z
_atom_site_occupancy
Ag1   Ag   0.500000000000000   0.000000000000000   0.000000000000000   1.0
Ag2   Ag   0.000000000000000   0.000000000000000   0.000000000000000   1.0
Ag3   Ag   0.499999948000003   0.500000000999997   0.500000008000001   1.0
Ag4   Ag   0.999999996000000   0.500000000999997   0.500000008000001   1.0
F5    F    0.285890170000002   0.279773198000001   0.932839244000000   1.0
F6    F    0.785890276000003   0.279773198000001   0.932839244000000   1.0
F7    F    0.214109792000002   0.720226981000003   0.067160713000000   1.0
F8    F    0.714109770000000   0.720226981000003   0.067160713000000   1.0
F9    F    0.214528604999998   0.220782051000001   0.432444699999998   1.0
F10   F    0.714528669000003   0.220782051000001   0.432444699999998   1.0
F11   F    0.285471329000003   0.779218186999998   0.567555341999999   1.0
F12   F    0.785471371000000   0.779218186999998   0.567555341999999   1.0
F13   F    0.291425881000002   0.090235044000003   0.176099977000000   1.0
F14   F    0.791425859999997   0.090235044000003   0.176099977000000   1.0
F15   F    0.208573973999997   0.909765145000001   0.823900051999999   1.0
F16   F    0.708573932000000   0.909765145000001   0.823900051999999   1.0
F17   F    0.208323548000003   0.408816655999999   0.676118428999999   1.0
F18   F    0.708323548000003   0.408816655999999   0.676118428999999   1.0
F19   F    0.291676471000002   0.591183385000001   0.323881534000002   1.0
F20   F    0.791676428999999   0.591183385000001   0.323881534000002   1.0
K21   K    0.028787004000002   0.413811877999997   0.151373503000002   1.0
K22   K    0.528786994000001   0.413811877999997   0.151373503000002   1.0
K23   K    0.471213114000001   0.586188124000003   0.848626525999997   1.0
K24   K    0.971213114000001   0.586188124000003   0.848626525999997   1.0
K25   K    0.471052761000003   0.085728349999997   0.650921746999998   1.0
K26   K    0.971052761000000   0.085728349999997   0.650921746999998   1.0
K27   K    0.028947260999999   0.914271573999997   0.349078243000001   1.0
K28   K    0.528947229000003   0.914271573999997   0.349078243000001   1.0

## DFT(PBEsol)+U K2AgF4 B p=40GPa magn.: af2
data_1
_audit_creation_method   'vasp2cif'
_cell_length_a    5.52279400358
_cell_length_b    6.0058881533
_cell_length_c    9.01232363269
_cell_angle_alpha    94.6616404134
_cell_angle_beta     90.9229141812
_cell_angle_gamma    84.4870849436

_symmetry_space_group_name_H-M    'P 1'
loop_
_atom_site_label
_atom_site_type_symbol
_atom_site_fract_x
_atom_site_fract_y
_atom_site_fract_z
_atom_site_occupancy
Ag1   Ag   0.500000000000000    -0.000000000000000   -0.000000000000000   1.0
Ag2   Ag   0.000000000000000     0.000000000000000   -0.000000000000000   1.0
Ag3   Ag   0.499999957999997     0.499999987000002    0.500000018000001   1.0
Ag4   Ag   0.000000000999996     0.499999987000002    0.500000018000001   1.0
F5    F    0.275920590398761     0.262841388644291    0.917873927321692   1.0
F6    F    0.775920547398761     0.262841388644291    0.917873927321692   1.0
F7    F    0.224079418601244     0.737158721355704    0.082126075678311   1.0
F8    F    0.724079396601242     0.737158721355704    0.082126075678311   1.0
```

```
F9    F    0.214357915472256    0.205022574220400    0.443066471113342    1.0
F10   F    0.714357915472256    0.205022574220400    0.443066471113342    1.0
F11   F    0.285642064527743    0.794977615779601    0.556933537886656    1.0
F12   F    0.785641934527746    0.794977615779601    0.556933537886656    1.0
F13   F    0.295373565976419    0.118754828702182    0.179827833127647    1.0
F14   F    0.795373522976419    0.118754828702182    0.179827833127647    1.0
F15   F    0.204626313023581    0.881245296297817    0.820172215872354    1.0
F16   F    0.704626248023579    0.881245296297817    0.820172215872354    1.0
F17   F    0.217663672523814    0.434129002346953    0.677347993479299    1.0
F18   F    0.717663715523814    0.434129002346953    0.677347993479299    1.0
F19   F    0.282336299476190    0.565871047653044    0.322652016520702    1.0
F20   F    0.782336321476185    0.565871047653044    0.322652016520702    1.0
K21   K    0.026788957307279    0.425318216920075    0.137341720370977    1.0
K22   K    0.526788930307280    0.425318216920075    0.137341720370977    1.0
K23   K    0.473211200692720    0.574681755079922    0.862658302629022    1.0
K24   K    0.973211157692720    0.574681755079922    0.862658302629022    1.0
K25   K    0.479907242984908    0.098245890222455    0.662555063627771    1.0
K26   K    0.979907156984908    0.098245890222455    0.662555063627771    1.0
K27   K    0.020092767015093    0.901754005777544    0.337444946372229    1.0
K28   K    0.520092810015093    0.901754005777544    0.337444946372229    1.0

## DFT(PBEsol)+U K2AgF4 B p=50GPa magn.: af1
data_1
_audit_creation_method    'vasp2cif'
_cell_length_a      5.43088971092
_cell_length_b      5.94098657332
_cell_length_c      8.86450604736
_cell_angle_alpha    84.2887194245
_cell_angle_beta     91.1952507418
_cell_angle_gamma    95.0705461439

_symmetry_space_group_name_H-M     'P 1'
loop_
_atom_site_label
_atom_site_type_symbol
_atom_site_fract_x
_atom_site_fract_y
_atom_site_fract_z
_atom_site_occupancy
Ag1   Ag   -0.000000000000000    0.000000000000000   -0.000000000000000    1.0
Ag2   Ag    0.500000000000000    0.000000000000000   -0.000000000000000    1.0
Ag3   Ag   -0.000000000000000    0.500000016999998    0.499999977000002    1.0
Ag4   Ag    0.500000004999997    0.500000016999998    0.499999977000002    1.0
F5    F    0.228511754114886    0.259564300806500    0.086277512995239    1.0
F6    F    0.728511776114888    0.259564300806500    0.086277512995239    1.0
F7    F    0.271488237885113    0.740435773193499    0.913722449004758    1.0
F8    F    0.771488193885116    0.740435773193499    0.913722449004758    1.0
F9    F    0.282944442015762    0.201874002060202    0.555912529248081    1.0
F10   F    0.782944398015766    0.201874002060202    0.555912529248081    1.0
F11   F    0.217055588984237    0.798126152939801    0.444087452751918    1.0
F12   F    0.717055577984239    0.798126152939801    0.444087452751918    1.0
F13   F    0.205181633196724    0.125013507488225    0.817682586412115    1.0
F14   F    0.705181611196729    0.125013507488225    0.817682586412115    1.0
F15   F    0.294818503803277    0.874986727511777    0.182317327587885    1.0
F16   F    0.794818546803277    0.874986727511777    0.182317327587885    1.0
F17   F    0.279859288508738    0.438052553591239    0.321557988512010    1.0
F18   F    0.779859267508733    0.438052553591239    0.321557988512010    1.0
F19   F    0.220140718491266    0.561947440408761    0.678442019487991    1.0
F20   F    0.720140696491264    0.561947440408761    0.678442019487991    1.0
K21   K    0.474551655330325    0.426492076714474    0.866047624990257    1.0
K22   K    0.974551699330321    0.426492076714474    0.866047624990257    1.0
K23   K    0.025448222669679    0.573507917285525    0.133952330009742    1.0
K24   K    0.525448220669679    0.573507917285525    0.133952330009742    1.0
K25   K    0.016490405473622    0.101267784760587    0.335702673588598    1.0
K26   K    0.516490366473622    0.101267784760587    0.335702673588598    1.0
K27   K    0.483509542526381    0.898732090239413    0.664297281411402    1.0
K28   K    0.983509586526378    0.898732090239413    0.664297281411402    1.0
```

```
## DFT(PBEsol)+U K2AgF4 B p=60GPa magn.: af2
data_1
_audit_creation_method   'vasp2cif'
_cell_length_a    5.35532117074
_cell_length_b    5.88154734445
_cell_length_c    8.74249952677
_cell_angle_alpha    96.2380262316
_cell_angle_beta    88.7294060388
_cell_angle_gamma    94.6518655779

_symmetry_space_group_name_H-M    'P 1'
loop_
_atom_site_label
_atom_site_type_symbol
_atom_site_fract_x
_atom_site_fract_y
_atom_site_fract_z
_atom_site_occupancy
Ag1   Ag   -0.000000000000000    0.000000000000000    0.000000000000000    1.0
Ag2   Ag    0.500000000000000    0.000000000000000    0.000000000000000    1.0
Ag3   Ag    0.999999999000003    0.499999989000003    0.499999997000003    1.0
Ag4   Ag    0.500000032999999    0.499999989000003    0.499999997000003    1.0
F5    F    0.219836791679525    0.299784302021039    0.055344130025403    1.0
F6    F    0.719836813679520    0.299784302021039    0.055344130025403    1.0
F7    F    0.280163190320481    0.700215768978964    0.944655931974595    1.0
F8    F    0.780163213320479    0.700215768978964    0.944655931974595    1.0
F9    F    0.268218855374384    0.241947611878555    0.588625786074426    1.0
F10   F    0.768421877374379    0.241947611878555    0.588625786074426    1.0
F11   F    0.231578197625617    0.758052474121445    0.411374179925571    1.0
F12   F    0.731578242625617    0.758052474121445    0.411374179925571    1.0
F13   F    0.221516189514924    0.060229939170225    0.820801965033463    1.0
F14   F    0.721516189514924    0.060229939170225    0.820801965033463    1.0
F15   F    0.278483967485079    0.939770289829776    0.179197972966539    1.0
F16   F    0.778483900485077    0.939770289829776    0.179197972966539    1.0
F17   F    0.293738321979140    0.371939576655348    0.315502787445723    1.0
F18   F    0.793738299979138    0.371939576655348    0.315502787445723    1.0
F19   F    0.206261700020862    0.628060426344649    0.684497233554282    1.0
F20   F    0.706261678020860    0.628060426344649    0.684497233554282    1.0
K21   K    0.485082300233443    0.397082979418370    0.834825230873466    1.0
K22   K    0.985082300233443    0.397082979418370    0.834825230873466    1.0
K23   K    0.014917599766556    0.602917007581633    0.165174817126538    1.0
K24   K    0.514917630766555    0.602917007581633    0.165174817126538    1.0
K25   K    0.024624471002731    0.073651572033620    0.367974360195543    1.0
K26   K    0.524624444002725    0.073651572033620    0.367974360195543    1.0
K27   K    0.475375518997275    0.926348263966380    0.632025660804456    1.0
K28   K    0.975375518997275    0.926348263966380    0.632025660804456    1.0

## DFT(PBEsol)+U K2AgF4 B p=70GPa magn.: af2
data_1
_audit_creation_method   'vasp2cif'
_cell_length_a    5.29090322805
_cell_length_b    6.34828005505
_cell_length_c    7.90830324044
_cell_angle_alpha    90.0003137565
_cell_angle_beta    89.9993641269
_cell_angle_gamma    86.7392774437

_symmetry_space_group_name_H-M    'P 1'
loop_
_atom_site_label
_atom_site_type_symbol
_atom_site_fract_x
_atom_site_fract_y
_atom_site_fract_z
_atom_site_occupancy
```

```
Ag1   Ag   -0.000000000000000    0.000000000000000    0.000000000000000    1.0
Ag2   Ag    0.500000000000000    0.000000000000000    0.000000000000000    1.0
Ag3   Ag    0.999999999000003    0.500000006999997    0.500000000000000    1.0
Ag4   Ag    0.499999997000003    0.500000006999997    0.500000000000000    1.0
F5    F     0.166292392978118    0.251406693715839    0.086225140062583    1.0
F6    F     0.666292403978116    0.251406693715839    0.086225140062583    1.0
F7    F     0.333707630021880    0.748593377284156    0.913774912937418    1.0
F8    F     0.833707630021880    0.748593377284156    0.913774912937418    1.0
F9    F     0.333689614719142    0.248599275569819    0.586226634583382    1.0
F10   F     0.833689547719140    0.248599275569819    0.586226634583382    1.0
F11   F     0.166310435280862    0.751400890430181    0.413773335416622    1.0
F12   F     0.666310435280862    0.751400890430181    0.413773335416622    1.0
F13   F     0.155767925602210    0.103725450915415    0.794958371086293    1.0
F14   F     0.655767948602215    0.103725450915415    0.794958371086293    1.0
F15   F     0.344232217397791    0.896274667084581    0.205041628913707    1.0
F16   F     0.844232172397791    0.896274667084581    0.205041628913707    1.0
F17   F     0.344230884223577    0.396268102008135    0.294959466349425    1.0
F18   F     0.844230906223579    0.396268102008135    0.294959466349425    1.0
F19   F     0.155769134776421    0.603731874991867    0.705040653650578    1.0
F20   F     0.655769134776421    0.603731874991867    0.705040653650578    1.0
K21   K     0.403722186234813    0.391801554738975    0.864042010683542    1.0
K22   K     0.903722231234813    0.391801554738975    0.864042010683542    1.0
K23   K     0.096277705765185    0.608198460261029    0.135958004316455    1.0
K24   K     0.596277761765183    0.608198460261029    0.135958004316455    1.0
K25   K     0.096271572606900    0.108196055709026    0.364046470440538    1.0
K26   K     0.596271617606900    0.108196055709026    0.364046470440538    1.0
K27   K     0.403728430393104    0.891803817290974    0.635953499559460    1.0
K28   K     0.903728430393104    0.891803817290974    0.635953499559460    1.0

## DFT(PBEsol)+U K2AgF4 B p=80GPa magn.: af2
data_1
_audit_creation_method    'vasp2cif'
_cell_length_a     5.23207592596
_cell_length_b     6.32886644472
_cell_length_c     7.78802609232
_cell_angle_alpha    89.999731689
_cell_angle_beta     90.0004627961
_cell_angle_gamma    86.0991952525

_symmetry_space_group_name_H-M    'P 1'
loop_
_atom_site_label
_atom_site_type_symbol
_atom_site_fract_x
_atom_site_fract_y
_atom_site_fract_z
_atom_site_occupancy
Ag1   Ag   -0.000000000000000   -0.000000000000000    0.000000000000000    1.0
Ag2   Ag    0.500000000000000   -0.000000000000000    0.000000000000000    1.0
Ag3   Ag    0.000000000999996    0.499999994000000    0.500000000999997    1.0
Ag4   Ag    0.499999973000001    0.499999994000000    0.500000000999997    1.0
F5    F     0.165196990157148    0.249585068649277    0.088069370359987    1.0
F6    F     0.665196967157149    0.249585068649277    0.088069370359987    1.0
F7    F     0.334802971842849    0.750414994350724    0.911930707640012    1.0
F8    F     0.834803017842853    0.750414994350724    0.911930707640012    1.0
F9    F     0.334822666365904    0.250409244911377    0.588068543434411    1.0
F10   F     0.834822689365902    0.250409244911377    0.588068543434411    1.0
F11   F     0.165177383634100    0.749590857088624    0.411931458565589    1.0
F12   F     0.665177383634100    0.749590857088624    0.411931458565589    1.0
F13   F     0.153405176569488    0.105779068061535    0.793201850344566    1.0
F14   F     0.653405199569487    0.105779068061535    0.793201850344566    1.0
F15   F     0.346594962430512    0.894221061938461    0.206798090655433    1.0
F16   F     0.846595008430509    0.894221061938461    0.206798090655433    1.0
F17   F     0.346599352065272    0.394226557818092    0.293200417454839    1.0
F18   F     0.846599306065276    0.394226557818092    0.293200417454839    1.0
F19   F     0.153400638934723    0.605773430181908    0.706799645545163    1.0
F20   F     0.653400638934723    0.605773430181908    0.706799645545163    1.0
```

```
K21  K   0.400852574624033    0.392455832840455    0.866342673538894   1.0
K22  K   0.900852528624029    0.392455832840455    0.866342673538894   1.0
K23  K   0.099147359375969    0.607544155159544    0.133657358461109   1.0
K24  K   0.599147371375970    0.607544155159544    0.133657358461109   1.0
K25  K   0.099155303042889    0.107546538480851    0.366339359061387   1.0
K26  K   0.599155268042889    0.107546538480851    0.366339359061387   1.0
K27  K   0.400844687957114    0.892453316519155    0.633660580938615   1.0
K28  K   0.900844687957114    0.892453316519155    0.633660580938615   1.0

## DFT(PBEsol)+U K2AgF4 B p=90GPa magn.: af2
data_1
_audit_creation_method   'vasp2cif'
_cell_length_a    5.17801508924
_cell_length_b    6.30532849176
_cell_length_c    7.69332671973
_cell_angle_alpha    89.9997791537
_cell_angle_beta     90.0005076835
_cell_angle_gamma    85.6672551559

_symmetry_space_group_name_H-M    'P 1'
loop_
_atom_site_label
_atom_site_type_symbol
_atom_site_fract_x
_atom_site_fract_y
_atom_site_fract_z
_atom_site_occupancy
Ag1  Ag   0.000000013117052   -0.000000030715845   -0.000000039093649   1.0
Ag2  Ag   0.500000011222850   -0.000000025708357   -0.000000044375240   1.0
Ag3  Ag   0.000000031241019    0.500000030101124    0.499999997753140   1.0
Ag4  Ag   0.499999960321301    0.500000094843020    0.500000005132886   1.0
F5   F    0.164652008902884    0.248456882369069    0.089483966252565   1.0
F6   F    0.664652033237610    0.248456901613340    0.089483986603641   1.0
F7   F    0.335347970766096    0.751543052327551    0.910516076389277   1.0
F8   F    0.835347968898103    0.751543038982531    0.910516069835621   1.0
F9   F    0.335353072006177    0.251540186410068    0.589483269512316   1.0
F10  F    0.835353102131122    0.251540186400400    0.589483283583186   1.0
F11  F    0.164646995030500    0.748459931632087    0.410516730906082   1.0
F12  F    0.664646964694651    0.748459931571618    0.410516731832228   1.0
F13  F    0.152326745614881    0.107108216547140    0.792048992443070   1.0
F14  F    0.652326743520454    0.107108247743575    0.792048957790495   1.0
F15  F    0.347673232967641    0.892891873309487    0.207951008359861   1.0
F16  F    0.847673242172248    0.892891859386132    0.207951025204425   1.0
F17  F    0.347676484360454    0.392893280145659    0.292051038443727   1.0
F18  F    0.847676473916137    0.392893287127247    0.292051009548273   1.0
F19  F    0.152323462517634    0.607106740616074    0.707949015110432   1.0
F20  F    0.652323453845034    0.607106724353080    0.707949013578141   1.0
K21  K    0.399446468467518    0.392839139243701    0.868025148602139   1.0
K22  K    0.899446464435240    0.392839122964111    0.868025153207470   1.0
K23  K    0.100553520026905    0.607160818818139    0.131974912774664   1.0
K24  K    0.600553495019093    0.607160832760934    0.131974907659881   1.0
K25  K    0.100556583910755    0.107160539150369    0.368023991614178   1.0
K26  K    0.600556583263983    0.107160547819451    0.368023964352451   1.0
K27  K    0.399443443068607    0.892839456356657    0.631976034790470   1.0
K28  K    0.899443472324032    0.892839440391641    0.631976018188255   1.0

## DFT(PBEsol)+U K2AgF4 B p=100GPa magn.: af1
data_1
_audit_creation_method   'vasp2cif'
_cell_length_a    5.12990963188
_cell_length_b    6.28066052336
_cell_length_c    7.60965154182
_cell_angle_alpha    89.9983249427
_cell_angle_beta     89.9997406305
_cell_angle_gamma    94.6985660816
```

```
_symmetry_space_group_name_H-M   'P 1'
loop_
_atom_site_label
_atom_site_type_symbol
_atom_site_fract_x
_atom_site_fract_y
_atom_site_fract_z
_atom_site_occupancy
Ag1   Ag   0.500000000000000    -0.000000000000000    -0.000000000000000    1.0
Ag2   Ag   0.000000000000000    -0.000000000000000    -0.000000000000000    1.0
Ag3   Ag   0.499999987000002     0.499999995000003     0.500000000000000    1.0
Ag4   Ag   0.999999998000000     0.499999995000003     0.500000000000000    1.0
F5    F    0.335582422580979     0.247637614338459     0.909346535239078    1.0
F6    F    0.835582468580976     0.247637614338459     0.909346535239078    1.0
F7    F    0.164417578419024     0.752362527661545     0.090653402760924    1.0
F8    F    0.664417566419023     0.752362527661545     0.090653402760924    1.0
F9    F    0.164418745889761     0.252360041297856     0.409344938803026    1.0
F10   F    0.664418722889756     0.252360041297856     0.409344938803026    1.0
F11   F    0.335581204110242     0.747640176702141     0.590655030196976    1.0
F12   F    0.835581274110241     0.747640176702141     0.590655030196976    1.0
F13   F    0.348355342001807     0.108128070127581     0.208846246508553    1.0
F14   F    0.848355273001805     0.108128070127581     0.208846246508553    1.0
F15   F    0.151644566998196     0.891872091872418     0.791153815491445    1.0
F16   F    0.651644578998197     0.891872091872418     0.791153815491445    1.0
F17   F    0.151642114801024     0.391870217187361     0.708847685023668    1.0
F18   F    0.651642091801019     0.391870217187361     0.708847685023668    1.0
F19   F    0.348357886198980     0.608129810812641     0.291152251976330    1.0
F20   F    0.848358025198980     0.608129810812641     0.291152251976330    1.0
K21   K    0.101550314981519     0.393169195735844     0.130626276679147    1.0
K22   K    0.601550297981521     0.393169195735844     0.130626276679147    1.0
K23   K    0.398449819018485     0.606830831264155     0.869373739320854    1.0
K24   K    0.898449772018485     0.606830831264155     0.869373739320854    1.0
K25   K    0.398449078501186     0.106831285946487     0.630627448759581    1.0
K26   K    0.898449101501184     0.106831285946487     0.630627448759581    1.0
K27   K    0.101550906498817     0.893168628053512     0.369372551240419    1.0
K28   K    0.601550910498817     0.893168628053512     0.369372551240419    1.0

## DFT(PBEsol)+U K2AgF4 C p=000GPa magn.: af1
data_1
_audit_creation_method   'vasp2cif'
_cell_length_a    6.03294182594
_cell_length_b    7.09772272292
_cell_length_c    7.08267455366
_cell_angle_alpha    66.0603848604
_cell_angle_beta     67.2553361129
_cell_angle_gamma    63.0869234026

_symmetry_space_group_name_H-M   'P 1'
loop_
_atom_site_label
_atom_site_type_symbol
_atom_site_fract_x
_atom_site_fract_y
_atom_site_fract_z
_atom_site_occupancy
Ag1   Ag   0.673409917601150     0.209095697838217     0.138370321258237    1.0
Ag2   Ag   0.414095795844345     0.293931928131945     0.611689178863406    1.0
F3    F    0.797939056664354     0.869105851125474     0.186731227363181    1.0
F4    F    0.289400292850642     0.633992965769111     0.563413651725370    1.0
F5    F    0.854966085433798     0.187738476254667     0.356039705345162    1.0
F6    F    0.543761264884316     0.251504940190291     0.875045223051448    1.0
F7    F    0.232427340548911     0.315404150117190     0.394003836492142    1.0
F8    F    0.043716853937960     0.251481981225319     0.875097058502045    1.0
F9    F    0.522097859472275     0.955857096824289     0.702416286529598    1.0
F10   F    0.565361859975211     0.547183015389689     0.047605729938393    1.0
K11   K    0.323271422913445     0.905701739453290     0.138634992676638    1.0
K12   K    0.046660802822641     0.007445912753590     0.680420503996528    1.0
```

```
K13   K   0.040824132158792    0.495622480032266    0.069624426361725   1.0
K14   K   0.764247734892162    0.597319384894665    0.611444397896119   1.0

## DFT(PBEsol)+U K2AgF4 C p=5GPa magn.: af1
data_1
_audit_creation_method   'vasp2cif'
_cell_length_a    5.67347173553
_cell_length_b    6.92599316359
_cell_length_c    6.91074913849
_cell_angle_alpha    64.3259769006
_cell_angle_beta    66.237320589
_cell_angle_gamma    63.221791696

_symmetry_space_group_name_H-M    'P 1'
loop_
_atom_site_label
_atom_site_type_symbol
_atom_site_fract_x
_atom_site_fract_y
_atom_site_fract_z
_atom_site_occupancy
Ag1   Ag   0.662457880076017    0.203340027677845    0.150522430001661   1.0
Ag2   Ag   0.424943383900326    0.299717565973584    0.599569740933319   1.0
F3    F    0.807323950727364    0.857140013657078    0.189352890181827   1.0
F4    F    0.280136057482622    0.645893421574517    0.560732350697721   1.0
F5    F    0.874758734710240    0.182456870786834    0.354157256321161   1.0
F6    F    0.543702217412797    0.251524757718723    0.875035058503865   1.0
F7    F    0.212733194086935    0.320586809003303    0.395936943300077   1.0
F8    F    0.043762083244554    0.251536060313681    0.874999713984800   1.0
F9    F    0.543349688770457    0.950787598583886    0.701607900761864   1.0
F10   F    0.544094359024521    0.552286042672024    0.048481947069018   1.0
K11   K    0.317738910528787    0.906352509390755    0.146170305668899   1.0
K12   K    0.042642872143943    -0.005414062777867   0.699815707563671   1.0
K13   K    0.044815368265686    0.508476174387982    0.050257116245437   1.0
K14   K    0.769721719625753    0.596701831037658    0.603897178766672   1.0

## DFT(PBEsol)+U K2AgF4 C p=10GPa magn.: af1
data_1
_audit_creation_method   'vasp2cif'
_cell_length_a    5.47508872249
_cell_length_b    6.97496263879
_cell_length_c    6.82850004518
_cell_angle_alpha    62.1417245436
_cell_angle_beta    67.2780835702
_cell_angle_gamma    58.5678405555

_symmetry_space_group_name_H-M    'P 1'
loop_
_atom_site_label
_atom_site_type_symbol
_atom_site_fract_x
_atom_site_fract_y
_atom_site_fract_z
_atom_site_occupancy
Ag1   Ag   0.655429745869612    0.209617337834960    0.151255911335075   1.0
Ag2   Ag   0.431986260725651    0.293457977026614    0.598769358367270   1.0
F3    F    0.834044762222887    0.850839735193117    0.194647508537037   1.0
F4    F    0.253372879454315    0.652137976681228    0.555573708920865   1.0
F5    F    0.880355220863418    0.193225944729262    0.350283062436089   1.0
F6    F    0.543554483517975    0.251528107337282    0.875051012048496   1.0
F7    F    0.206666451982923    0.309768807297682    0.399767571913875   1.0
F8    F    0.043766148656279    0.251646871214682    0.874969820353655   1.0
F9    F    0.618111037534409    0.924318521176306    0.723231371374745   1.0
F10   F    0.469271456557783    0.578760560626416    0.026748617638610   1.0
K11   K    0.336448880725425    0.902533733536274    0.154120109321009   1.0
K12   K    0.082080553675923    0.973982924970872    0.718200108084335   1.0
```

```
K13   K   0.005406506247410   0.529025230284406   0.032025795545090   1.0
K14   K   0.751106031965985   0.600541892090895   0.595892584123863   1.0

## DFT(PBEsol)+U K2AgF4 C p=20GPa magn.: af1
data_1
_audit_creation_method   'vasp2cif'
_cell_length_a    5.20657120292
_cell_length_b    6.91998552494
_cell_length_c    6.65155084058
_cell_angle_alpha    59.671537201
_cell_angle_beta     66.2431521174
_cell_angle_gamma    57.8698275175

_symmetry_space_group_name_H-M    'P 1'
loop_
_atom_site_label
_atom_site_type_symbol
_atom_site_fract_x
_atom_site_fract_y
_atom_site_fract_z
_atom_site_occupancy
Ag1   Ag   0.651365523495639   0.191196616135052   0.169747356242096   1.0
Ag2   Ag   0.436195114286757   0.311893876291158   0.580527738288312   1.0
F3    F    0.864002301551025   0.835902946236333   0.189870213139614   1.0
F4    F    0.223499439043668   0.667127646813665   0.560040703297999   1.0
F5    F    0.914430212166290   0.171064585337390   0.349504618576681   1.0
F6    F    0.543134655202927   0.251361387442029   0.875584920392963   1.0
F7    F    0.173510678530 40   0.331951216339572   0.400646835475972   1.0
F8    F    0.043996376286720   0.251688497672041   0.874911129131702   1.0
F9    F    0.620128074318176   0.930925892841295   0.715373729257581   1.0
F10   F    0.467712308600028   0.572207406196258   0.034330469839260   1.0
K11   K    0.358930652205328   0.892407096447816   0.147920290681436   1.0
K12   K    0.090535509076978   0.970094531018844   0.729910343132531   1.0
K13   K    0.996908494038900   0.532931029150765   0.020321172818861   1.0
K14   K    0.728340691874532   0.610632892077793   0.601847019724989   1.0

## DFT(PBEsol)+U K2AgF4 C p=30GPa magn.: af1
data_1
_audit_creation_method   'vasp2cif'
_cell_length_a    5.01548866811
_cell_length_b    6.88841166503
_cell_length_c    6.52869620901
_cell_angle_alpha    58.0502705109
_cell_angle_beta     65.5137012344
_cell_angle_gamma    57.4531008618

_symmetry_space_group_name_H-M    'P 1'
loop_
_atom_site_label
_atom_site_type_symbol
_atom_site_fract_x
_atom_site_fract_y
_atom_site_fract_z
_atom_site_occupancy
Ag1   Ag   0.646074242273357   0.176887250630981   0.185703742503100   1.0
Ag2   Ag   0.441386532188265   0.326227490266691   0.564461424309548   1.0
F3    F    0.882637349039347   0.827238634253716   0.190531748556656   1.0
F4    F    0.204744464681249   0.675866098729179   0.559603510057344   1.0
F5    F    0.941815821796305   0.158103779182596   0.345029777938797   1.0
F6    F    0.543294409888935   0.251525754863408   0.875149383946281   1.0
F7    F    0.145005992132794   0.344905951488024   0.405001375478888   1.0
F8    F    0.043770942449154   0.251512773130073   0.875096148972609   1.0
F9    F    0.620342704517795   0.938520360651468   0.705370524338849   1.0
F10   F    0.467198041635901   0.564668101697004   0.044307985861950   1.0
K11   K    0.376902071023159   0.885421231001090   0.142085663468969   1.0
K12   K    0.091685208110412   0.969157572344194   0.735990480008248   1.0
```

```
K13    K    0.995669562897294    0.533943586570613    0.014249711366529    1.0
K14    K    0.710618077366029    0.617407035190975    0.607955063192238    1.0

## DFT(PBEsol)+U K2AgF4 C p=40GPa magn.: f
data_1
_audit_creation_method   'vasp2cif'
_cell_length_a    4.75729022204
_cell_length_b    6.86518718462
_cell_length_c    6.22039669057
_cell_angle_alpha    55.3422715498
_cell_angle_beta     73.4761733117
_cell_angle_gamma    60.2950594903

_symmetry_space_group_name_H-M   'P 1'
loop_
_atom_site_label
_atom_site_type_symbol
_atom_site_fract_x
_atom_site_fract_y
_atom_site_fract_z
_atom_site_occupancy
Ag1    Ag    0.754576176693057    0.115027784792048    0.225932741246652    1.0
Ag2    Ag    0.317712694145222    0.392273213405517    0.524214294130104    1.0
F3     F     0.960310573396171    0.784054317990815    0.193547511016249    1.0
F4     F     0.124778867165470    0.728006137306835    0.545185498327079    1.0
F5     F     0.304425344975911    0.089553389804848    0.324765718906676    1.0
F6     F     0.529456984967389    0.259723803105487    0.853825335429157    1.0
F7     F     0.828419057698063    0.411359005204794    0.424616502736100    1.0
F8     F     0.029563113937392    0.239526308078979    0.901812385512272    1.0
F9     F     0.589174198255435    0.995453528518809    0.662337504405613    1.0
F10    F     0.492526428214056    0.507190418673183    0.092749320523673    1.0
K11    K     0.457632625466900    0.848213549519564    0.122932894870267    1.0
K12    K     0.058191176754988    0.986258414704764    0.738054473966301    1.0
K13    K     0.029534597788185    0.518701712472698    0.006425111571623    1.0
K14    K     0.635679580541769    0.646044036421685    0.634137247358225    1.0

## DFT(PBEsol)+U K2AgF4 C p=50GPa magn.: f
data_1
_audit_creation_method   'vasp2cif'
_cell_length_a    4.69670784246
_cell_length_b    6.02013594382
_cell_length_c    6.02133741057
_cell_angle_alpha    113.573914887
_cell_angle_beta     103.389524188
_cell_angle_gamma    105.602075545

_symmetry_space_group_name_H-M   'P 1'
loop_
_atom_site_label
_atom_site_type_symbol
_atom_site_fract_x
_atom_site_fract_y
_atom_site_fract_z
_atom_site_occupancy
Ag1    Ag    0.398267036119565    0.227538158094896    0.340856175756235    1.0
Ag2    Ag    0.517073105716672    0.518666272544502    0.903896847408478    1.0
F3     F     0.225510635252058    0.206187358767089    0.978813321863925    1.0
F4     F     0.690537525327310    0.553422837930355    0.276593637229969    1.0
F5     F     0.980102691681857    0.327370465905180    0.425943749442384    1.0
F6     F     0.456583428443511    0.850092955632474    0.112870164173081    1.0
F7     F     0.900111272640481    0.424313780923986    0.831117148517302    1.0
F8     F     0.981150901733908    0.893275065413324    0.137586029074449    1.0
F9     F     0.541972506850388    0.668494821272716    0.656152001414099    1.0
F10    F     0.373197687097540    0.080480170428085    0.595486073245248    1.0
K11    K     0.713197372411108    0.122001857222311    0.978815154654632    1.0
K12    K     0.076254353898724    0.747126108742241    0.726657747753781    1.0
```

```
K13   K   0.837741566893305   0.007568952721957   0.529143084523659   1.0
K14   K   0.196119515933566   0.623997764400886   0.277991054942770   1.0

## DFT(PBEsol)+U K2AgF4 C p=60GPa magn.: f
data_1
_audit_creation_method   'vasp2cif'
_cell_length_a    4.63686754809
_cell_length_b    5.95226560859
_cell_length_c    5.95432605191
_cell_angle_alpha    113.504709613
_cell_angle_beta    103.711545242
_cell_angle_gamma    105.639008089

_symmetry_space_group_name_H-M    'P 1'
loop_
_atom_site_label
_atom_site_type_symbol
_atom_site_fract_x
_atom_site_fract_y
_atom_site_fract_z
_atom_site_occupancy
Ag1   Ag   0.408151359081366   0.237073958414797   0.356275747785517   1.0
Ag2   Ag   0.508654782592281   0.521274712016741   0.909809948479293   1.0
F3    F    0.220748977027185   0.197273807817432   0.975364487831534   1.0
F4    F    0.687982592273890   0.543572946339154   0.275087422545532   1.0
F5    F    0.014526131816979   0.325543656124594   0.425308874047801   1.0
F6    F    0.455209085047549   0.899038288341572   0.139724432444964   1.0
F7    F    0.924045479563624   0.421998435922717   0.822790435323547   1.0
F8    F    0.933419010710431   0.858610856049573   0.116649134653450   1.0
F9    F    0.541952991679715   0.673082153552812   0.657464224127807   1.0
F10   F    0.366627745444689   0.075389417300357   0.594796181586652   1.0
K11   K    0.716995211642697   0.127358158515314   0.976608212900267   1.0
K12   K    0.077883977171013   0.744738513058046   0.723663630530929   1.0
K13   K    0.834134236942937   0.001032705596614   0.527069167696055   1.0
K14   K    0.197488019005640   0.624548960950264   0.271310290046650   1.0

## DFT(PBEsol)+U K2AgF4 C p=70GPa magn.: f
data_1
_audit_creation_method   'vasp2cif'
_cell_length_a    4.58393947221
_cell_length_b    5.89317023134
_cell_length_c    5.89860871193
_cell_angle_alpha    113.471684318
_cell_angle_beta    103.954188382
_cell_angle_gamma    105.651066274

_symmetry_space_group_name_H-M    'P 1'
loop_
_atom_site_label
_atom_site_type_symbol
_atom_site_fract_x
_atom_site_fract_y
_atom_site_fract_z
_atom_site_occupancy
Ag1   Ag   0.408879208607352   0.230188399037608   0.345650278788682   1.0
Ag2   Ag   0.493536644205264   0.507975802584063   0.890973758022405   1.0
F3    F    0.223904640372851   0.206737453950417   0.977771637659018   1.0
F4    F    0.693045728364126   0.553145845539149   0.278882886747525   1.0
F5    F    0.993322306269898   0.328309025043865   0.433038476946180   1.0
F6    F    0.458702310190411   0.851857146498617   0.113972917385169   1.0
F7    F    0.896895290196397   0.424795470063621   0.825536723640106   1.0
F8    F    0.977997637813900   0.890118113241904   0.135562244225184   1.0
F9    F    0.548959629284916   0.678660886546062   0.660102843262990   1.0
F10   F    0.368408315289784   0.074746858465481   0.595856279059710   1.0
K11   K    0.716061510591381   0.127857404226226   0.983526612153526   1.0
K12   K    0.080255355862057   0.750544046877714   0.725768784615816   1.0
```

```
K13   K   0.832306602835729   0.003806975591864   0.529764286623646   1.0
K14   K   0.195334420115920   0.621793142333402   0.275514460870049   1.0

## DFT(PBEsol)+U K2AgF4 C p=80GPa magn.: f
data_1
_audit_creation_method   'vasp2cif'
_cell_length_a    4.53803511503
_cell_length_b    5.8408889786
_cell_length_c    5.84923665869
_cell_angle_alpha    113.470947969
_cell_angle_beta    104.100682979
_cell_angle_gamma    105.651329476

_symmetry_space_group_name_H-M    'P 1'
loop_
_atom_site_label
_atom_site_type_symbol
_atom_site_fract_x
_atom_site_fract_y
_atom_site_fract_z
_atom_site_occupancy
Ag1   Ag   0.413097309616901   0.231487520678861   0.347852291518770   1.0
Ag2   Ag   0.486162459463252   0.504585167546324   0.886980753002661   1.0
F3    F    0.223647156641171   0.207050107988380   0.977554237697710   1.0
F4    F    0.694022207107558   0.553361562523612   0.279686683129491   1.0
F5    F    0.996606176743505   0.328199515363934   0.434623844679997   1.0
F6    F    0.459053921179965   0.852508425006384   0.114356483392342   1.0
F7    F    0.895994488047677   0.425074358888786   0.823910534434397   1.0
F8    F    0.976434435019525   0.889170574251345   0.134749781792309   1.0
F9    F    0.551146716775683   0.681568025173483   0.661408753225700   1.0
F10   F    0.366529796639540   0.072537936442217   0.595641267627173   1.0
K11   K    0.716824705830604   0.129669167361335   0.984747204254660   1.0
K12   K    0.081732841432419   0.751704727480480   0.725582729811713   1.0
K13   K    0.830492132803045   0.002593967992532   0.529965232596396   1.0
K14   K    0.195486252699148   0.621025513302321   0.274862392836702   1.0

## DFT(PBEsol)+U K2AgF4 C p=90GPa magn.: af1
data_1
_audit_creation_method   'vasp2cif'
_cell_length_a    4.49860099421
_cell_length_b    5.73922939859
_cell_length_c    5.77321194311
_cell_angle_alpha    113.016212257
_cell_angle_beta    103.793719414
_cell_angle_gamma    105.382731076

_symmetry_space_group_name_H-M    'P 1'
loop_
_atom_site_label
_atom_site_type_symbol
_atom_site_fract_x
_atom_site_fract_y
_atom_site_fract_z
_atom_site_occupancy
Ag1   Ag   0.457193577981313   0.254471126702698   0.378158429473345   1.0
Ag2   Ag   0.455934267033396   0.495852605764583   0.875352062011463   1.0
F3    F    0.225009977607044   0.212123394344335   0.966059586230199   1.0
F4    F    0.687187613093289   0.537854147234804   0.287094737998163   1.0
F5    F    0.032417452648927   0.331490416187363   0.441071632329804   1.0
F6    F    0.456161528970063   0.874756175808507   0.126451089929629   1.0
F7    F    0.880027288428986   0.418526778163973   0.811968357513307   1.0
F8    F    0.956517087478305   0.875278913938930   0.126602669634523   1.0
F9    F    0.553672358087977   0.693330875860510   0.668335498742858   1.0
F10   F    0.358925797919256   0.056829166893633   0.584711202503938   1.0
K11   K    0.725389392916843   0.133263048716351   0.986779728098887   1.0
K12   K    0.084904451403723   0.758159933959871   0.733031468510742   1.0
```

```
K13   K   0.827711415208317   0.991903397978015   0.520013046950641   1.0
K14   K   0.186767391222567   0.616696588446429   0.266292680072498   1.0

## DFT(PBEsol)+U K2AgF4 C p=100GPa magn.: af1
data_1
_audit_creation_method    'vasp2cif'
_cell_length_a     4.46059922009
_cell_length_b     5.69900371671
_cell_length_c     5.82610045646
_cell_angle_alpha     112.954379784
_cell_angle_beta      105.751027396
_cell_angle_gamma     105.342821034

_symmetry_space_group_name_H-M    'P 1'
loop_
_atom_site_label
_atom_site_type_symbol
_atom_site_fract_x
_atom_site_fract_y
_atom_site_fract_z
_atom_site_occupancy
Ag1   Ag   0.920059748660868   0.123213332563819   0.378306097259380   1.0
Ag2   Ag   0.420797795060215   0.379755350257880   0.874626531356601   1.0
F3    F    0.742271546789800   0.755553737181670   0.966805229945816   1.0
F4    F    0.597821991434894   0.747414740013048   0.286273117325046   1.0
F5    F    0.409724467452631   0.109731267869910   0.440531947948704   1.0
F6    F    0.670442417899843   0.251242221387018   0.126705850847482   1.0
F7    F    0.931242693683076   0.393367695249568   0.812707676415972   1.0
F8    F    0.170638255864885   0.251808263458185   0.126549821689809   1.0
F9    F    0.113937842419306   0.974929130491784   0.668090077871254   1.0
F10   F    0.226737544237016   0.528313873547354   0.585089659323738   1.0
K11   K    0.261562203977990   0.853082126140135   0.986528330558793   1.0
K12   K    0.646884782401549   0.974218415988766   0.732290798359257   1.0
K13   K    0.693781371072451   0.528834254981184   0.520817405462622   1.0
K14   K    0.078551939045490   0.649921230869693   0.266599645635531   1.0

## DFT(PBEsol)+U K2AgF4 D p=000GPa magn.: f
data_1
_audit_creation_method    'vasp2cif'
_cell_length_a     4.90613313321
_cell_length_b     7.52063247457
_cell_length_c     7.5299786112
_cell_angle_alpha     100.797300003
_cell_angle_beta      108.894371377
_cell_angle_gamma     109.003038006

_symmetry_space_group_name_H-M    'P 1'
loop_
_atom_site_label
_atom_site_type_symbol
_atom_site_fract_x
_atom_site_fract_y
_atom_site_fract_z
_atom_site_occupancy
Ag1   Ag   0.981171002196058   0.680802539693155   0.407064272332176   1.0
Ag2   Ag   0.468136637045686   0.180052899419889   0.906707817936935   1.0
F3    F    0.462190875117354   0.356138604525132   0.731879789611325   1.0
F4    F    0.161011145181227   0.737022922697658   0.712118456011360   1.0
F5    F    0.965162843790149   0.180710301431445   0.906998421514967   1.0
F6    F    0.800196051675528   0.375270467235382   0.350131866254996   1.0
F7    F    0.481001377454015   0.681069237639102   0.406809404878531   1.0
F8    F    0.160993852503871   0.986336456005439   0.464339634150773   1.0
F9    F    0.798773512123050   0.624660892919247   0.102151199730578   1.0
F10   F    0.456209711865095   0.004727589012881   0.082363142671219   1.0
K11   K    0.728576660657952   0.774587233667831   0.809992205339421   1.0
K12   K    0.231309473474049   0.276308138849338   0.311209213656970   1.0
```

```
K13  K  0.231440694606426   0.587158679363679   0.004199196243683   1.0
K14  K  0.729559842309551   0.085556447539835   0.502801869667055   1.0

## DFT(PBEsol)+U K2AgF4 D p=5GPa magn.: f
data_1
_audit_creation_method   'vasp2cif'
_cell_length_a    4.76227725881
_cell_length_b    7.26482340535
_cell_length_c    7.26565077189
_cell_angle_alpha    101.202380457
_cell_angle_beta    109.047320999
_cell_angle_gamma    109.172912656

_symmetry_space_group_name_H-M    'P 1'
loop_
_atom_site_label
_atom_site_type_symbol
_atom_site_fract_x
_atom_site_fract_y
_atom_site_fract_z
_atom_site_occupancy
Ag1  Ag  0.981133097507813   0.680746337251978   0.407076849150092   1.0
Ag2  Ag  0.468640854398579   0.180638508502877   0.907022879497733   1.0
F3   F   0.462402028598916   0.363801767123885   0.724374626793925   1.0
F4   F   0.169251796848867   0.744486686837398   0.719783044764366   1.0
F5   F   0.966655855996845   0.180679598442975   0.907028333478400   1.0
F6   F   0.792638317416717   0.367643812890301   0.342323789555216   1.0
F7   F   0.480994473137795   0.680768661346114   0.407032758950021   1.0
F8   F   0.167939066620229   0.993864505828109   0.471765582140149   1.0
F9   F   0.790863961350732   0.617211486809017   0.094453393813851   1.0
F10  F   0.455264524383615  -0.002325010600323   0.089784530677446   1.0
K11  K   0.735686324042920   0.780509419493713   0.817434963162948   1.0
K12  K   0.225187950734126   0.269315920673796   0.306321957108845   1.0
K13  K   0.223923795025609   0.580825544158980  -0.003420467978616   1.0
K14  K   0.735241633937250   0.092235171241194   0.507784248885613   1.0

## DFT(PBEsol)+U K2AgF4 D p=10GPa magn.: af1
data_1
_audit_creation_method   'vasp2cif'
_cell_length_a    4.65827333147
_cell_length_b    7.29358329734
_cell_length_c    7.29278491311
_cell_angle_alpha    99.1973140273
_cell_angle_beta    113.633509601
_cell_angle_gamma    113.628092751

_symmetry_space_group_name_H-M    'P 1'
loop_
_atom_site_label
_atom_site_type_symbol
_atom_site_fract_x
_atom_site_fract_y
_atom_site_fract_z
_atom_site_occupancy
Ag1  Ag  0.975765539730934   0.680909860925513   0.407145636129810   1.0
Ag2  Ag  0.475713688931243   0.180786534745826   0.907144402421617   1.0
F3   F   0.482889405796682   0.303665654539281   0.677181134541240   1.0
F4   F   0.223675873225477   0.846103656349175   0.741342327404955   1.0
F5   F   0.974032236055097   0.180037121014709   0.906939159544042   1.0
F6   F   0.982797988935387   0.450868882893075   0.529976342736996   1.0
F7   F   0.474877132495626   0.680652785194807   0.406399198277065   1.0
F8   F   0.970026864996262   0.911065081090149   0.284429228858059   1.0
F9   F   0.723684045816360   0.515056165454095   0.072438960058495   1.0
F10  F   0.469882643711933   0.058092443933115   0.137367767988401   1.0
K11  K   0.724276409071417   0.763521988939209   0.824472572347314   1.0
K12  K   0.226759006387307   0.283081053049651   0.305070989581260   1.0
```

```
K13   K   0.223969841609149   0.597857656944920   0.989561667979582   1.0
K14   K   0.726702003237117   0.078703524926492   0.509297102131161   1.0
```

## DFT(PBEsol)+U K2AgF4 D p=20GPa magn.: af1
```
data_1
_audit_creation_method   'vasp2cif'
_cell_length_a    4.52910056368
_cell_length_b    6.97156519213
_cell_length_c    6.97062123804
_cell_angle_alpha    106.85931494
_cell_angle_beta    108.954497962
_cell_angle_gamma    109.006580669

_symmetry_space_group_name_H-M    'P 1'
loop_
_atom_site_label
_atom_site_type_symbol
_atom_site_fract_x
_atom_site_fract_y
_atom_site_fract_z
_atom_site_occupancy
Ag1   Ag   0.975622712552226   0.680726808510253   0.407039661190718   1.0
Ag2   Ag   0.475177739935575   0.180726790582565   0.907028518235516   1.0
F3    F    0.475456126283771   0.392789103267975   0.695046240534658   1.0
F4    F    0.167914977038548   0.753322127750519   0.718663176650850   1.0
F5    F    0.975120413197512   0.180774550396036   0.907082155761253   1.0
F6    F    0.782702487914903   0.368676218329063   0.332878133878898   1.0
F7    F    0.475652383841617   0.680755760217275   0.407097292573796   1.0
F8    F    0.168378382527566   -0.007218245874586  0.481196680491835   1.0
F9    F    0.783170560584245   0.608133715809385   0.095419911081275   1.0
F10   F    0.474475057207956   0.968711437274422   0.119073777565073   1.0
K11   K    0.756173675047079   0.799392451345449   0.849284894166080   1.0
K12   K    0.194708857483233   0.237795330035319   0.288297428891516   1.0
K13   K    0.194909536428556   0.562126827599282   0.964841918458783   1.0
K14   K    0.756360769957202   0.123689534757050   0.525816700519744   1.0
```

## DFT(PBEsol)+U K2AgF4 D p=30GPa magn.: af1
```
data_1
_audit_creation_method   'vasp2cif'
_cell_length_a    4.42221645251
_cell_length_b    6.85613617315
_cell_length_c    6.81941708933
_cell_angle_alpha    107.484374964
_cell_angle_beta    108.413710125
_cell_angle_gamma    109.335015044

_symmetry_space_group_name_H-M    'P 1'
loop_
_atom_site_label
_atom_site_type_symbol
_atom_site_fract_x
_atom_site_fract_y
_atom_site_fract_z
_atom_site_occupancy
Ag1   Ag   0.975413272400692   0.680754833273448   0.407069715652079   1.0
Ag2   Ag   0.475501470980699   0.180766131245364   0.907075896820549   1.0
F3    F    0.476317649486969   0.396326984404470   0.691510405117556   1.0
F4    F    0.173163426857332   0.754796000761909   0.722303480927458   1.0
F5    F    0.975887259476199   0.180749625195804   0.907052045518240   1.0
F6    F    0.783539309939549   0.365258619468521   0.332197659143895   1.0
F7    F    0.475363618843379   0.680685961521422   0.407035936104107   1.0
F8    F    0.167327364290136   -0.003764328739390  0.481958296783498   1.0
F9    F    0.777639423482225   0.606732199144917   0.091804892379170   1.0
F10   F    0.474489392976141   0.965163188534242   0.122593151204887   1.0
K11   K    0.761906452392912   0.802565002495840   0.854851424962611   1.0
K12   K    0.192124850672823   0.232503158922680   0.285169671014830   1.0
```

```
K13  K   0.188860111577330   0.558920787918655   0.959223217940328   1.0
K14  K   0.758690076623598   0.128944245852142   0.528920696430802   1.0

## DFT(PBEsol)+U K2AgF4 D p=40GPa magn.: f
data_1
_audit_creation_method   'vasp2cif'
_cell_length_a    4.68054241982
_cell_length_b    6.93859875308
_cell_length_c    5.86099037487
_cell_angle_alpha    95.8779215011
_cell_angle_beta    98.5344986968
_cell_angle_gamma    123.366098637

_symmetry_space_group_name_H-M    'P 1'
loop_
_atom_site_label
_atom_site_type_symbol
_atom_site_fract_x
_atom_site_fract_y
_atom_site_fract_z
_atom_site_occupancy
Ag1  Ag  0.975310688831861   0.680570912779707   0.407204174875169   1.0
Ag2  Ag  0.475528254863683   0.180852544618233   0.906871379986519   1.0
F3   F   0.679322385553611   0.404152457943894   0.683581001069243   1.0
F4   F   0.298321974859050   0.812513677638790   0.737091919239248   1.0
F5   F   0.975754035667343   0.181056739093208   0.906785896074233   1.0
F6   F   0.836435413421192   0.350878892255700   0.275562204971870   1.0
F7   F   0.475136876423491   0.680453514203990   0.407299750667200   1.0
F8   F   0.114043542728915   0.010322403214786   0.538387206343153   1.0
F9   F   0.652697093504628   0.549070112941990   0.077251707001693   1.0
F10  F   0.271691855765120   0.957607649570043   0.130373279277627   1.0
K11  K   0.760091208046504   0.776452363495572   0.806916457685958   1.0
K12  K   0.264123399338928   0.280703585758822   0.311079301454576   1.0
K13  K   0.191137520954300   0.585407697981012   0.007639117610417   1.0
K14  K   0.686229430041372   0.080359858504266   0.502723093743100   1.0

## DFT(PBEsol)+U K2AgF4 D p=50GPa magn.: af1
data_1
_audit_creation_method   'vasp2cif'
_cell_length_a    4.36644319958
_cell_length_b    6.55461964231
_cell_length_c    6.276685092
_cell_angle_alpha    116.761343284
_cell_angle_beta    95.517848744
_cell_angle_gamma    111.027775708

_symmetry_space_group_name_H-M    'P 1'
loop_
_atom_site_label
_atom_site_type_symbol
_atom_site_fract_x
_atom_site_fract_y
_atom_site_fract_z
_atom_site_occupancy
Ag1  Ag  0.712118273622815   0.744220162087384   0.255040538500923   1.0
Ag2  Ag  0.698006994229697   0.188764269301419   0.292294724566219   1.0
F3   F   0.892692427299597   0.408469962289481   0.657631531555859   1.0
F4   F   0.517942631210117   0.524400628328955   0.889721515739506   1.0
F5   F   0.117771166945407   0.123021351365922   0.333850645735804   1.0
F6   F   0.421156563639804   0.301423250244050   0.112865817119579   1.0
F7   F   0.292850735120234   0.809904692131069   0.213539825719250   1.0
F8   F   0.808740466230443   0.976285656361046   0.616920849005466   1.0
F9   F   0.989552527000364   0.631585009055618   0.434611448699403   1.0
F10  F   0.601832316731267   0.956528619871395   0.930441416817955   1.0
K11  K   0.037307434147949   0.840999013742923   0.889678130885602   1.0
K12  K   0.980063210157233   0.373757273541986   0.999772689665002   1.0
```

```
K13   K   0.430597278049113   0.559127503752413   0.547575988597692   1.0
K14   K   0.373346725615965   0.091915057926338   0.657690847391741   1.0

## DFT(PBEsol)+U K2AgF4 D p=60GPa magn.: af1
data_1
_audit_creation_method   'vasp2cif'
_cell_length_a    4.36748776554
_cell_length_b    6.46161580204
_cell_length_c    6.10765258211
_cell_angle_alpha    116.114562825
_cell_angle_beta    96.1825965681
_cell_angle_gamma    111.294931323

_symmetry_space_group_name_H-M    'P 1'
loop_
_atom_site_label
_atom_site_type_symbol
_atom_site_fract_x
_atom_site_fract_y
_atom_site_fract_z
_atom_site_occupancy
Ag1   Ag   0.715495676053985   0.746249901490913   0.242451481345869   1.0
Ag2   Ag   0.695164291125460   0.186636753080424   0.304929764474931   1.0
F3    F    0.907528718037279   0.419039552326410   0.673288123353892   1.0
F4    F    0.503020564145818   0.513902372711853   0.874109669292048   1.0
F5    F    0.130033923265106   0.132271705703584   0.336325645114344   1.0
F6    F    0.434993068111348   0.301587237994436   0.108295456564969   1.0
F7    F    0.280439506425959   0.800606539728565   0.211020855256193   1.0
F8    F    0.817775368145449   0.985105016182657   0.612790916621559   1.0
F9    F    0.976308143500260   0.631373303765439   0.439138408513194   1.0
F10   F    0.592809541740350   0.947801460504644   0.934538199373301   1.0
K11   K    0.032864192469578   0.845389363280420   0.889135101895274   1.0
K12   K    0.982748213683276   0.365340526334101   0.009260708870510   1.0
K13   K    0.427947461035831   0.567528383627753   0.538089637630135   1.0
K14   K    0.377750082260318   0.087570333268796   0.658262001693783   1.0

## DFT(PBEsol)+U K2AgF4 D p=70GPa magn.: af1
data_1
_audit_creation_method   'vasp2cif'
_cell_length_a    4.38343356914
_cell_length_b    6.34844726874
_cell_length_c    5.95890979103
_cell_angle_alpha    115.332613602
_cell_angle_beta    97.337801947
_cell_angle_gamma    111.295350815

_symmetry_space_group_name_H-M    'P 1'
loop_
_atom_site_label
_atom_site_type_symbol
_atom_site_fract_x
_atom_site_fract_y
_atom_site_fract_z
_atom_site_occupancy
Ag1   Ag   0.717439215798294   0.745091058106541   0.225578653830176   1.0
Ag2   Ag   0.693273307093855   0.187790556224567   0.321812485266975   1.0
F3    F    0.922612852250668   0.433185755246204   0.693004999927261   1.0
F4    F    0.487757760098076   0.499752696426002   0.854411767409876   1.0
F5    F    0.140762498150627   0.142807550303874   0.338846216806264   1.0
F6    F    0.448555466049908   0.302883751124237   0.104602188833613   1.0
F7    F    0.269951584061828   0.790062255923347   0.208439328483770   1.0
F8    F    0.828416510367685   0.994459013894991   0.605246508692398   1.0
F9    F    0.962345033689492   0.630085611130780   0.442844290840969   1.0
F10   F    0.582105081028215   0.938437046132357   0.942081911799861   1.0
K11   K    0.029321434438817   0.853621567060144   0.890246655847660   1.0
K12   K    0.985173624031806   0.353862809639792   0.018278746194541   1.0
```

```
K13   K   0.425600907312806    0.578986831248195    0.529105374126695    1.0
K14   K   0.381263475627934    0.079375947538953    0.657136841939948    1.0

## DFT(PBEsol)+U K2AgF4 D p=80GPa magn.: af1
data_1
_audit_creation_method   'vasp2cif'
_cell_length_a    4.37218832545
_cell_length_b    6.25618938211
_cell_length_c    5.86989948305
_cell_angle_alpha    115.006091407
_cell_angle_beta    97.9636587135
_cell_angle_gamma    111.191247283

_symmetry_space_group_name_H-M    'P 1'
loop_
_atom_site_label
_atom_site_type_symbol
_atom_site_fract_x
_atom_site_fract_y
_atom_site_fract_z
_atom_site_occupancy
Ag1   Ag   0.717265088640230    0.742508996465642    0.214732790525779    1.0
Ag2   Ag   0.693001444371980    0.190527074825966    0.332802972671666    1.0
F3    F    0.932856914782475    0.442896424781336    0.706005598239799    1.0
F4    F    0.479213396819333    0.490182721836765    0.841385902872807    1.0
F5    F    0.146996564310331    0.148790913380591    0.340944502541645    1.0
F6    F    0.454554739255337    0.303937196567801    0.102275099848364    1.0
F7    F    0.263737379286281    0.784128592561217    0.206380224366838    1.0
F8    F    0.835648351760954   -0.000011540182070    0.600464630979400    1.0
F9    F    0.955616762387407    0.628868903741368    0.444944736706177    1.0
F10   F    0.574916957380143    0.932884623873894    0.947049598823530    1.0
K11   K    0.027479778700336    0.859978766959021    0.891462232472696    1.0
K12   K    0.984380475030834    0.346594154072391    0.022423497225579    1.0
K13   K    0.425905820464048    0.586341509478262    0.524972259978117    1.0
K14   K    0.382997076810323    0.072774111637797    0.655791922747618    1.0

## DFT(PBEsol)+U K2AgF4 D p=90GPa magn.: af1
data_1
_audit_creation_method   'vasp2cif'
_cell_length_a    4.35205244864
_cell_length_b    6.18417881155
_cell_length_c    5.80465209888
_cell_angle_alpha    114.871652973
_cell_angle_beta    98.3166255241
_cell_angle_gamma    111.134450988

_symmetry_space_group_name_H-M    'P 1'
loop_
_atom_site_label
_atom_site_type_symbol
_atom_site_fract_x
_atom_site_fract_y
_atom_site_fract_z
_atom_site_occupancy
Ag1   Ag   0.717953908600773    0.740390250595090    0.207652420461259    1.0
Ag2   Ag   0.692811986659894    0.192462954154223    0.339632456308880    1.0
F3    F    0.937476126447425    0.449176661444715    0.714971899869650    1.0
F4    F    0.472254322020844    0.483622138741458    0.832389507802993    1.0
F5    F    0.150066622597152    0.152386863809445    0.342180517446048    1.0
F6    F    0.458075991957010    0.305009354463064    0.100914688809328    1.0
F7    F    0.260713949362299    0.780542445834286    0.205222652327372    1.0
F8    F    0.841045748359026    0.003679334030828    0.597903156344319    1.0
F9    F    0.952786441182595    0.627945656361220    0.446511054118095    1.0
F10   F    0.569724032618673    0.929248889866803    0.949429289919987    1.0
K11   K    0.026348203178082    0.864370602983366    0.892364484682104    1.0
K12   K    0.983541149847808    0.341853240526525    0.024995990191543    1.0
```

```
K13   K   0.427313414926015   0.591082901959621   0.522417083744709   1.0
K14   K   0.384466852242409   0.068631155229350   0.655050767973720   1.0

## DFT(PBEsol)+U K2AgF4 D p=100GPa magn.: af1
data_1
_audit_creation_method   'vasp2cif'
_cell_length_a    4.46063020429
_cell_length_b    5.73288033334
_cell_length_c    5.82656981224
_cell_angle_alpha    114.671407494
_cell_angle_beta    105.752434609
_cell_angle_gamma    103.838224661

_symmetry_space_group_name_H-M    'P 1'
loop_
_atom_site_label
_atom_site_type_symbol
_atom_site_fract_x
_atom_site_fract_y
_atom_site_fract_z
_atom_site_occupancy
Ag1   Ag   0.481734752006825   0.435410876123862   0.153811087556267   1.0
Ag2   Ag   0.725792625644151   0.178899422544729   0.393764933532057   1.0
F3    F    0.603974837243718   0.307725303796986   0.774199870304816   1.0
F4    F    0.103325754266123   0.306723821621337   0.772947362996957   1.0
F5    F    0.223145882372065   0.165457624384597   0.317894850108775   1.0
F6    F    0.672641860253648   0.803153166038101   0.109902569614392   1.0
F7    F    0.985016446242332   0.449088018309156   0.229352893463409   1.0
F8    F    0.824418217628248   0.584107145979598   0.591967737559335   1.0
F9    F    0.536217017410060   0.811150965750048   0.437415543534397   1.0
F10   F    0.383902118889145   0.030689524574104   0.956669166517778   1.0
K11   K    0.850131383929952   0.029896225035966   0.890632215192537   1.0
K12   K    0.094379473396147   0.705613621559630   0.032036822721320   1.0
K13   K    0.114525526220054   0.908764482252682   0.515266597426622   1.0
K14   K    0.357851334497543   0.584553322029195   0.656774299471321   1.0

## DFT(PBEsol)+U K2AgF4 E p=000GPa magn.: f
data_1
_audit_creation_method   'vasp2cif'
_cell_length_a    7.29552613033
_cell_length_b    7.29559583315
_cell_length_c    8.20470904155
_cell_angle_alpha    90.0
_cell_angle_beta    90.0
_cell_angle_gamma    90.0000117875

_symmetry_space_group_name_H-M    'P 1'
loop_
_atom_site_label
_atom_site_type_symbol
_atom_site_fract_x
_atom_site_fract_y
_atom_site_fract_z
_atom_site_occupancy
Ag1   Ag   0.000000350000000   0.999999490000000   0.000000260000000   1.0
Ag2   Ag   0.000000129999997   0.499999740000000   0.750018114977003   1.0
Ag3   Ag   0.499999299999999   0.499999740000000   0.500001339999997   1.0
Ag4   Ag   0.499999410000001   0.999999490000000   0.249979735022997   1.0
F5    F    0.431323123396055   0.277080934817619   0.203149824573970   1.0
F6    F    0.568676846603942   0.722918545182381   0.203144994573968   1.0
F7    F    0.277049864817741   0.568656206604182   0.796852855426032   1.0
F8    F    0.722950095182256   0.431343283395819   0.796850435426030   1.0
F9    F    0.431623748460957   0.222902646008904   0.546436133760665   1.0
F10   F    0.568376211539046   0.777098193991094   0.546434933760665   1.0
F11   F    0.277120238874957   0.931621806510236   0.953551707552116   1.0
F12   F    0.722879721125040   0.068377543489767   0.953556537552110   1.0
```

```
F13  F  0.931350749902897  0.777046567774441  0.703165504627073  1.0
F14  F  0.068648530097107  0.222952912225558  0.703163084627079  1.0
F15  F  0.777075887774565  0.068669230097349  0.296837175372921  1.0
F16  F  0.222924072225439  0.931330659902656  0.296835965372928  1.0
F17  F  0.931636466510454  0.722867241125130  0.046456162447889  1.0
F18  F  0.068364313489546  0.277132238874877  0.046454592447887  1.0
F19  F  0.777105883991148  0.431638488460734  0.453553256239331  1.0
F20  F  0.222892716008850  0.568362351539264  0.453558086239333  1.0
K21  K  0.749932835690714  0.386455260708431  0.125128768930388  1.0
K22  K  0.250067124309283  0.613544219291569  0.125128768930388  1.0
K23  K  0.386468590708326  0.250061844309279  0.874873901069613  1.0
K24  K  0.613532729291670  0.749938995690719  0.874875111069613  1.0
K25  K  0.750065012405145  0.113524427443139  0.625121742512725  1.0
K26  K  0.249934947594859  0.886475192556858  0.625121742512725  1.0
K27  K  0.886463662556757  0.750071232405126  0.374874887487280  1.0
K28  K  0.113535087443246  0.249929607594872  0.374874887487280  1.0

## DFT(PBEsol)+U K2AgF4 E p=5GPa magn.: f
data_1
_audit_creation_method   'vasp2cif'
_cell_length_a    7.04805638551
_cell_length_b    7.0481678944
_cell_length_c    7.86214549337
_cell_angle_alpha    90.0
_cell_angle_beta     90.0
_cell_angle_gamma    90.0000195193

_symmetry_space_group_name_H-M    'P 1'
loop_
_atom_site_label
_atom_site_type_symbol
_atom_site_fract_x
_atom_site_fract_y
_atom_site_fract_z
_atom_site_occupancy
Ag1  Ag  0.000000350000000  0.999999490000000  0.000000260000000  1.0
Ag2  Ag  0.000000129999997  0.499999740000000  0.750018114977003  1.0
Ag3  Ag  0.499999299999999  0.499999740000000  0.500001339999997  1.0
Ag4  Ag  0.499999410000001  0.999999490000000  0.249979735022997  1.0
F5   F   0.428033069672981  0.286351806694919  0.207369288533380  1.0
F6   F   0.571966900327016  0.713647673305080  0.207364458533378  1.0
F7   F   0.286320743455274  0.571946258184811  0.792633391466622  1.0
F8   F   0.713679216544723  0.428053231815190  0.792630971466620  1.0
F9   F   0.428183499955689  0.213626361186451  0.542443843405678  1.0
F10  F   0.571816460044315  0.786374478813548  0.542442643405678  1.0
F11  F   0.286396530253757  0.928181560534682  0.957543997907102  1.0
F12  F   0.713603429746240  0.071817789465322  0.957548827907097  1.0
F13  F   0.928060696179823  0.786317439651742  0.707384968586483  1.0
F14  F   0.071938583820181  0.213682040348258  0.707382548586489  1.0
F15  F   0.786346766412098  0.071959281677978  0.292617711413511  1.0
F16  F   0.213653193587906  0.928040608322027  0.292616501413518  1.0
F17  F   0.928196218005186  0.713590956302675  0.042463872092902  1.0
F18  F   0.071804561994814  0.286408523697331  0.042462302092901  1.0
F19  F   0.786382175369947  0.428198242485180  0.457545546594318  1.0
F20  F   0.213616424630051  0.571802597514819  0.457550376594320  1.0
K21  K   0.749970359694185  0.395485355023882  0.125078727797480  1.0
K22  K   0.250029600305812  0.604514124976118  0.125078727797480  1.0
K23  K   0.395498693304102  0.250024320503472  0.874923942202521  1.0
K24  K   0.604502626695893  0.749976519496526  0.874925152202521  1.0
K25  K   0.750027488401674  0.104494333127687  0.625071701379817  1.0
K26  K   0.249972471598329  0.895505286872309  0.625071701379817  1.0
K27  K   0.895493765152533  0.750033708599320  0.374924928620188  1.0
K28  K   0.104504984847470  0.249967131400679  0.374924928620188  1.0

## DFT(PBEsol)+U K2AgF4 E p=10GPa magn.: f
data_1
```

```
_audit_creation_method   'vasp2cif'
_cell_length_a    6.91499294007
_cell_length_b    6.91513645352
_cell_length_c    7.63456028221
_cell_angle_alpha    90.0
_cell_angle_beta     90.0
_cell_angle_gamma    90.000025605

_symmetry_space_group_name_H-M    'P 1'
loop_
_atom_site_label
_atom_site_type_symbol
_atom_site_fract_x
_atom_site_fract_y
_atom_site_fract_z
_atom_site_occupancy
Ag1    Ag    0.000000350000000    0.999999490000000    0.000000260000000    1.0
Ag2    Ag    0.000000129999997    0.499999740000000    0.750018114977003    1.0
Ag3    Ag    0.499999299999999    0.499999740000000    0.500001339999997    1.0
Ag4    Ag    0.499999410000001    0.999999490000000    0.249979735022997    1.0
F5     F     0.426464721564651    0.290711304774392    0.208986031457294    1.0
F6     F     0.573535248435346    0.709288175225608    0.208981201457292    1.0
F7     F     0.290680247253805    0.573514604466359    0.791016648542708    1.0
F8     F     0.709319712746191    0.426484885533641    0.791014228542707    1.0
F9     F     0.426618297494413    0.209272089610027    0.540793734620813    1.0
F10    F     0.573381662505591    0.790728750389971    0.540792534620813    1.0
F11    F     0.290750807491928    0.926616360189930    0.959194106691967    1.0
F12    F     0.709249152508069    0.073382989810073    0.959198936691962    1.0
F13    F     0.926492348071494    0.790676937731214    0.709001711510397    1.0
F14    F     0.073506931928511    0.209322542268785    0.708999291510402    1.0
F15    F     0.790706270210630    0.073527627959526    0.291000968489598    1.0
F16    F     0.209293689789374    0.926472262040479    0.290999758489604    1.0
F17    F     0.926631015543909    0.709236684726252    0.040813763308038    1.0
F18    F     0.073369764456091    0.290762795273755    0.040812193308036    1.0
F19    F     0.790736452608118    0.426633042140428    0.459195655379183    1.0
F20    F     0.209262147391880    0.573367797859570    0.459200485379185    1.0
K21    K     0.749976471367559    0.397416371640815    0.125087079133688    1.0
K22    K     0.250023488632437    0.602583108549185    0.125087079133688    1.0
K23    K     0.397429714314018    0.250018208949826    0.874915590866313    1.0
K24    K     0.602571605685978    0.749982631050172    0.874916800866314    1.0
K25    K     0.750021376728300    0.102563316700754    0.625080052716025    1.0
K26    K     0.249978583271704    0.897436303299243    0.625080052716025    1.0
K27    K     0.897424786162449    0.750027597045674    0.374916577283980    1.0
K28    K     0.102573963837554    0.249973242954325    0.374916577283980    1.0

## DFT(PBEsol)+U K2AgF4 E p=20GPa magn.: f
data_1
_audit_creation_method   'vasp2cif'
_cell_length_a    6.74624748247
_cell_length_b    6.74645957172
_cell_length_c    7.33811590226
_cell_angle_alpha    90.0
_cell_angle_beta     90.0
_cell_angle_gamma    90.0000387863

_symmetry_space_group_name_H-M    'P 1'
loop_
_atom_site_label
_atom_site_type_symbol
_atom_site_fract_x
_atom_site_fract_y
_atom_site_fract_z
_atom_site_occupancy
Ag1    Ag    0.000000350000000    0.999999490000000    0.000000260000000    1.0
Ag2    Ag    0.000000129999997    0.499999740000000    0.750018114977003    1.0
Ag3    Ag    0.499999299999999    0.499999740000000    0.500001339999997    1.0
Ag4    Ag    0.499999410000001    0.999999490000000    0.249979735022997    1.0
```

```
F5    F    0.424863289812914    0.295784010084622    0.210542809256925    1.0
F6    F    0.575136680187083    0.704215469915377    0.210537979256923    1.0
F7    F    0.295752957781416    0.575116034794150    0.789459870743077    1.0
F8    F    0.704247002218581    0.424883455205851    0.789457450743075    1.0
F9    F    0.424999758338024    0.204207628338680    0.539216040593443    1.0
F10   F    0.575000201661980    0.795793211661319    0.539214840593443    1.0
F11   F    0.295815273911451    0.924997822706006    0.960771800719337    1.0
F12   F    0.704184686088545    0.075001527293997    0.960776630719332    1.0
F13   F    0.924890916319756    0.795749643041445    0.710558489310028    1.0
F14   F    0.075108363680248    0.204249836958555    0.710556069310034    1.0
F15   F    0.795778980738239    0.075129058287317    0.289444190689966    1.0
F16   F    0.204220979261764    0.924870831712688    0.289442980689973    1.0
F17   F    0.925012476387521    0.704172223454905    0.039236069280668    1.0
F18   F    0.074988303612480    0.295827256545102    0.039234499280665    1.0
F19   F    0.795800919027642    0.425014504656505    0.460773349406553    1.0
F20   F    0.204197680972356    0.574986335343494    0.460778179406555    1.0
K21   K    0.749989798846583    0.398144276245032    0.125093127380587    1.0
K22   K    0.250010161153414    0.601855203754968    0.125093127380587    1.0
K23   K    0.398157620129348    0.250004881509995    0.874909542619413    1.0
K24   K    0.601843699870647    0.749995958490004    0.874910752619414    1.0
K25   K    0.750008049249276    0.101835411906538    0.625086100962924    1.0
K26   K    0.249991910750727    0.898164208093459    0.625086100962924    1.0
K27   K    0.898152691977779    0.750014269605842    0.374910529037081    1.0
K28   K    0.101846058022224    0.249986570394156    0.374910529037081    1.0

## DFT(PBEsol)+U K2AgF4 E p=30GPa magn.: f
data_1
_audit_creation_method   'vasp2cif'
_cell_length_a    6.62626697711
_cell_length_b    6.62653483829
_cell_length_c    7.14047185947
_cell_angle_alpha    90.0
_cell_angle_beta    90.0
_cell_angle_gamma    90.0000498724

_symmetry_space_group_name_H-M    'P 1'
loop_
_atom_site_label
_atom_site_type_symbol
_atom_site_fract_x
_atom_site_fract_y
_atom_site_fract_z
_atom_site_occupancy
Ag1   Ag   0.000000350000000    0.999999490000000    0.000000260000000    1.0
Ag2   Ag   0.000000129999997    0.499999740000000    0.750018114977003    1.0
Ag3   Ag   0.499999299999999    0.499999740000000    0.500001339999997    1.0
Ag4   Ag   0.499999410000001    0.999999490000000    0.249979735022997    1.0
F5    F    0.423887057428475    0.298986319941485    0.211374925332251    1.0
F6    F    0.576112912571522    0.701013160058515    0.211370095332249    1.0
F7    F    0.298955266745638    0.576092268011926    0.788627754667751    1.0
F8    F    0.701044693254358    0.423907221988075    0.788625334667749    1.0
F9    F    0.424003954495883    0.201012127834478    0.538375071024165    1.0
F10   F    0.575996005504121    0.798988712165520    0.538373871024165    1.0
F11   F    0.299010773531614    0.924002018049623    0.961112770288615    1.0
F12   F    0.700989186468382    0.075997331950380    0.961117600288610    1.0
F13   F    0.923914683935317    0.798951952898307    0.711390605385354    1.0
F14   F    0.076084596064687    0.201047527101692    0.711388185385360    1.0
F15   F    0.798981289702463    0.076105291505093    0.288612074614641    1.0
F16   F    0.201018670297541    0.923894598494912    0.288610864614647    1.0
F17   F    0.924016672545380    0.700976722950703    0.038395099711389    1.0
F18   F    0.075984107454621    0.299022757049303    0.038393529711387    1.0
F19   F    0.798996418647805    0.424187000000121    0.461614318975831    1.0
F20   F    0.201002181352193    0.575982139999877    0.461619148975833    1.0
K21   K    0.750002229929628    0.398033930112648    0.125097596900221    1.0
K22   K    0.249997730070369    0.601965549887352    0.125097596900221    1.0
K23   K    0.398047271769312    0.249992450402554    0.874905073099779    1.0
K24   K    0.601954048230684    0.750008389597445    0.874906283099780    1.0
```

```
K25   K    0.749995618166231    0.101945758038921    0.625090570482558    1.0
K26   K    0.250004341833773    0.898053861961076    0.625090570482558    1.0
K27   K    0.898042343617743    0.750001838498401    0.374906059517447    1.0
K28   K    0.101956406382260    0.249999001501597    0.374906059517447    1.0

## DFT(PBEsol)+U K2AgF4 E p=40GPa magn.: f
data_1
_audit_creation_method   'vasp2cif'
_cell_length_a    6.53020644913
_cell_length_b    6.53052237033
_cell_length_c    6.99228012213
_cell_angle_alpha    90.0
_cell_angle_beta     90.0
_cell_angle_gamma    90.0000596856

_symmetry_space_group_name_H-M    'P 1'
loop_
_atom_site_label
_atom_site_type_symbol
_atom_site_fract_x
_atom_site_fract_y
_atom_site_fract_z
_atom_site_occupancy
Ag1   Ag   0.000000350000000    0.999999490000000    0.000000260000000    1.0
Ag2   Ag   0.000000129999997    0.499999740000000    0.750018114977003    1.0
Ag3   Ag   0.499999299999999    0.499999740000000    0.500001339999997    1.0
Ag4   Ag   0.499999410000001    0.999999490000000    0.249979735022997    1.0
F5    F    0.423211616105331    0.301251480872866    0.211834981750256    1.0
F6    F    0.576788353894667    0.698747999127134    0.211830151750254    1.0
F7    F    0.301220423408234    0.576767711481875    0.788167698249745    1.0
F8    F    0.698779536591763    0.423231778518125    0.788165278249744    1.0
F9    F    0.423311588947992    0.198756619960896    0.537907186859775    1.0
F10   F    0.576688371052012    0.801244220039103    0.537905986859775    1.0
F11   F    0.301266277173201    0.923309650262508    0.962080654453005    1.0
F12   F    0.698733682826795    0.076689699737495    0.962085484453000    1.0
F13   F    0.923239242612173    0.801217113829688    0.711850661803359    1.0
F14   F    0.076760037387832    0.198782366170312    0.711848241803365    1.0
F15   F    0.801246446365058    0.076780734975042    0.288152018196635    1.0
F16   F    0.198753513634946    0.923219155024963    0.288150808196642    1.0
F17   F    0.923324306997489    0.698721215077121    0.037927215546999    1.0
F18   F    0.076676473002511    0.301278264922886    0.037925645546997    1.0
F19   F    0.801251922289392    0.423326332213006    0.462082203140221    1.0
F20   F    0.198746677710606    0.576674507786992    0.462087033140223    1.0
K21   K    0.750012778273828    0.397714224627297    0.125103393347306    1.0
K22   K    0.249987181726168    0.602285255372702    0.125103393347306    1.0
K23   K    0.397727562808045    0.249981902023109    0.874899276652695    1.0
K24   K    0.602273757191950    0.750018937976890    0.874900486652695    1.0
K25   K    0.749985069822031    0.102265463524272    0.625096366929643    1.0
K26   K    0.250014890177973    0.897734156475725    0.625096366929643    1.0
K27   K    0.897722634656477    0.749991290118956    0.374900263070362    1.0
K28   K    0.102276115343526    0.250009549881042    0.374900263070362    1.0

## DFT(PBEsol)+U K2AgF4 E p=50GPa magn.: f
data_1
_audit_creation_method   'vasp2cif'
_cell_length_a    6.45023004
_cell_length_b    6.45023004
_cell_length_c    6.87494378
_cell_angle_alpha    90.0
_cell_angle_beta     90.0
_cell_angle_gamma    90.0

_symmetry_space_group_name_H-M    'P 1'
loop_
_atom_site_label
_atom_site_type_symbol
```

```
_atom_site_fract_x
_atom_site_fract_y
_atom_site_fract_z
_atom_site_occupancy
Ag1   Ag   0.999999279999997   0.999998830000003   0.000000499999999   1.0
Ag2   Ag   0.499999189999997   0.000000620000002   0.749999170000002   1.0
Ag3   Ag   0.499994400000000   0.500000829999998   0.500000219999997   1.0
Ag4   Ag   0.999999840000001   0.499999789999997   0.249998769999998   1.0
F5    F    0.302972750000002   0.577325860000002   0.212178250000001   1.0
F6    F    0.697027120000001   0.422674440000001   0.212176829999997   1.0
F7    F    0.577324490000002   0.697027679999998   0.787824469999997   1.0
F8    F    0.422673820000000   0.302973229999999   0.787819599999999   1.0
F9    F    0.197026760000000   0.577325819999999   0.537825259999998   1.0
F10   F    0.802972609999998   0.422674340000000   0.537820490000001   1.0
F11   F    0.922674120000003   0.697027319999997   0.962176780000000   1.0
F12   F    0.077324740000002   0.302973190000003   0.962177720000000   1.0
F13   F    0.802972760000003   0.077325930000001   0.712177539999999   1.0
F14   F    0.197027329999997   0.922674350000001   0.712176550000002   1.0
F15   F    0.077324699999998   0.197027079999998   0.287824350000001   1.0
F16   F    0.922675320000003   0.802973170000001   0.287819399999997   1.0
F17   F    0.697026530000002   0.077324519999998   0.037825329999997   1.0
F18   F    0.302972789999998   0.922674080000000   0.037820590000003   1.0
F19   F    0.422674989999997   0.197026909999998   0.462176630000002   1.0
F20   F    0.577324099999998   0.802973379999997   0.462178160000001   1.0
K21   K    0.397370100000003   0.249997780000001   0.124998939999998   1.0
K22   K    0.602632880000002   0.749998810000001   0.125000530000001   1.0
K23   K    0.249999610000003   0.602629839999999   0.874999940000002   1.0
K24   K    0.750000149999998   0.397376790000003   0.875003139999997   1.0
K25   K    0.102633140000002   0.249998900000001   0.624997620000002   1.0
K26   K    0.897370770000002   0.749999889999998   0.625000010000001   1.0
K27   K    0.749999570000000   0.102622719999999   0.375000120000003   1.0
K28   K    0.249999860000003   0.897372660000002   0.375001159999997   1.0

## DFT(PBEsol)+U K2AgF4 E p=60GPa magn.: f
data_1
_audit_creation_method   'vasp2cif'
_cell_length_a     6.38157727788
_cell_length_b     6.38157727788
_cell_length_c     6.77677544535
_cell_angle_alpha     90.0
_cell_angle_beta      90.0
_cell_angle_gamma     90.0

_symmetry_space_group_name_H-M    'P 1'
loop_
_atom_site_label
_atom_site_type_symbol
_atom_site_fract_x
_atom_site_fract_y
_atom_site_fract_z
_atom_site_occupancy
Ag1   Ag   0.999999279999997   0.999998830000003   0.000000499999999   1.0
Ag2   Ag   0.499999189999997   0.000000620000002   0.749999170000002   1.0
Ag3   Ag   0.499994400000000   0.500000829999998   0.500000219999997   1.0
Ag4   Ag   0.999999840000001   0.499999789999997   0.249998769999998   1.0
F5    F    0.304379286826447   0.577875046924722   0.212457256311097   1.0
F6    F    0.695620583173557   0.422125253075282   0.212455836311093   1.0
F7    F    0.577873676924722   0.695621143173553   0.787545463688901   1.0
F8    F    0.422124633075280   0.304379766826444   0.787540593688903   1.0
F9    F    0.195620223173555   0.577875006924718   0.537546253688903   1.0
F10   F    0.804379146826442   0.422125153075281   0.537541483688906   1.0
F11   F    0.922124933075284   0.695620783173552   0.962455786311096   1.0
F12   F    0.077873926924721   0.304379726826448   0.962456726311095   1.0
F13   F    0.804379296826448   0.077875116924720   0.712456546311095   1.0
F14   F    0.195620793173553   0.922125163075281   0.712455556311098   1.0
F15   F    0.077873886924718   0.195620543173553   0.287545343688905   1.0
F16   F    0.922126133075284   0.804379706826446   0.287540393688901   1.0
```

```
F17   F   0.695619993173557   0.077873706924717   0.037546323688901   1.0
F18   F   0.304379326826443   0.922124893075280   0.037541583688907   1.0
F19   F   0.422125803075278   0.195620373173553   0.462455636311098   1.0
F20   F   0.577873286924718   0.804379916826442   0.462457166311096   1.0
K21   K   0.396940258455496   0.249997780000001   0.124998939999998   1.0
K22   K   0.603062721544509   0.749998810000001   0.125000530000001   1.0
K23   K   0.249999610000003   0.603059681544507   0.874999940000002   1.0
K24   K   0.750000149999998   0.396946948455495   0.875003139999997   1.0
K25   K   0.103062981544510   0.249998900000001   0.624997620000002   1.0
K26   K   0.896940928455494   0.749999889999998   0.625000010000001   1.0
K27   K   0.749999570000000   0.103052561544507   0.375000120000003   1.0
K28   K   0.249999860000003   0.896942818455494   0.375001159999997   1.0

## DFT(PBEsol)+U K2AgF4 E p=70GPa magn.: f
data_1
_audit_creation_method    'vasp2cif'
_cell_length_a    6.32172145287
_cell_length_b    6.32172145287
_cell_length_c    6.69318356557
_cell_angle_alpha    90.0
_cell_angle_beta    90.0
_cell_angle_gamma    90.0

_symmetry_space_group_name_H-M    'P 1'
loop_
_atom_site_label
_atom_site_type_symbol
_atom_site_fract_x
_atom_site_fract_y
_atom_site_fract_z
_atom_site_occupancy
Ag1   Ag  0.999999279999997   0.999998830000003   0.000000499999999   1.0
Ag2   Ag  0.499999189999997   0.000000620000002   0.749999170000002   1.0
Ag3   Ag  0.499999400000000   0.500000829999998   0.500000219999997   1.0
Ag4   Ag  0.999999840000001   0.499999789999997   0.249998769999998   1.0
F5    F   0.305560242885351   0.578287680693276   0.212591294701972   1.0
F6    F   0.694439627114652   0.421712619306728   0.212589874701969   1.0
F7    F   0.578286310693276   0.694440187114649   0.787411425298025   1.0
F8    F   0.421711999306726   0.305560722885348   0.787406555298027   1.0
F9    F   0.194439267114651   0.578287640693273   0.537412215298027   1.0
F10   F   0.805560102885347   0.421712519306727   0.537407445298030   1.0
F11   F   0.921712299306730   0.694439827114648   0.962589824701972   1.0
F12   F   0.078286560693275   0.305560682885352   0.962590764701971   1.0
F13   F   0.805560252885352   0.078287750693274   0.712590584701971   1.0
F14   F   0.194439837114649   0.921712529306727   0.712589594701974   1.0
F15   F   0.078286520693272   0.194439587114649   0.287411305298030   1.0
F16   F   0.921713499306729   0.805560662885350   0.287406355298025   1.0
F17   F   0.694439037114653   0.078286340693272   0.037412285298025   1.0
F18   F   0.305560282885347   0.921712259306726   0.037407545298031   1.0
F19   F   0.421713169306723   0.194439417114649   0.462589674701973   1.0
F20   F   0.578285920693272   0.805560872885346   0.462591204701972   1.0
K21   K   0.396605945551140   0.249997780000001   0.124998939999998   1.0
K22   K   0.603397034448865   0.749998810000001   0.125000530000001   1.0
K23   K   0.249999610000003   0.603393994448863   0.874999940000002   1.0
K24   K   0.750000149999998   0.396612635551139   0.875003139999997   1.0
K25   K   0.103397294448866   0.249998900000001   0.624997620000002   1.0
K26   K   0.896606615551138   0.749999889999998   0.625000010000001   1.0
K27   K   0.749999570000000   0.103386874448863   0.375000120000003   1.0
K28   K   0.249999860000003   0.896608505551138   0.375001159999997   1.0

## DFT(PBEsol)+U K2AgF4 E p=80GPa magn.: f
data_1
_audit_creation_method    'vasp2cif'
_cell_length_a    6.26796462945
_cell_length_b    6.26796462945
_cell_length_c    6.6197549436
```

```
_cell_angle_alpha     90.0
_cell_angle_beta      90.0
_cell_angle_gamma     90.0

_symmetry_space_group_name_H-M    'P 1'
loop_
_atom_site_label
_atom_site_type_symbol
_atom_site_fract_x
_atom_site_fract_y
_atom_site_fract_z
_atom_site_occupancy
Ag1   Ag   0.999999279999997   0.999998830000003   0.000000499999999   1.0
Ag2   Ag   0.499999189999997   0.000000620000002   0.749999170000002   1.0
Ag3   Ag   0.499994400000000   0.500000829999998   0.500000219999997   1.0
Ag4   Ag   0.999999840000001   0.499999789999997   0.249998769999998   1.0
F5    F    0.306541431792701   0.578662174448761   0.212704204385709   1.0
F6    F    0.693458438207302   0.421338125551243   0.212702784385705   1.0
F7    F    0.578660804448761   0.693458998207299   0.787298515614289   1.0
F8    F    0.421337505551241   0.306541911792699   0.787293645614291   1.0
F9    F    0.193458078207301   0.578662134448757   0.537299305614291   1.0
F10   F    0.806541291792697   0.421338025551242   0.537294535614294   1.0
F11   F    0.921337805551245   0.693458638207297   0.962702734385708   1.0
F12   F    0.078661054448760   0.306541871792702   0.962703674385707   1.0
F13   F    0.806541441792702   0.078662244448759   0.712703494385707   1.0
F14   F    0.193458648207298   0.921338035551242   0.712702504385710   1.0
F15   F    0.078661014448757   0.193458398207299   0.287298395614293   1.0
F16   F    0.921339005551244   0.806541851792701   0.287293445614289   1.0
F17   F    0.693457848207303   0.078660834448756   0.037299375614289   1.0
F18   F    0.306541471792698   0.921337765551241   0.037294635614295   1.0
F19   F    0.421338675551239   0.193458228207299   0.462702584385709   1.0
F20   F    0.578660414448757   0.806542061792697   0.462704114385708   1.0
K21   K    0.396309509034550   0.249997780000001   0.124998939999998   1.0
K22   K    0.603693470965455   0.749998810000001   0.125000530000001   1.0
K23   K    0.249999610000003   0.603690430965453   0.874999940000002   1.0
K24   K    0.750000149999998   0.396316199034549   0.875003139999997   1.0
K25   K    0.103693730965456   0.249998900000001   0.624997620000002   1.0
K26   K    0.896310179034548   0.749998899999998   0.625000010000001   1.0
K27   K    0.749999570000000   0.103683310965453   0.375000120000003   1.0
K28   K    0.249999860000003   0.896312069034548   0.375001159999997   1.0

## DFT(PBEsol)+U K2AgF4 E p=90GPa magn.: f
data_1
_audit_creation_method   'vasp2cif'
_cell_length_a    6.21945438976
_cell_length_b    6.21945438976
_cell_length_c    6.55330490116
_cell_angle_alpha     90.0
_cell_angle_beta      90.0
_cell_angle_gamma     90.0

_symmetry_space_group_name_H-M    'P 1'
loop_
_atom_site_label
_atom_site_type_symbol
_atom_site_fract_x
_atom_site_fract_y
_atom_site_fract_z
_atom_site_occupancy
Ag1   Ag   0.999999279999997   0.999998830000003   0.000000499999999   1.0
Ag2   Ag   0.499999189999997   0.000000620000002   0.749999170000002   1.0
Ag3   Ag   0.499994400000000   0.500000829999998   0.500000219999997   1.0
Ag4   Ag   0.999999840000001   0.499999789999997   0.249998769999998   1.0
F5    F    0.307365625253413   0.578951197591636   0.212785133669533   1.0
F6    F    0.692634244746590   0.421049102408367   0.212783713669529   1.0
F7    F    0.578949827591637   0.692634804746587   0.787217586330464   1.0
F8    F    0.421048482408365   0.307366105253411   0.787212716330466   1.0
```

```
F9    F   0.192633884746589   0.578951157591633   0.537218376330466   1.0
F10   F   0.807365485253409   0.421049002408366   0.537213606330469   1.0
F11   F   0.921048782408369   0.692634444746585   0.962783663669532   1.0
F12   F   0.078950077591636   0.307366065253414   0.962784603669532   1.0
F13   F   0.807365635253414   0.078951267591635   0.712784423669531   1.0
F14   F   0.192634454746586   0.921049012408367   0.712783433669535   1.0
F15   F   0.078950037591633   0.192634204746587   0.287217466330469   1.0
F16   F   0.921049982408369   0.807366045253413   0.287212516330464   1.0
F17   F   0.692633654746591   0.078949857591632   0.037218446330465   1.0
F18   F   0.307365665253410   0.921048742408366   0.037213706330470   1.0
F19   F   0.421049652408363   0.192634034746587   0.462783513669534   1.0
F20   F   0.578949437591633   0.807366255253409   0.462785043669533   1.0
K21   K   0.395967305886173   0.249997780000001   0.124998939999998   1.0
K22   K   0.604035674113832   0.749998810000001   0.125000530000001   1.0
K23   K   0.249999610000003   0.604032634113829   0.874999940000002   1.0
K24   K   0.750000149999998   0.395973995886172   0.875003139999997   1.0
K25   K   0.104035934113832   0.249998900000001   0.624997620000002   1.0
K26   K   0.895967975886171   0.749999889999998   0.625000010000001   1.0
K27   K   0.749999570000000   0.104025514113830   0.375000120000003   1.0
K28   K   0.249999860000003   0.895969865886172   0.375001159999997   1.0

## DFT(PBEsol)+U K2AgF4 E p=100GPa magn.: f
data_1
_audit_creation_method   'vasp2cif'
_cell_length_a    6.1753690001
_cell_length_b    6.1753690001
_cell_length_c    6.49468370629
_cell_angle_alpha    90.0
_cell_angle_beta     90.0
_cell_angle_gamma    90.0

_symmetry_space_group_name_H-M    'P 1'
loop_
_atom_site_label
_atom_site_type_symbol
_atom_site_fract_x
_atom_site_fract_y
_atom_site_fract_z
_atom_site_occupancy
Ag1   Ag  0.999999279999997   0.999998830000003   0.000000499999999   1.0
Ag2   Ag  0.499999189999997   0.000000620000002   0.749999170000002   1.0
Ag3   Ag  0.499994400000000   0.500000829999998   0.500000219999997   1.0
Ag4   Ag  0.999999840000001   0.499999789999997   0.249998769999998   1.0
F5    F   0.308091862510026   0.579234673149415   0.212845228924799   1.0
F6    F   0.691908007489977   0.420765626850588   0.212843808924796   1.0
F7    F   0.579233303149416   0.691908567489973   0.787157491075198   1.0
F8    F   0.420765006850587   0.308092342510023   0.787152621075200   1.0
F9    F   0.191907647489976   0.579234633149412   0.537158281075200   1.0
F10   F   0.808091722510022   0.420765526850587   0.537153511075203   1.0
F11   F   0.920765306850590   0.691908207489972   0.962843758924799   1.0
F12   F   0.079233553149415   0.308092302510027   0.962844698924798   1.0
F13   F   0.808091872510027   0.079234743149414   0.712844518924798   1.0
F14   F   0.191908217489973   0.920765536850588   0.712843528924801   1.0
F15   F   0.079233513149412   0.191907967489974   0.287157371075202   1.0
F16   F   0.920766506850590   0.808092282510026   0.287152421075198   1.0
F17   F   0.691907417489978   0.079233333149411   0.037158351075198   1.0
F18   F   0.308091902510023   0.920765266850587   0.037153611075204   1.0
F19   F   0.420766176850584   0.191907797489974   0.462843608924800   1.0
F20   F   0.579232913149412   0.808092492510022   0.462845138924799   1.0
K21   K   0.395691747326412   0.249997780000001   0.124998939999998   1.0
K22   K   0.604311232673593   0.749998810000001   0.125000530000001   1.0
K23   K   0.249999610000003   0.604308192673590   0.874999940000002   1.0
K24   K   0.750000149999998   0.395698437326412   0.875003139999997   1.0
K25   K   0.104311492673593   0.249998900000001   0.624997620000002   1.0
K26   K   0.895692417326411   0.749999889999998   0.625000010000001   1.0
K27   K   0.749999570000000   0.104301072673591   0.375000120000003   1.0
K28   K   0.249999860000003   0.895694307326411   0.375001159999997   1.0
```